\documentclass[aps,prd,superscriptaddress,twocolumn,preprintnumbers,nofootinbib,amsmath,amssymb]{revtex4-2}
\usepackage{graphicx}
\usepackage{xcolor}
\usepackage[normalem] {ulem}

\usepackage[pdftitle={Center vortices and localized Dirac
  modes in the deconfined phase of (2+1)-dimensional lattice $Z_2$ gauge theory}]{hyperref}

\usepackage{mathrsfs,multirow}  
\newcommand{\f}[2]{\frac{#1}{#2}}
\newcommand{\tf}[2]{{\textstyle\f{#1}{#2}}}

\newcommand{\la}{\langle}

\newcommand{\ra}{\rangle}

\newcommand{\tr}{{\rm tr}\,}

\begin{document}

\title{Center vortices and localized Dirac modes in the deconfined
phase of (2+1)-dimensional lattice $\mathbb{Z}_2$ gauge theory}

\author{Gy{\"o}rgy Baranka}
\email{barankagy@caesar.elte.hu}
\affiliation{Institute of Physics and Astronomy, ELTE E\"otv\"os
  Lor\'and University, P\'azm\'any P\'eter s\'et\'any 1/A, H-1117,
  Budapest, Hungary}

\author{D\'enes Berta}
\email{denes.berta@ttk.elte.hu}
\affiliation{Institute of Physics and Astronomy, ELTE E\"otv\"os
  Lor\'and University, P\'azm\'any P\'eter s\'et\'any 1/A, H-1117,
  Budapest, Hungary}

\author{Matteo Giordano}
\email{giordano@bodri.elte.hu}
\affiliation{Institute of Physics and Astronomy, ELTE E\"otv\"os
  Lor\'and University, P\'azm\'any P\'eter s\'et\'any 1/A, H-1117,
  Budapest, Hungary}

\date{\today}
\begin{abstract}
  We study the deconfinement transition in (2+1)-dimensional lattice
  $\mathbb{Z}_2$ gauge theory both as a percolation transition of
  center vortices and as a localization transition for the low-lying
  Dirac modes. We study in detail the critical properties of the
  Anderson transition in the Dirac spectrum in the deconfined phase,
  showing that it is of BKT type; and the critical properties of the
  center-vortex percolation transition, showing that they differ from
  those of ordinary two-dimensional percolation. We then study the
  relation between localized modes and center vortices in the
  deconfined phase, identifying the simple center-vortex structures
  that mainly support the localized Dirac modes. As the system
  transitions to the confined phase, center vortices merge together
  into an infinite cluster, causing the low Dirac modes to delocalize.
\end{abstract}

\maketitle

\section{Introduction}
\label{sec:intro}

A growing amount of evidence shows that the localization properties of
low-lying Dirac modes in a gauge theory are strongly connected with
its confining properties: Low Dirac modes, while extended in the
confined phase of a gauge theory, become localized in the deconfined
phase, doing so exactly at the critical temperature if the phase
transition is a genuine thermodynamic transition, and within the
crossover region otherwise~\cite{GarciaGarcia:2005vj,
  GarciaGarcia:2006gr,Kovacs:2009zj,Kovacs:2010wx,Bruckmann:2011cc,
  Kovacs:2012zq,Giordano:2013taa,Nishigaki:2013uya,Ujfalusi:2015nha,
  Cossu:2016scb,Giordano:2016nuu,Kovacs:2017uiz,
  Holicki:2018sms,Giordano:2019pvc,Vig:2020pgq,Bonati:2020lal,
  Kovacs:2021fwq,Cardinali:2021fpu,Baranka:2021san,Baranka:2022dib,
  Kehr:2023wrs,Baranka:2023ani,Bonanno:2023mzj} (see
Ref.~\cite{Giordano:2021qav} for a recent review). More precisely, in
the deconfined phase the localized low Dirac modes are separated from
the delocalized bulk modes by a critical point in the spectrum known
as ``mobility edge''. At the mobility edge the localization length
diverges and a phase transition along the spectrum, known as Anderson
transition~\cite{Anderson:1958vr,thouless1974electrons,
  lee1985disordered,kramer1993localization,Evers:2008zz}, takes
place. In the cases investigated so far, as the temperature decreases
the mobility edge is pushed towards the origin of the spectrum, and
localized modes disappear from the spectrum (possibly
abruptly~\cite{Kovacs:2021fwq,Bonati:2020lal}) as one crosses over to
the confined phase. While the existence of a relation between
localization and deconfinement is by now well established, the nature
of this relation has not been fully elucidated yet.

The connection between deconfinement and localization of low Dirac
modes in gauge theories is qualitatively understood in terms of the
``sea/islands'' picture proposed in Ref.~\cite{Bruckmann:2011cc},
further developed in Refs.~\cite{Giordano:2015vla,Giordano:2016cjs,
  Giordano:2016vhx,Giordano:2021qav}, and refined in
Ref.~\cite{Baranka:2022dib}. In the deconfined phase, in the
``physical sector'' where Polyakov loops order along 1, selected by
dynamical fermions or, in the case of pure gauge theories, by
(infinitely) heavy external fermionic probes, a pseudogap opens in the
Dirac spectrum due to the ordering of the Polyakov loop.  In the
resulting ``sea'' of ordered local Polyakov loops, gauge-field
fluctuations that reduce the correlation across time slices, such as
Polyakov-loop fluctuations away from 1, are localized on ``islands'',
typically well-separated in space. These fluctuations reduce the
ability of Dirac modes to diffuse in space, and can ``trap'' the Dirac
modes if these develop an eigenvalue below the gap, localizing the
corresponding eigenvectors. This mechanism applies in the deconfined
phase of a generic gauge theory, explaining the observed universal
nature of the connection between localization and deconfinement; and,
more generally, in phases where the Polyakov loop gets ordered, e.g.,
in the Higgs phase of the $\mathrm{SU}(2)$ Higgs
model~\cite{Baranka:2023ani}.

The connection between localization and deconfinement shows up even in
the simplest gauge theory with a deconfining transition, namely pure
$\mathbb{Z}_2$ gauge theory in (2+1) dimensions~\cite{Baranka:2021san}.
This model is dual to an anisotropic three-dimensional Ising
model~\cite{Wipf:2013vp} (summed over boundary conditions: see
Refs.~\cite{Caselle:1995wn,vonSmekal:2012vx} and references therein,
and Ref.~\cite{Baranka:2022dib} for a detailed derivation), which
allows for an accurate determination of the critical value of the
gauge coupling~\cite{Caselle:1995wn} through the use of a cluster
algorithm~\cite{Swendsen:1987ce,Wolff:1988uh}. While localization of
low Dirac modes and its appearance at deconfinement were clearly
demonstrated in Ref.~\cite{Baranka:2021san}, a precise determination
of the mobility edge and of its temperature dependence, as well as a
detailed study of the corresponding Anderson transition, were not
undertaken.

Gauge configurations in an Abelian model like the $\mathbb{Z}_2$ gauge
theory are entirely determined by the simplest plaquettes and by the
Polyakov loops in the various directions (with spatial Polyakov loops
playing no important dynamical role in the large-volume limit). In the
$\mathbb{Z}_2$ gauge theory, plaquette configurations on the direct
lattice are in one-to-one correspondence with configurations of closed
loops on the dual lattice, constructed by associating an ``active''
dual link to each nontrivial plaquette.  Dual loops can further
intersect each other, and are conveniently grouped into disconnected
clusters of various sizes. Since the gauge group is Abelian and
therefore equal to its own center, these clusters are trivially the
same as the center vortices~\cite{Greensite:2003bk} of this model, and
so we will refer to them also as center vortices in the
following. Center vortices can be constructed for more general gauge
groups as well, and have been proposed as the origin of confinement in
gauge theories in Ref.~\cite{tHooft:1977nqb}. The center-vortex
picture of confinement has been the subject of intense research
activity~\cite{DelDebbio:1996lih,Engelhardt:1998wu,Langfeld:1998cz,Engelhardt:1999fd,
  Engelhardt:2003wm,Gattnar:2004gx,Bornyakov:2007fz,Hollwieser:2008tq,Bornyakov:2008bg}
that is still ongoing (see, e.g.,
Refs.~\cite{Junior:2022bol,Sale:2022qfn,Biddle:2023lod,Kamleh:2023gho,Dehghan:2024rly}).
We could do no justice to the vast related literature: we refer the
interested reader to Ref.~\cite{Greensite:2003bk} for an extensive
discussion of this picture, and to
Refs.~\cite{Junior:2022bol,Sale:2022qfn,Biddle:2023lod,
  Kamleh:2023gho,Dehghan:2024rly} for recent developments and updated
lists of references.
  
The center-vortex picture of confinement is particularly simple in the
case of $\mathbb{Z}_2$ gauge theory, where the sum of the sizes of the
center-vortex clusters equals the number of negative plaquettes of the
gauge configuration on the direct lattice, and one can exactly recast
the partition function in terms of dual clusters. The dynamics of
$\mathbb{Z}_2$ gauge fields is then the dynamics of center
vortices. In this language, confinement can be understood as
center-vortex percolation, i.e., as the formation of a percolating
cluster of dual active links, that leads to a nonvanishing string
tension. Conversely, deconfinement corresponds to the disappearance of
the percolating cluster, with only clusters of finite size surviving
the large-volume limit.\footnote{Notice that since a plaquette
  configuration is left unchanged by a center transformation, the
  center sector selected by the Polyakov loop, as it gets ordered and
  breaks center symmetry, plays no role in this dynamics.} The idea of
looking at the confinement-deconfinement transition as a center-vortex
percolation transition was proposed in
Refs.~\cite{Langfeld:1998cz,Engelhardt:1999fd}, where
finite-temperature pure $\mathrm{SU}(2)$ gauge theory in (3+1)
dimensions was investigated. Reference~\cite{Gliozzi:2002ht} studied
instead zero-temperature $\mathbb{Z}_2$ gauge theory in (2+1)
dimensions, showing the presence of an infinite center-cluster in
typical gauge configurations in the confined phase, and demonstrating
its crucial role in establishing confinement. On the other hand, the
behavior of clusters across the (bulk) deconfinement
transition~\cite{Balian:1974ir,Bhanot:1980pc} was not investigated.
This was done more recently and in a different context in
Ref.~\cite{Agrawal:2024mra}, that reports on a detailed study of
percolation and its relation with the thermodynamic transition.

Pure $\mathbb{Z}_2$ gauge theory in (2+1) dimensions is clearly the
easiest setup in which one can study the nature of the gauge-field
fluctuations relevant to localization, and how these relate to the
confining properties of the theory. The connection between
deconfinement and localization on the one hand, and that between
confinement and percolation on the other, suggest that the
nonpercolating center clusters expected in the deconfined phase
should play an important role for localization.  In this paper we show
that this is indeed the case. Of course, establishing a connection
between localized Dirac modes and nonpercolating center clusters in
the present model has no direct bearing on the role played by vortices
in the confinement mechanism in the physically relevant case of QCD.
Nonetheless, it shows that at least in this model the features of
gauge-field configurations responsible for the
confinement-deconfinement transition are exactly the same responsible
for the corresponding change in the localization properties of Dirac
modes. Identifying the gauge-field structures that drive this change
across the deconfinement transition in QCD and in other gauge theories
could then be of guidance in the search for the correct microscopic
mechanism behind confinement in realistic theories.

The relation between center vortices and the Dirac spectrum has been
studied in pure $\mathrm{SU}(2)$ gauge theory in (3+1) dimensions both
at zero~\cite{Gattnar:2004gx,Hollwieser:2008tq} and nonzero
temperature~\cite{Bornyakov:2007fz,Hollwieser:2008tq,Bornyakov:2008bg},
making use of the overlap discretization of the Dirac
operator~\cite{Neuberger:1997fp} (as well as of asqtad
fermions~\cite{MILC:2009mpl} in Ref.~\cite{Hollwieser:2008tq}). As the
main focus of those studies is the spectral density of the modes
(although localization properties were also investigated in
Refs.~\cite{Hollwieser:2008tq,Bornyakov:2008bg}) and its connection
with chiral symmetry breaking, using overlap fermions makes a
difference due to their superior chiral properties. Here we are
interested instead in the localization properties of Dirac modes and
their relation with confinement, and for this kind of study the use of
a discretization with good chiral properties is not required. Indeed,
localization of low Dirac modes in the deconfined phase has been
observed with a variety of fermion discretizations, with the same
picture resulting from the use of chiral and nonchiral
discretizations (see Ref.~\cite{Giordano:2021qav} for a review). A
particularly clear example is pure $\mathrm{SU}(3)$ gauge theory in
(3+1) dimensions, where localized modes appear exactly at the
first-order deconfinement transition both in the
staggered~\cite{Kovacs:2017uiz} and in the overlap
spectrum~\cite{Vig:2020pgq,Kovacs:2021fwq}.

In this paper we investigate in detail the localization of Dirac
modes, the percolation of center vortices, and how they affect each
other in finite-temperature pure $\mathbb{Z}_2$ gauge theory in (2+1)
dimensions on the lattice, probed with external staggered fermions. In
Sec.~\ref{sec:z2recap}, after briefly introducing the model in
Sec.~\ref{sec:z2model} and staggered fermions in
Sec.~\ref{sec:stagDspec}, we discuss a few generalities about center
vortices in Sec.~\ref{sec:cv}, and about localization and the
sea/islands picture in Secs.~\ref{sec:eloc} and \ref{sec:refsi}. In
Sec.~\ref{sec:num} we present our numerical results. We first
determine the nature of the Anderson transition and the temperature
dependence of the mobility edge in Sec.~\ref{sec:me}. We then study
the percolation of center vortices across the deconfinement transition
in Sec.~\ref{sec:cp} and the correlation between center clusters and
localized modes in Sec.~\ref{sec:eccorr}, and test the refined
sea/islands picture of Ref.~\cite{Baranka:2022dib} in
Sec.~\ref{sec:numrefsi}. We then identify the features of center
vortices mostly relevant to localization in
Sec.~\ref{sec:spconf}. Finally, in Sec.~\ref{sec:concl} we present our
conclusions and discuss prospects for future studies. A brief
discussion of some unexplained features of the staggered spectrum
noted in Ref.~\cite{Baranka:2021san} is reported in
Appendix~\ref{sec:vhsing}.

\section{Localization and center vortices in $\mathbb{Z}_2$ lattice
  gauge theory at finite temperature}
\label{sec:z2recap}

In this section, after briefly describing (2+1)-dimensional pure
$\mathbb{Z}_2$ gauge theory on the lattice and how to probe it with
external staggered fermions, we summarize the relevant aspects of
center vortices and their percolation, and of Dirac mode localization.

\subsection{Finite-temperature $\mathbb{Z}_2$ lattice gauge theory in
  (2+1) dimensions}
\label{sec:z2model}

We consider pure $\mathbb{Z}_2$ gauge theory on a (2+1)-dimensional
hypercubic $N_t\times N_s^2$ lattice. Lattice sites are denoted by
$n=(n_1,n_2,n_3)=(t,\vec{x})$, where $0\le t=n_1\le N_t-1$,
$0\le x_{1,2}=n_{2,3}\le N_s-1$. The dynamical variables of the model
are the link variables $U_\mu(n)=\pm 1$, taking values in
$\mathbb{Z}_2$, and associated with the oriented lattice links
connecting a site $n$ with its neighbors $n+\hat{\mu}$, where
$\mu=1,2,3$ and $\hat{\mu}$ is the unit lattice vector in direction
$\mu$. Periodic boundary conditions are imposed in all directions.

The relevant gauge-invariant observables are the plaquette variables
$U_{\mu\nu}(n)$ associated with the elementary lattice squares,
\begin{equation}
  \label{eq:z2_plaq}
  \begin{aligned}
  U_{\mu\nu}(n)&\equiv U_\mu(n)U_\nu(n+\hat{\mu})U_\mu(n+\hat{\nu})^{-1}U_\nu(n)^{-1}
\\ & =U_\mu(n)U_\nu(n+\hat{\mu})U_\mu(n+\hat{\nu})U_\nu(n)\,, 
  \end{aligned}
\end{equation}
and the Polyakov loops winding around the temporal direction, 
\begin{equation}
  \label{eq:z2_pol}
  P(\vec{x})\equiv\prod_{t=0}^{N_t-1}U_1(t,\vec{x})\,.
\end{equation}
Together with the analogs of $P(\vec{x})$ winding around the spatial
directions, these quantities determine the link configuration up to a
gauge transformation.

Expectation values of observables $O[U]$ are defined as
\begin{equation}
  \label{eq:z2rec3bis}
  \begin{aligned}
  \la O[U] \ra &=  Z^{-1} \sum_{\{U_\mu(n)=\pm 1\}} e^{-S[U]} \,O[U]\,,\\
  Z &= \sum_{\{U_\mu(n)=\pm 1\}} e^{-S[U]}\,,
  \end{aligned}
\end{equation}
where $Z$ is the partition function, the sum is over all possible
gauge configurations $\{U_\mu(n)\}$, and $S$ is taken to be the Wilson
action, which for this model reads (up to an irrelevant additive
constant)
\begin{equation}
  \label{eq:z2rec1}
  S[U] = -\beta\sum_n \sum_{\substack{\mu,\nu=1\\ \mu<\nu}}^3 U_{\mu\nu}(n)\,,
\end{equation}
where $\beta$ is the dimensionless lattice coupling,
$\beta= 1/(e^2 a)$, with $e$ the coupling constant of mass dimension
$1/2$ and $a$ the lattice spacing.

We work at finite temperature $T=1/(a N_t)$, taking the thermodynamic
limit by sending the lattice volume $V=N_s^2\to\infty$ at fixed
$N_t$. We set the temperature by varying $\beta$ at fixed $N_t$, so in
this approach the lattice coupling and the temperature are essentially
identified, i.e., $T/e^2 = \beta /N_t$ (up to scaling violations that
we ignore for simplicity). The system undergoes a second-order phase
transition at a critical coupling $\beta_c=\beta_c(N_t)$, from a
low-temperature confined phase ($\beta<\beta_c$) to a high-temperature
deconfined phase ($\beta>\beta_c$). The values of $\beta_c$ for
several temporal sizes in lattice units have been determined in
Ref.~\cite{Caselle:1995wn} exploiting the duality with the
three-dimensional Ising model. In the deconfined phase the
$\mathbb{Z}_2$ center symmetry is spontaneously broken; we select the
sector where $P(\vec{x})$ prefers the value 1 (physical sector), i.e.,
configurations where the spatially averaged Polykov loop
$\bar{P}\equiv\f{1}{V}\sum_{\vec{x}}P(\vec{x})$ is positive.

\subsection{Staggered Dirac spectrum}
\label{sec:stagDspec}

We probe the gauge-field configurations using external fermion fields
by studying the spectrum of the staggered Dirac operator, which for
(2+1)-dimensional gauge theories reads
\begin{equation}
  \label{eq:z2rec5}
  D^{\rm stag} = \f{1}{2}\sum_{\mu=1}^3
  \eta_\mu(U_\mu T_\mu - T_\mu^\dag U_\mu^\dag)\,, 
\end{equation}
where $T_\mu$ are translation operators,
$(T_\mu)_{n,n'}=\delta_{n+\hat{\mu},n'}$, with periodic boundary
conditions in the spatial directions and antiperiodic boundary
conditions in the temporal direction,
$(U_\mu)_{n,n'}=U_\mu(n)\delta_{n,n'}$, and
$(\eta_\mu)_{n,n'}=\eta_\mu(n)\delta_{n,n'}$ with
$\eta_\mu(n) = (-1)^{\sum_{\nu<\mu}n_\nu}$.  Of course, for
$\mathbb{Z}_2$ gauge fields $U_\mu^\dag = U_\mu^*= U_\mu$. The
spectrum of $D^{\rm stag}$ is purely imaginary thanks to the
anti-Hermiticity of $D^{\rm stag}$,
\begin{equation}
  \label{eq:z2rec5bis}
  D^{\rm stag} \psi_l = i\lambda_l \psi_l\,, \qquad
  \lambda_l\in\mathbb{R}\,, 
\end{equation}
where the eigenvectors $\psi_l$ are normalized as usual,
$\sum_n |\psi_l(n)|^2= 1$. The spectrum is also symmetric about zero
since
\begin{equation}
  \label{eq:z2rec5ter}
  \begin{aligned}
  D^{\rm stag} \varepsilon\psi_l &= -i\lambda_l
  \varepsilon\psi_l\,,\\
  (\varepsilon)_{n'n}&=  \varepsilon(n)\delta_{n'n}\,,\qquad
  \varepsilon(n)\equiv (-1)^{\sum_{\mu=1}^3 n_\mu}\,,
  \end{aligned}
\end{equation}
so it suffices to focus on the positive spectrum only. Notice that for
free staggered fermions the eigenvalues are
\begin{equation}
  \label{eq:spec_free}
   \lambda^{\mathrm{free}}(\omega,\vec{p}\,) = \pm\sqrt{ (\sin\omega)^2 + (\sin p_1)^2+ (\sin
  p_2)^2}\,,
\end{equation}
where $\omega = \f{(2k_0+1)\pi}{N_t}$ with $k_0=0,\ldots,N_t-1$ are
the Matsu\-ba\-ra frequencies of the system, and
$\vec{p}=(p_1,p_2) = \f{2\pi \vec{k}}{N_s}$ with
$k_{1,2}=0,\ldots,N_s-1$ are the allowed spatial momenta (in lattice
units). The positive spectrum is then entirely contained in the finite
interval $[\lambda_{(0)},\lambda_{(1)}]$ with
$\lambda_{(0)} = \sin\frac{\pi}{N_t}$ the lowest lattice Matsubara
frequency, and
$\lambda_{(1)} = \sqrt{\max_\omega (\sin \omega)^2 + 2}$. We will
refer to this region as the bulk of the spectrum, and to modes in the
regions $0\le \lambda<\lambda_{(0)}$ and $\lambda_{(1)}<\lambda\le 3$
as low and high modes, respectively.

Observables related to the spectrum and the eigenvectors of
$D^{\mathrm{stag}}$ are computed locally in the spectrum by averaging
within (in principle infinitesimal) spectral bins and over
configurations,
\begin{equation}
  \label{eq:genav}
  \overline{O}(\lambda;N_s) \equiv \f{\la \sum_l \delta(\lambda - \lambda_l)
    O_l \ra}{ \la \sum_l \delta(\lambda - \lambda_l) \ra}\,,
\end{equation}
where $O_l$ denotes any eigenmode observable evaluated for mode $l$,
and we have made explicit the dependence of the average on the spatial
size $N_s$ of the system. The denominator in Eq.~\eqref{eq:genav} is
the spectral density, i.e., the local density of modes, that grows
proportionally to the lattice volume. The suitably normalized spectral
density,
\begin{equation}
  \label{eq:spdens_def}
\rho(\lambda) \equiv \f{1}{N_t V} \left\la{\textstyle\sum_l}
    \delta(\lambda-\lambda_l)\right\ra\,,  
\end{equation}
has then a well-defined thermodynamic limit. Finally, we define the
scaling dimension $d_O$ of the spectral quantities
$\overline{O}(\lambda;N_s)$ via
\begin{equation}
  \label{eq:genscaldim}
  \begin{aligned}
  d_O(N_s',N_s) &=
  \log\f{\overline{O}(\lambda;N_s')}{\overline{O}(\lambda;N_s)}\bigg/
  \log\frac{N_s^\prime}{N_s}\,, \\
  d_O&=\lim_{\substack{N_s,N_s'\to\infty\\N_s\neq 
      N_s'}} d_O(N_s',N_s)\,.
  \end{aligned}
\end{equation}

\subsection{Center vortices}
\label{sec:cv}

The sum over the configurations of link variables $U_\mu(n)$ defining
the partition function of $\mathbb{Z}_2$ gauge theory defines also a
percolation problem for the plaquette variables $U_{\mu\nu}(n)$. This
is most easily formulated on the dual lattice, where each dual link
corresponds to one elementary lattice square on the direct lattice,
namely the one it pierces in its center.  One then defines a dual
lattice link variable equal to the plaquette variable on the
corresponding direct lattice square. ``Active bonds'', i.e., negative
dual link variables, must form closed loops: since $\mathbb{Z}_2$ is
Abelian, the product of the plaquettes surrounding every elementary
direct-lattice cube must equal one, and so an even number of active
dual links must meet at each dual-lattice site. These closed, possibly
self-intersecting loops provide an unambiguous definition of clusters
of dual links, disconnected from each other. In turn, the sites
touched by the dual links in such a cluster form an unambiguously and
uniquely identified cluster of dual sites. We will then refer to the
union of a dual-link cluster and the associated dual-site cluster
simply as a cluster, or as a center vortex.

One can then ask if in typical gauge configurations one finds an
infinite center vortex, i.e., a cluster whose size grows linearly with
the lattice volume. This is the percolation problem we are interested
in. As argued in Ref.~\cite{Engelhardt:1999fd}, the presence of such a
cluster would lead to a finite string tension and so to a confining
theory. That this happens in the $\mathbb{Z}_2$ gauge theory was shown
in Ref.~\cite{Gliozzi:2002ht}. The cluster size can be measured by the
number of links, or alternatively by the number of sites it
contains. We choose the latter, and denote with $\mathcal{S}$ the
fraction of dual sites occupied by the largest cluster in a given
gauge configuration, i.e.,
\begin{equation}
  \label{eq:sclustsize}
  \mathcal{S} = \f{\text{no.~of dual sites in the largest
      cluster}}{N_t\times N_s^2}\,.
\end{equation}
The expectation value of $\mathcal{S}$ depends on the lattice coupling
$\beta$ through the probability $\alpha$ of activating a dual-lattice
link, that in turn equals the probability of finding a negative
plaquette on the direct-lattice gauge configuration, i.e.,
\begin{equation}
  \label{eq:probnegplaq}
  \alpha \equiv \f{1-\la U_{\mu\nu}\ra}{2}\,.
\end{equation}
This quantity sets the concentration of active bonds in the
corresponding percolation problem. Notice that the use of periodic
boundary conditions in space and the topological constraints on the
bond configurations, as well as the presence of an additional (but
finite) dimension, make it a different problem from usual percolation
in two dimensions (see Refs.~\cite{RevModPhys.54.235,
  percolation,smirnov2001,10.1214/11-AOP740}).

To study the correlation between center vortices, defined on the dual
lattice, and staggered eigenmodes, defined on the direct lattice, one
can proceed as follows. For each dual-link cluster $C$ we define a
corresponding size $S(C)$ on the direct lattice by counting the number
of sites of the direct lattice touched by the negative plaquettes
corresponding to the dual links belonging to $C$. If a site $n$ on the
direct lattice is touched by negative plaquettes belonging to $N$
different clusters, we add a contribution $s(n,C)=1/N$ to the size of
each of the $N$ clusters; otherwise we set $s(n,C)=0$. In formulas,
\begin{equation}
  \label{eq:dclsize}
  S(C)=\sum_n s(n,C)\,.  
\end{equation}
We then measure the weight $W(l,C)$ of the $l$th eigenmode on cluster
$C$ by adding up the amplitude squared $|\psi_l(n)|^2$, weighted by
the contribution $s(n,C)$ of $n$ to the cluster,
\begin{equation}
  \label{eq:clmw}
  W(l,C)=\sum_n|\psi_l(n)|^2s(n,C)\,.  
\end{equation}
For each mode $\psi_l$ we then identify the cluster
$C_{\mathrm{max}}(l)$ carrying the largest fraction of the mode's
weight, i.e.,
\begin{equation}
  \label{eq:clmaxw}
  \max_{C}W(l,C) = W(l,C_{\mathrm{max}}(l))=W_{\mathrm{max}}(l)\,. 
\end{equation}
The size of this mode is denoted with
$S_{\mathrm{max}}(l)=S(C_{\mathrm{max}}(l))$. The density of mode $l$
on this cluster is denoted with
\begin{equation}
  \label{eq:clmdens}
  D_{\mathrm{max}}(l) = \f{W_{\mathrm{max}}(l)}{S_{\mathrm{max}}(l)}\,.  
\end{equation}
To estimate on how many clusters a mode is concentrated, we use the
``cluster IPR'', defined as
\begin{equation}
  \label{eq:clipr}
  \mathrm{IPR}_{\mathrm{clust}}(l) = \f{\sum_C W(l,C)^2}{\left(\sum_C
      W(l,C)\right)^2}\,.   
\end{equation}
In fact, for a mode whose weight on negative plaquettes lies entirely
in a single cluster, $\mathrm{IPR}_{\mathrm{clust}}=1$. For a mode
whose weight on negative plaquettes is spread out evenly on $N$
clusters, one finds instead $\mathrm{IPR}_{\mathrm{clust}}=1/N$.

\subsection{Eigenmode localization}
\label{sec:eloc}

The staggered operator, used here to probe the pure $\mathbb{Z}_2$
gauge theory, is technically a random matrix with entries distributed
according to the probability distribution of the gauge links implied
by the partition function, Eq.~\eqref{eq:z2rec3bis}. Its eigenvalues
and eigenvectors are then also random variables, whose statistical
properties are studied by averaging over the ``disorder realizations''
provided by the gauge link configurations.

The localization properties of eigenmodes can be determined from the
volume scaling of their inverse participation ratio
(IPR)~\cite{thouless1974electrons,
  lee1985disordered,kramer1993localization,Evers:2008zz},
\begin{equation}
  \label{eq:PR}
  \mathrm{IPR}_l \equiv \sum_{n} |\psi_l(n)|^4\,,
\end{equation}
averaged over gauge configurations and over (infinitesimal) spectral
bins [see Eq.~\eqref{eq:genav}],
\begin{equation}
  \label{eq:PR2}
  \overline{\mathrm{IPR}}(\lambda,N_s)= \f{\la \sum_l
    \delta(\lambda-\lambda_l) \mathrm{IPR}_l\ra}{\la \sum_l
    \delta(\lambda-\lambda_l)\ra}\,. 
\end{equation}
If modes in a certain spectral region are delocalized over the entire
lattice volume, then $|\psi_l(n)|^2\sim \f{1}{VN_t}$ and so
$\overline{\mathrm{IPR}}\to 0$ as $V\to \infty$. If they are instead
localized in a finite spatial region whose typical size $V_0$ is
independent of the system size, then $|\psi_l(n)|^2\sim \f{1}{V_0N_t}$
inside that region and is negligible outside, and so
$\overline{\mathrm{IPR}}$ tends to a finite constant as $V\to \infty$.
Instead of $\overline{\mathrm{IPR}}$ it is sometimes convenient to use
the average mode size $\overline{\mathrm{IPR}^{-1}}$, or the average
participation ratio
$\overline{\mathrm{PR}}=\overline{\mathrm{IPR}^{-1}}/(N_tV)$, which
measures the fraction of the system occupied by the modes. Since
$|\psi_l(n)| = |\varepsilon(n)\psi_l(n)|$, the eigenmodes $\psi_l$ and
$\varepsilon\psi_l$ have the same $\mathrm{IPR}$, and so it suffices
to study the positive part of the spectrum.

The localization properties of the eigenvectors of a random matrix are
conveniently determined exploiting their connection with the
statistical properties of the corresponding
eigenvalues~\cite{altshuler1986repulsion}. Here we discuss the subject
only briefly; a detailed presentation can be found in
Ref.~\cite{Giordano:2021qav}. For delocalized modes, the universal
statistical properties of the spectrum are expected to be described by
the appropriate symmetry class of Random Matrix Theory
(RMT)~\cite{mehta2004random,Verbaarschot:2000dy}. For localized modes,
instead, eigenvalue fluctuations are expected to be described by
Poisson statistics~\cite{mehta2004random}. Analytic predictions are
available in both cases for the properties of the so-called {\it
  unfolded spectrum}, obtained from the original spectrum by means of
a monotonic mapping that makes the (unfolded) spectral density equal
to 1 throughout the spectrum, 
\begin{equation}
  \label{eq:unfolding}
  \lambda_i \to x_i =
  N_tV \int_{\lambda_{\mathrm{min}}}^{\lambda_i}d\lambda\,\rho(\lambda)\,,
\end{equation}
where the normalized spectral density
$\rho(\lambda)$ is defined in Eq.~\eqref{eq:spdens_def}. In
particular, the probability distribution
$p(s)$ of the unfolded level spacings
$s_i=x_{i+1}-x_i$ is known analytically. For Poisson statistics,
\begin{equation}
  \label{eq:unfolding2}
  p_{\rm P}(s) = e^{-s}\,,
\end{equation}
while for RMT statistics the corresponding $p_{\rm RMT}$ depends on
the symmetry class, and is not available in closed form.

The localization properties of modes in a given spectral region can
then be determined by comparing the spectral statistics of the
unfolded spectrum of the model, computed locally in the spectrum, with
the theoretical predictions: As the system size increases, these
observables will tend to the Poisson or RMT theoretical expectation
depending on whether the modes are localized or delocalized in the
spectral region under consideration. A point in the spectrum
separating localized and delocalized modes is known as
\textit{mobility edge} and usually denoted $\lambda_c$. At a mobility
edge the localization length $\xi$ diverges, typically as a power law,
$\xi \sim |\lambda-\lambda_c|^{-\nu}$, and the system displays a
second-order phase transition along the spectrum known as
\textit{Anderson transition}~\cite{Anderson:1958vr,
  thouless1974electrons,lee1985disordered,kramer1993localization,
  Evers:2008zz}.  This implies that at the mobility edge the spectral
statistics are volume independent. Moreover, one can precisely
determine the mobility edge, the critical statistics, and the critical
exponent $\nu$ by means of a finite-size-scaling analysis of spectral
statistics~\cite{Shklovskii:1993zz}.

In the case at hand, the random matrix of interest is
$H=-iD^{\rm stag}$, which is invariant under the antiunitary
transformation $T=\varepsilon K$, where $K$ denotes complex
conjugation and $\varepsilon$ is defined in
Eq.~\eqref{eq:z2rec5ter}. Since $T^2=1$, $H$ belongs to the orthogonal
class.\footnote{The random matrix of interest satisfies also the
  chiral property $\{\varepsilon,H\}=0$ with $[\varepsilon,T]=0$,
  which puts it in the chiral orthogonal class for what concerns the
  statistical behavior of eigenvalues near zero. This distinction does
  not, however, affect the statistical properties in the bulk of the
  spectrum.} The corresponding unfolded level spacing distribution can
be approximated quite accurately with the so-called orthogonal Wigner
surmise, $p_{\rm WS}$~\cite{mehta2004random}. An even better
approximation is obtained using the family of distributions
$p_{\rm B}(s,\omega)$, $\omega\in[0,1]$, known as Brody
distributions~\cite{brody1973statistical},
\begin{equation}
  \label{eq:brody}
  \begin{aligned}
    p_{\rm B}(s,\omega) &= a(\omega) s^\omega
    e^{-\f{a(\omega)}{1+\omega}
      s^{1+\omega}}\,,\\
    a(\omega)&=
    \left[\Gamma\left(\tf{1}{1+\omega}\right)\right]^{1+\omega}(1+\omega)^{-\omega}\,, 
  \end{aligned}
\end{equation}
which interpolates between $p_{\rm P}(s)=p_{\rm B}(s,0)$ and
$p_{\rm WS}(s)=p_{\rm B}(s,1)$. In
Ref.~\cite{scaramazza2016integrable} it was found that the optimal
$\omega$ to approximate $p_{\rm RMT}$ is
$\omega\approx 0.957\equiv\omega_0$ rather than $\omega=
1$. Correspondingly, one finds $ a(\omega_0) \simeq 1.546$.

To monitor the transition from localized to delocalized modes, it is
convenient to look at observables extracted from the unfolded level
spacing distribution, computed locally in the spectrum, and compare
them with the theoretical predictions. A useful observable in this
context is the integrated probability
density~\cite{Shklovskii:1993zz}, $I_{s_0}$, that for Brody
distributions reads
\begin{equation}
  \label{eq:brody4}
  I_{s_0}(\omega) = \int_0^{s_0}ds\,p_{\rm B}(s,\omega) = 1-
  e^{-\f{a(\omega)}{1+\omega}s_0^{1+\omega}}\,. 
\end{equation}
To maximize the difference between the values of $I_{s_0}$ in the
Poisson and RMT regimes, one takes $s_0\simeq 0.4687$, which is the
first crossing point of the exponential distribution and the Brody
distribution with $\omega=\omega_0$. Correspondingly, one finds
$I_{s_0\,{\rm RMT}}\simeq I_{s_0}(\omega_0)\simeq 0.1642$, and
$I_{s_0\,{\rm P}}=I_{s_0}(0)\simeq 0.3742$. These values should be
compared with the local estimate
\begin{equation}
  \label{eq:local_Is0}
  \overline{I}_{s_0}(\lambda,N_s) = \f{\la\sum_l \delta(\lambda-\lambda_l)
    \theta(s_0-s_l)  \ra}{\la\sum_l \delta(\lambda-\lambda_l)\ra}\,,
\end{equation}
where $\theta(x)$ is the Heaviside step function.

Anderson transitions observed in certain (spatially) two-dimensional
systems in the unitary and orthogonal classes are of Berezinski{\u
  \i}-Kosterlitz-Thouless (BKT)
type~\cite{PhysRevB.48.11095,PhysRevLett.72.1886,
  PhysRevB.60.5295,Wen-ShengLiu_1999,xie1998kosterlitz,
  zhang2009localization,Giordano:2019pvc}, with a localization length
diverging exponentially at the mobility edge as
\begin{equation}
  \label{eq:loclength}
  \xi \sim \xi_0\exp \left\{A|\lambda-\lambda_c|^{-\f{1}{2}}\right\}\,.
\end{equation}
Moreover, all points in the spectrum beyond the mobility edge are
critical, with volume-independent spectral statistical
properties. Indications of this behavior for the $\mathbb{Z}_2$ model
were found in Ref.~\cite{Baranka:2021san}. Similar indications were
found also in (2+1)-dimensional pure $\mathbb{Z}_3$ gauge theory in
Ref.~\cite{Baranka:2022dib}. The divergence of the localization length
shown in Eq.~\eqref{eq:loclength} leads one to expect the following
scaling behavior for $I_{s_0}$ near the mobility edge,
\begin{equation}
  \label{eq:scaling}
  \overline{I}_{s_0}(\lambda,L)  = f\left((\lambda-\lambda_c)(\log L/L_0)^{\f{1}{\nu}}\right)\,,
\end{equation}
with $f$ analytic, for $\nu=\f{1}{2}$, and some constant scale $L_0$.

\subsection{Sea/islands picture and Dirac-Anderson Hamiltonian}
\label{sec:refsi}

For the staggered Dirac operator the sea/islands picture briefly
sketched in the Introduction can be understood more precisely in the
language of the Dirac-Anderson
Hamiltonian~\cite{Giordano:2016cjs,Giordano:2016vhx,Giordano:2021qav,Baranka:2022dib}.
In ($d+1$) dimensions, in the basis where the temporal part of the
staggered operator is diagonal, the ``Hamiltonian''
$-iD^{\mathrm{stag}}$ reads
$H^{\rm DA}=-i\Omega^\dag D^{\mathrm{stag}}\Omega$, where $\Omega$ is
a suitable unitary change of basis and~\cite{Baranka:2022dib}
\begin{equation}
  \label{eq:DA_general}
  \begin{aligned}
    H^{\rm DA} &= \mathcal{E} 
    + \f{1}{2i}\sum_{j=1}^d \eta_j\left[\mathcal{V}_jT_j
      -T_j{}^\dag\mathcal{V}_j^\dag\right]\,,\\
\mathcal{E} &=        \begin{pmatrix}
      E & \mathbf{0} \\ \mathbf{0} & -E
    \end{pmatrix}\,,\quad
\mathcal{V}_j=      \begin{pmatrix}
        A_j & B_j \\ B_j & A_j
      \end{pmatrix}\,,
    \end{aligned}
\end{equation}
where $E$, $A_j$, and $B_j$ are $\vec{x}$-dependent
$\f{N_t}{2}\times \f{N_t}{2}$ matrices, with $E$ diagonal and
$\mathcal{V}_j$ unitary, and $T_j$ is the translation operator in
direction $j$ (with periodic boundary conditions understood), defined
under Eq.~\eqref{eq:z2rec5}. The matrix $E$ depends only on the local
Polyakov loops, while $A_j$ and $B_j$ are obtained by a discrete
temporal Fourier transform of the spatial links, with frequencies
related to the local Polyakov loops on neighboring sites.  The precise
form of the various matrices is given in
Ref.~\cite{Baranka:2022dib}. There, it is argued that locations where
the $A_j$s are larger are found where the correlation among gauge
fields on different time slices is reduced, and that they are
favorable places for low (and high) modes to live on. Fluctuations
reducing the temporal correlation become rare and typically separated
in space in the deconfined phase, so they become favorable places for
localization. In other words, locations with larger $A_j$s provide the
islands of the sea/islands picture. To detect them one can use the
quantity
\begin{equation}
  \label{eq:newsi_meas}
  \begin{aligned}
  {\cal A}(\vec{x}) &\equiv 
  \frac{1}{N_t d}
  \sum_{j=1}^d \tr\left\{
    A_j(\vec{x}){}^\dag A_j(\vec{x})  \right. \\ & \hphantom{\equiv 
  \frac{1}{N_t d}
  \sum_{j=1}^d \tr}\left. + 
    A_j(\vec{x}-\hat\jmath){}^\dag
    A_j(\vec{x}-\hat\jmath) \right\}\,, 
  \end{aligned}
\end{equation}
normalized so that $ 0 \le {\cal A}(\vec{x})\le 1$.
One can study their correlation with the staggered eigenmodes by
measuring ${\cal A}(\vec{x})$ ``as seen by the modes'' normalized by
its average, i.e.,
\begin{equation}
  \label{eq:modwA}
  {\cal A}_\psi(\lambda)\equiv
  \f{  \left\la \sum_l \delta(\lambda-\lambda_l) 
      \sum_{\vec{x}}   {\cal A}(\vec{x})
      \sum_{t=0}^{N_t-1}|\psi_l(t,\vec{x})|^2\right\ra}
  {  \left\la \sum_l \delta(\lambda-\lambda_l) \right\ra \la {\cal A}(\vec{x})\ra}
  \,.
\end{equation}
It is instructive to study also the correlation between islands and
Polyakov loops 
by using the quantity
\begin{equation}
  \label{eq:AP_alt2}
  C_{\mathcal{A}P}\equiv \f{\left\la
      \mathcal{A}(\vec{x})\f{1-\mathrm{sgn}(\bar{P})
        P(\vec{x})}{2}
    \right\ra}{\left\la\mathcal{A}(\vec{x})\right\ra 
    \f{1-\left\la|\bar{P}|\right\ra}{2}}\,. 
\end{equation}
This measures the average $\mathcal{A}$ on sites where the local
Polyakov loop, $P(\vec{x})$, has sign opposite to that of the
spatially averaged Polyakov loop, $\bar{P}$, normalized by the average
value of $\mathcal{A}$ and by the probability of finding such a
Polyakov-loop fluctuation.

\section{Numerical results}
\label{sec:num}

\begin{figure}[t]
  \centering
  \includegraphics[width=0.48\textwidth]{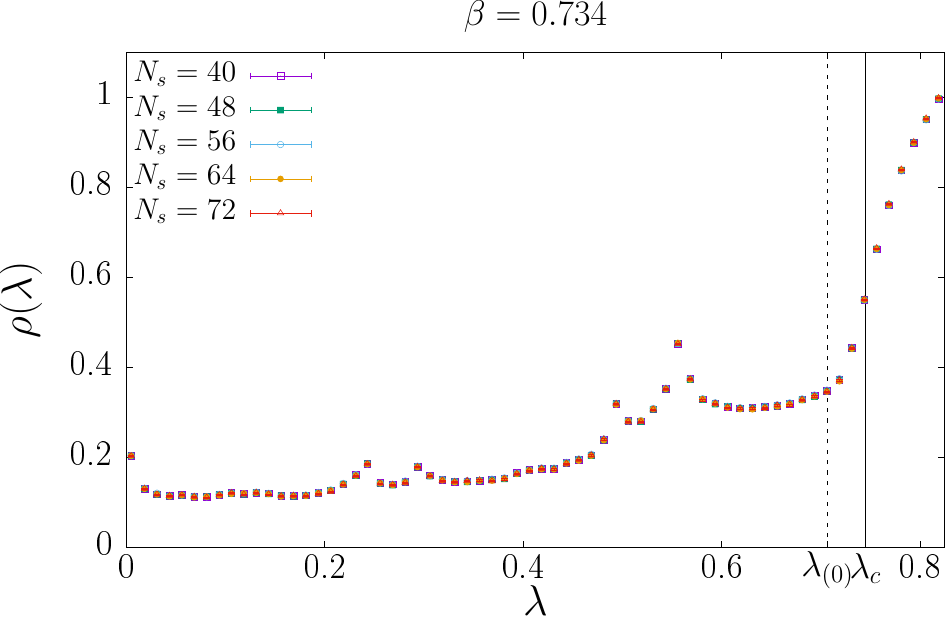}

  \vspace{0.5cm}
  \includegraphics[width=0.48\textwidth]{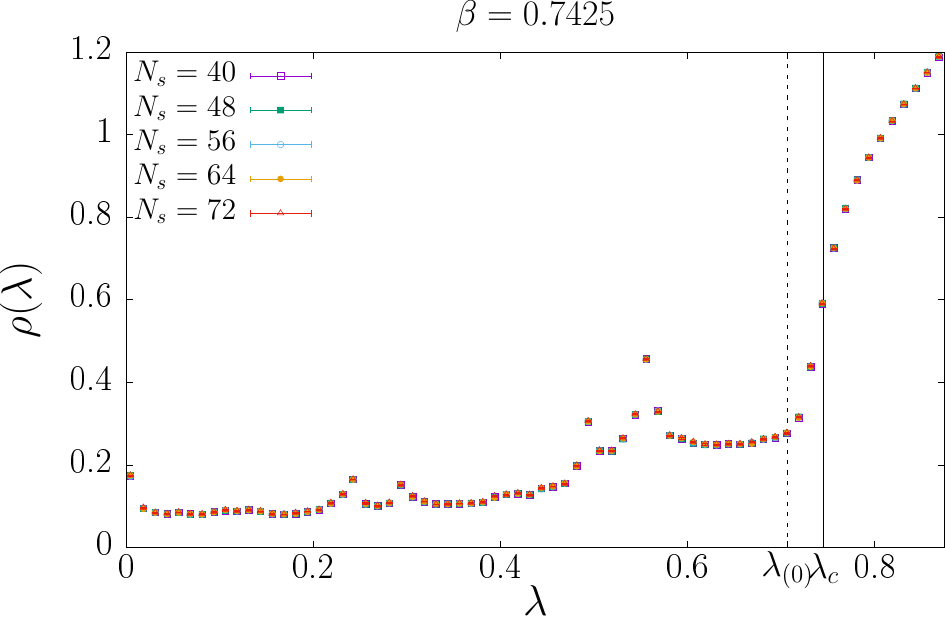}
  \caption{Normalized spectral density $\rho(\lambda)$,
    Eq.~\eqref{eq:spdens_def}, in the deconfined phase at
    $\beta=0.734$ (top panel) and $\beta=0.7425$ (bottom panel), for
    $N_t=4$ and different spatial system sizes. The low end of the
    bulk, $\lambda_{(0)}$, and the mobility edge, $\lambda_c$, are
    also shown.}
  \label{fig:spd}
\end{figure}
\begin{figure}[th]
  \centering
  \includegraphics[width=0.48\textwidth]{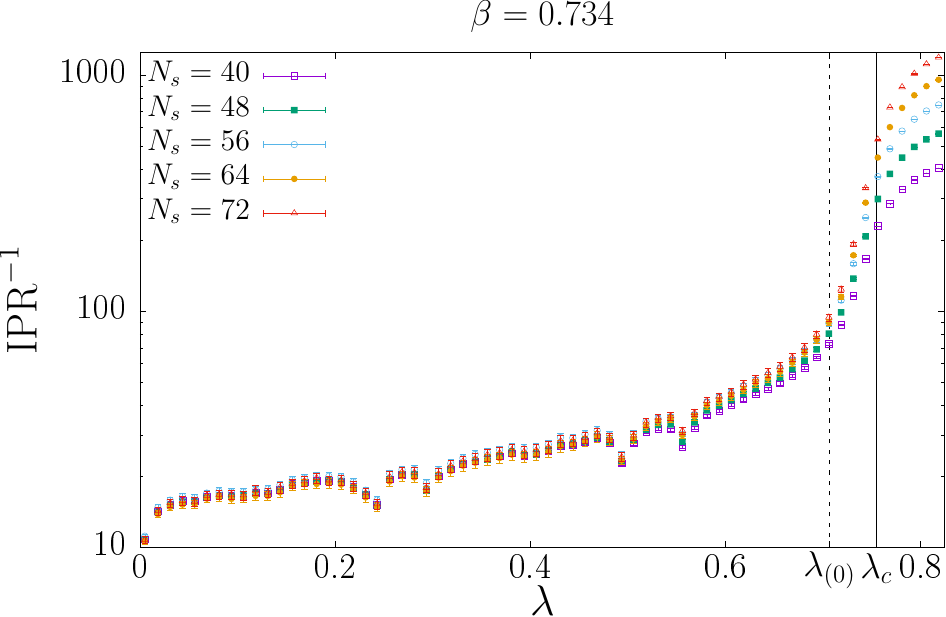}

  \vspace{0.5cm}
  \includegraphics[width=0.48\textwidth]{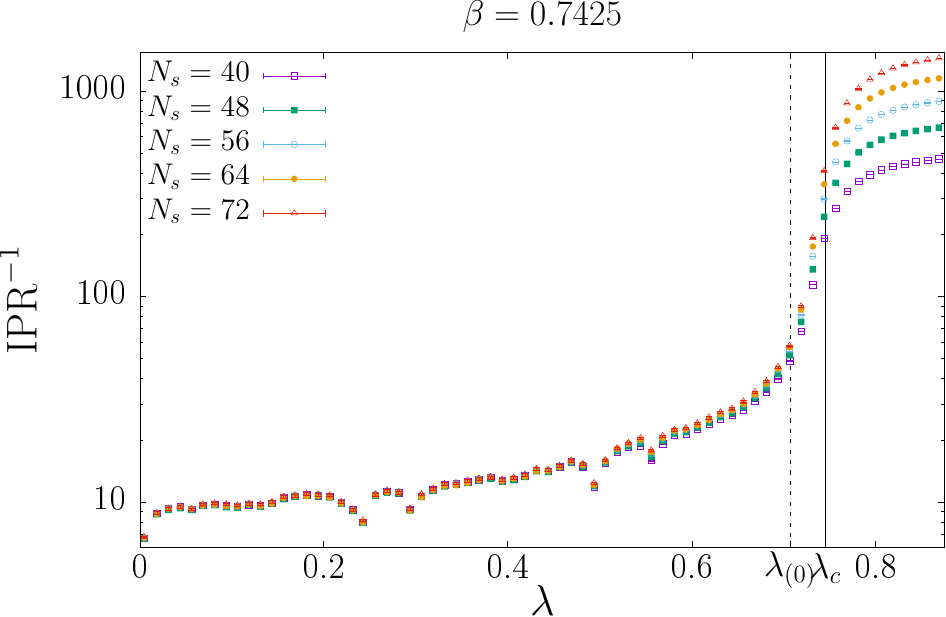}
  \caption{Mode size $\mathrm{IPR}^{-1}$ (in logarithmic scale) at
    $\beta=0.734$ (top panel) and $\beta=0.7425$ (bottom panel), for
    $N_t=4$ and different spatial system sizes. The low end of the
    bulk, $\lambda_{(0)}$, and the mobility edge, $\lambda_c$, are
    also shown.}
  \label{fig:PR}
\end{figure}

We performed numerical simulations of finite-temperature
$\mathbb{Z}_2$ gauge theory on hypercubic $N_t\times N_s^2$ lattices
with fixed $N_t=4$ and several values of $N_s$ and $\beta$, using a
standard Metropolis algorithm. Details about the values of $N_s$ and
$\beta$, as well as about the number of generated configurations, are
given below. For $N_t=4$ the critical coupling separating the confined
and deconfined phases is $\beta_c=0.73107(2)$~\cite{Caselle:1995wn},
and the bulk of the staggered spectrum is bounded by
$\lambda_{(0)}={1}/{\sqrt{2}}$ and $\lambda_{(1)}=\sqrt{{5}/{2}}$ (see
above in Sec.~\ref{sec:stagDspec}).

\subsection{Mobility edge}
\label{sec:me}

\begin{figure}[t]
  \centering
  \includegraphics[width=0.48\textwidth]{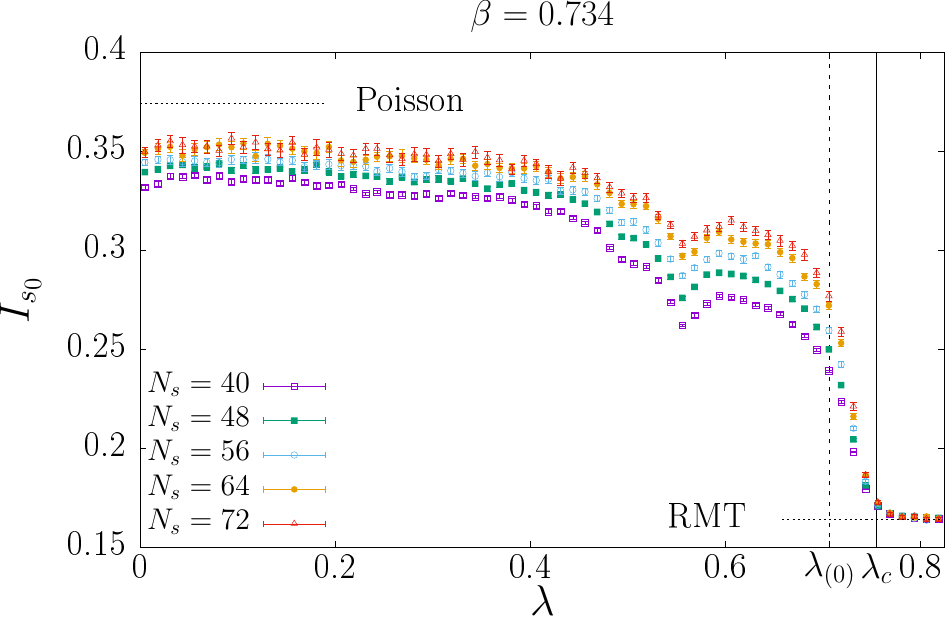}

  \vspace{0.5cm}
  \includegraphics[width=0.48\textwidth]{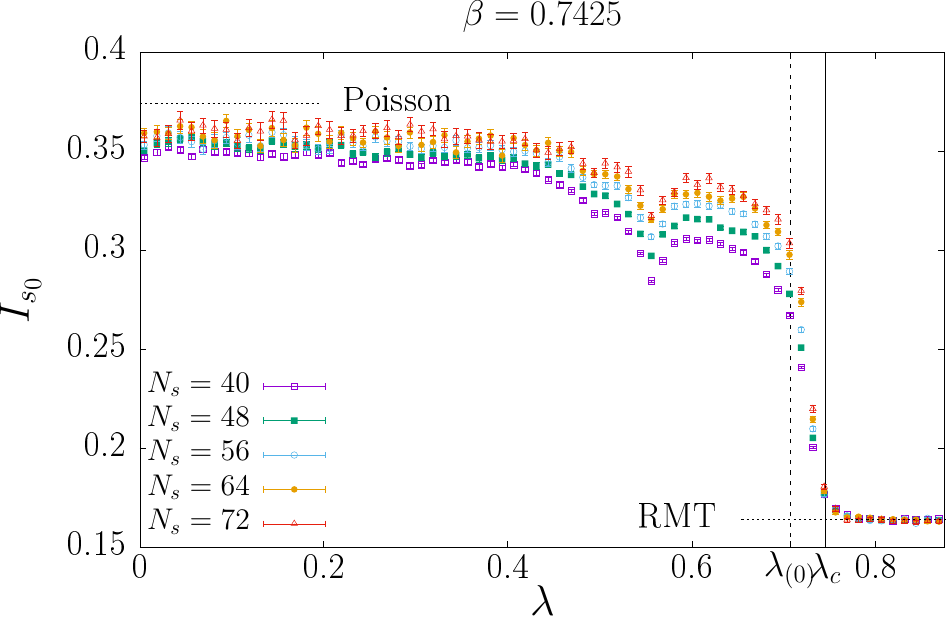}

  \caption{Spectral statistics $I_{s_0}$ at $\beta=0.734$ (top panel)
    and $\beta=0.7425$ (bottom panel), for $N_t=4$ and different
    spatial system sizes. The low end of the bulk, $\lambda_{(0)}$,
    the mobility edge, $\lambda_c$, and the expectations for Poisson
    and RMT statistics are also shown.}
  \label{fig:9}
\end{figure}

To establish the BKT nature of the Anderson transition and measure the
temperature dependence of the mobility edge we performed numerical
simulations of the model in the deconfined phase using spatial sizes
$N_s=40,48,56,64,72$, for
$\beta=0.733, 0.734, 0.735, 0.7375, 0.74, 0.7425$. We generated
approximately 86k, 31k, 12k, 6k, and 2.5k configurations for the
$N_s=40,48,56,64,72$ lattices, respectively.  We selected
configurations in the physical sector,
$\bar{P} = \f{1}{V}\sum_{\vec{x}}P(\vec{x})> 0$; configurations with
$\bar{P}<0$ were not discarded, but transformed to a configuration of
equal weight with $\bar{P}>0$ by multiplying by $-1$ all the temporal
links in the last time slice. We then computed the positive low-lying
eigenvalues of the staggered Dirac operator and the corresponding
eigenvectors using the ARPACK library~\cite{lehoucq1998arpack}. To
keep approximately fixed the spectral range and to cover the low-mode
region, we increased the number of computed eigenmodes with the system
size, obtaining $472,672,920,1200,1520$ modes for
$N_s=40,48,56,64,72$, respectively.

Numerical estimates $\overline{O}^{\mathrm{num}}$ of the local
averages $\overline{O}(\lambda;N_s)$ of spectral observables $O_l$,
Eq.~\eqref{eq:genav}, are obtained by dividing the spectrum in
disjoint intervals ${\cal I}$ of size $\Delta\lambda$ (in lattice
units), first averaging $O_l$ inside ${\cal I}$ and over
configurations to obtain $\overline{O}({\cal I},N_s)$,
\begin{equation}
  \label{eq:local_av_num}
  \begin{aligned}
    \overline{O}({\cal I},N_s) &\equiv 
    \f{\la\sum_{\lambda_l \in {\cal I}} O_l \ra}{ \mathcal{N}({\cal
        I},N_s)}
    \,, \\
    \mathcal{N}({\cal I},N_s) &\equiv \la \textstyle\sum_{\lambda_l
      \in {\cal I}} 1\ra\,,
  \end{aligned}
\end{equation}
and then assigning the result to the average eigenvalue in the
interval,
\begin{equation}
  \label{eq:local_av_num2}
  \begin{aligned}
    \overline{O}^{\mathrm{num}}(\overline{\lambda}({\cal I},N_s))
    &\equiv
    \overline{O}({\cal I},N_s) \,,\\
    \overline{\lambda}({\cal I},N_s) &= \f{\la\sum_{\lambda_l \in
        {\cal I}} \lambda_l \ra}{ \mathcal{N}({\cal I},N_s) }\,.
  \end{aligned}
\end{equation}
The spectral density is estimated numerically as
\begin{equation}
  \label{eq:local_av_num3}
  \rho^{\mathrm{num}}(\overline{\lambda}({\cal I},N_s)) \equiv
  \f{\mathcal{N}({\cal I},N_s)}{N_t V \Delta 
    \lambda}\,.
\end{equation}
We estimated statistical errors on these quantities with the jackknife
method using 100 jackknife samples.

In Figs.~\ref{fig:spd} and \ref{fig:PR} we show the normalized
spectral density $\rho(\lambda)$, Eq.~\eqref{eq:spdens_def}, and the
average mode size $\overline{\mathrm{IPR}^{-1}}$, see under
Eq.~\eqref{eq:PR2}, along the low-lying spectrum. The spectral density
is small but nonzero near $\lambda=0$, and starts to increase rapidly
in the bulk, above $\lambda_{(0)}$. For low modes the mode size is
essentially volume independent, indicating localization.  Peak
structures are present in $\rho$ at specific values of $\lambda$ below
$\lambda_{(0)}$, where one correspondingly finds dips in the mode
size. We will return on this feature in Sec.~\ref{sec:spconf}.

We then studied the statistical properties of the unfolded spectrum.
Since degenerate eigenvalues are absent in the spectral region and for
the volumes under consideration, we have obtained the unfolded
eigenvalues numerically by sorting in ascending order all the
eigenvalues on all the configurations for a given choice of $\beta$
and $N_s$, and assigning to each eigenvalue its rank divided by the
number of configurations. We then used them to numerically estimate
$\overline{I}_{s_0}(\lambda,N_s)$, Eq.~\eqref{eq:local_Is0}, as
explained above, and studied its volume dependence.  The behavior
expected for a BKT-type Anderson transitions is clearly visible in
Fig.~\ref{fig:9}, where we show $\overline{I}_{s_0}$ for different
spatial volumes at $\beta=0.734$ and $\beta=0.7425$ (here
$\Delta\lambda=0.0125$). While the low modes tend towards Poissonian
statistics as the system size increases, higher up in the spectrum the
spectral statistics does not change with $N_s$.  The lowest lattice
Matsubara frequency, $\lambda_{(0)}$, and the mobility edge,
$\lambda_c$, determined via finite size scaling as discussed later,
are also shown. The collapse of the $\overline{I}_{s_0}$ curves for
different spatial sizes on top of each other starts precisely at
$\lambda_c$.

\begin{figure}[t]
  \centering
  \includegraphics[width=0.48\textwidth]{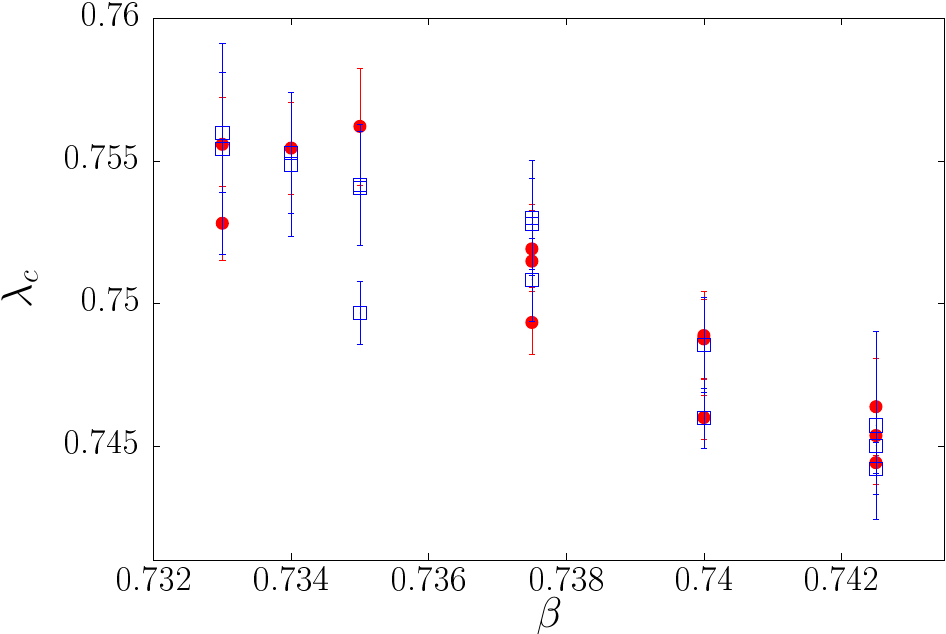}

   \vspace{0.52cm} 
  
  \includegraphics[width=0.48\textwidth]{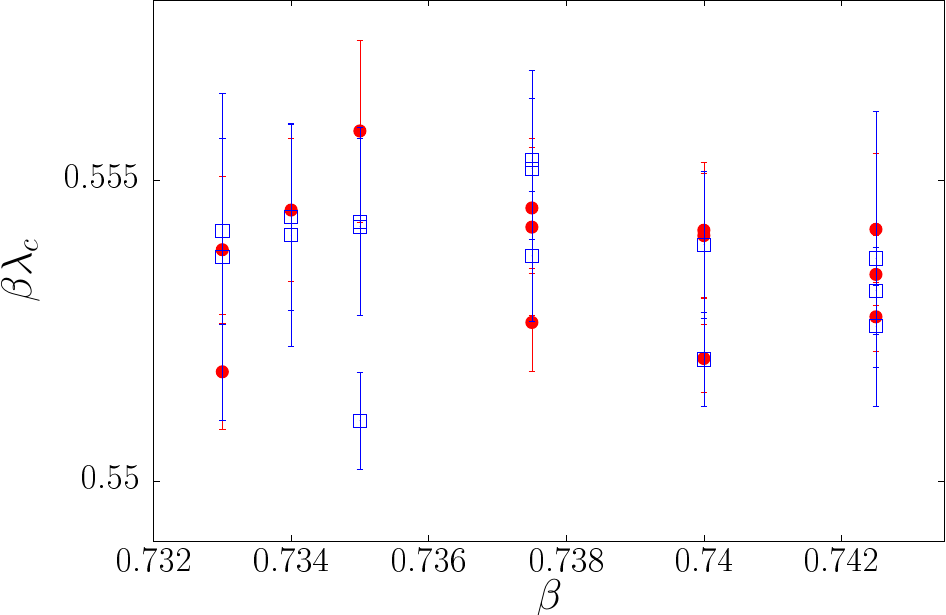}

  \caption{Mobility edge obtained from $I_{s_0}$ via a finite size
    scaling study, in lattice units (top panel) and physical units
    (bottom panel). Red circles correspond to fits including data from
    all system sizes, and blue squares to fits excluding data from the
    smallest size, $L=40$. Different data points at the same $\beta$
    correspond to different fitting intervals around the mobility
    edge.}
  \label{fig:10}
\end{figure}

\begin{figure}[th]
  \centering
  \includegraphics[width=0.48\textwidth]{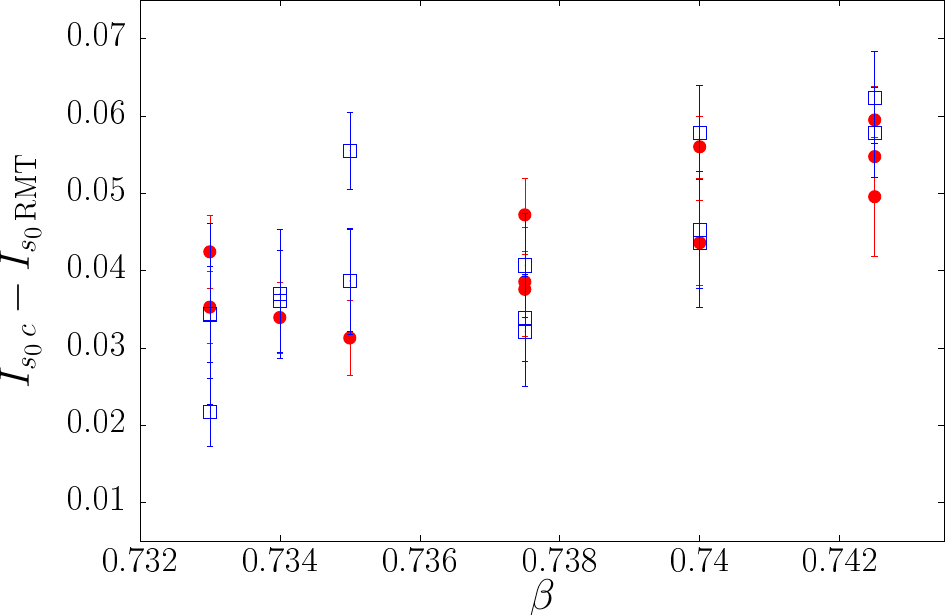}

    \vspace{0.52cm} 
  \includegraphics[width=0.48\textwidth]{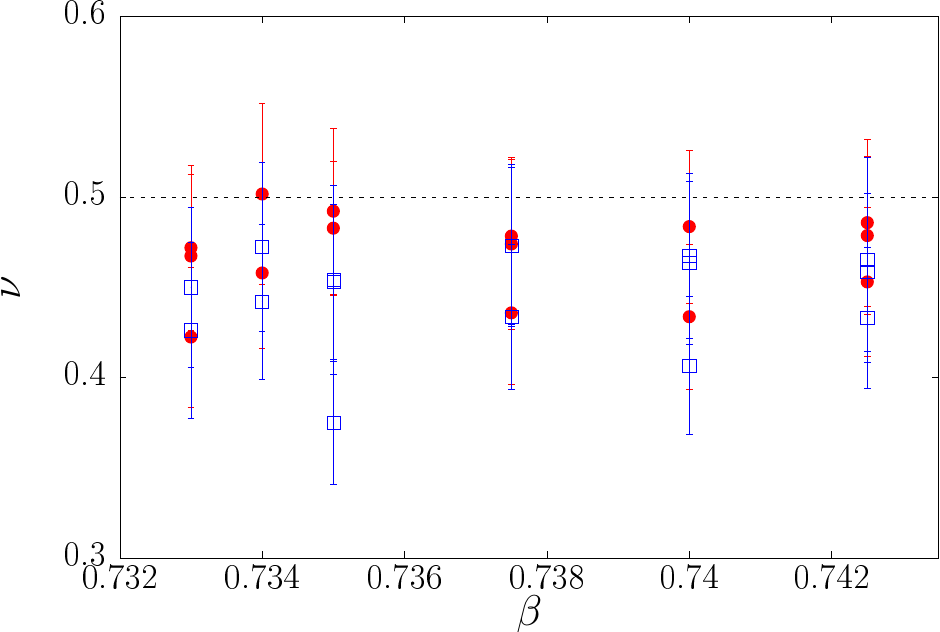}
  \caption{Critical value of $I_{s_0}$ at the mobility edge minus the
    RMT expectation $I_{s_0\,{\rm RMT}}\simeq 0.1642$ (top panel), and
    critical exponent $\nu$ (bottom panel), obtained from $I_{s_0}$
    via a finite size scaling study. Red circles correspond to fits
    including data from all system sizes, and blue squares to fits
    exluding data from the smallest size, $L=40$. Different data
    points at the same $\beta$ correspond to different fitting
    intervals around the mobility edge.}
  \label{fig:11}
\end{figure}

\begin{figure}[th]
  \centering
  \includegraphics[width=0.48\textwidth]{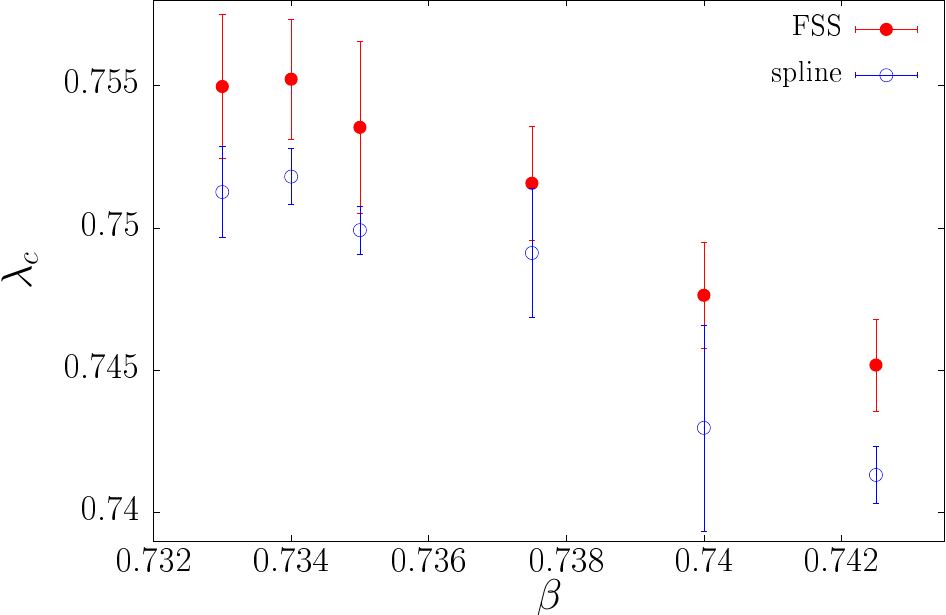}
  \caption{Comparison of the finite size scaling determination of the
    mobility edge, $\lambda_c^{\mathrm{FSS}}$, with the
    infinite-volume extrapolation of the estimate
    $\lambda_c^{\mathrm{spline}}$, obtained as the intercept of a
    spline interpolation of numerical data with a prescribed value, as
    discussed in the text.}
  \label{fig:12}
\end{figure}

We fitted our data for $\overline{I}_{s_0}$ (with
$\Delta\lambda=0.0075$) to a polynomial truncation of the scaling
function Eq.~\eqref{eq:scaling}, obviously setting $L=N_s$, and using
the constrained fitting technique of Ref.~\cite{Lepage:2001ym}. This
amounts in practice to minimizing an augmented $\chi^2$, i.e.,
$\chi^2_{\mathrm{aug}}=\chi^2+\chi^2_{\mathrm{prior}}$, where
\begin{equation}
  \label{eq:priors}
  \chi^2_{\mathrm{prior}} = \sum_n \f{(c_n - \bar{c}_n)^2}{\bar{\sigma}_n^2}\,,
\end{equation}
where the sum extends over the fitting parameters $c_n$, and the
priors $\bar{c}_n$ and $\bar{\sigma}_n$ encode our expectations on
their value.  We used loose priors with a large variance for most of
the coefficients, with a few noteworthy exceptions. For $\nu$ we chose
$\bar{c}_\nu=0.5$ and $\bar{\sigma}_\nu=0.05$, which roughly speaking
amounts to allowing a 10\% deviation from the theoretical
expectation. For $\lambda_c$ we chose $\bar{c}_{\lambda_c}=0.75$ and
$\bar{\sigma}_{\lambda_c}=0.05$: this is justified by a a qualitative
estimate of the point where the $\overline{I}_{s_0}$ curves for
different spatial sizes collapse on top of each other, that places the
mobility edge somewhere in the interval $[0.7,0.8]$ for all the
$\beta$ values considered here. Finally, for $L_0$ we took both the
central value and the variance to be
$\bar{c}_{L_0}=\bar{\sigma}_{L_0}= 1$. We then increased the order of
the polynomial until the errors on the fit parameters stabilized,
monitoring at the same time the value of $\chi^2_{\mathrm{aug}}$ per
data point: a value of order 1 indicates that the choice of priors is
reasonable. We repeated the fit using different fitting ranges around
the mobility edge, and including or excluding the smallest volume. We
used the MINUIT library~\cite{James:1975dr}, and discarded as
unsuccessful those fits for which the error on the fitting parameters
could not be estimated reliably with the MINOS routine.

Our results for $\lambda_c$ are shown in Fig.~\ref{fig:10}, top panel.
The first surprising aspect is that $\lambda_c$ exceeds the lowest
lattice Matsubara frequency $\lambda_{(0)}$. Localized modes then eat
into the bulk of the spectrum, with disorder in the link variables
being able to localize modes that one would expect to be extended,
instead of modes localized on favorable gauge-field fluctuations
becoming delocalized due to mixing with delocalized modes.  The
second, even more surprising aspect is that $\lambda_c$ in lattice
units decreases with $\beta$, while in physical units it is compatible
with a constant within 1\%, see Fig.~\ref{fig:10}, bottom panel. This
suggests that the mobility edge jumps from zero to a finite value at
the critical value of the coupling, $\beta_c$, unexpectedly since the
thermal transition is continuous.

The difference between the critical value of $I_{s_0}$ at the mobility
edge and the RMT value $I_{s_0\,{\rm RMT}}$ is shown in
Fig.~\ref{fig:11}, top panel, and the critical exponent $\nu$ is shown
in Fig.~\ref{fig:11}, bottom panel. Although compatible with a
constant behavior within errors, $I_{s_0}(\lambda_c)$ shows an
increasing trend with $\beta$. Our estimate of the critical exponent
is consistent with the expectation $\nu=\f{1}{2}$ within errors,
although all our estimates but one are below this value.

To check our surprising results for the $\beta$ dependence of
$\lambda_c$, we estimated the mobility edge using a different
approach, less rigorous but also less prone to fit instabilities due
to the peculiar nature of the Anderson transition in this system.  The
idea is that the value of $\lambda$ at which $\overline{I}_{s_0}$
takes any prescribed value between the Poisson expectation and the
critical value found at $\lambda_c$ will eventually tend to the
mobility edge as the volume is increased. We then did a spline
interpolation $I_{s_0}^{\mathrm{spline}}(\lambda)$ of the numerical
data for each $\beta$ and $L$, and estimated
$\lambda_c^{\mathrm{spline}}$ as the solution to
$I_{s_0}^{\mathrm{spline}}(\lambda_c^{\mathrm{spline}}(\beta,L))=0.185$. The
specific value on the right-hand side of the equation was chosen
safely above the (possibly $\beta$ dependent) critical value at the
mobility edge, estimated by observing where the curves for the various
volumes fall on top of each other at each $\beta$. We then obtained
our alternative estimate of the mobility edge by extrapolating to
infinite volume,
$\lambda_c^{\mathrm{spline}}(\beta)=\lim_{L\to\infty}\lambda_c^{\mathrm{spline}}(\beta,L)$,
assuming a linear dependence on $1/L$. We estimated the systematics
from finite size effects by repeating the extrapolation after
excluding the smallest volume. For $\beta=0.74$ the result for the
largest volume is clearly off of a linear trend, and its inclusion in
the fit leads to a very large $\chi^2/\mathrm{d.o.f.}$. We have then
exluded it from the fit in the determination of the central value,
while including the change due to its inclusion in the estimate of the
error. In Fig.~\ref{fig:12} the results are compared with the average
$\lambda_c^{\mathrm{FSS}}$ of the determinations of $\lambda_c$ via
finite size scaling shown in Fig.~\ref{fig:10}, with total error
estimated adding in quadrature the average of the statistical error
and the standard deviation over the set of determinations. The two
estimates $\lambda_c^{\mathrm{FSS}}$ and $\lambda_c^{\mathrm{spline}}$
are in agreement within errors, although the alternative estimate is
always smaller than the finite size scaling one. This is not entirely
surprising, as $\lambda_c^{\mathrm{spline}}(\beta,L)$ always
underestimates $\lambda_c$ by construction, and the linear
extrapolation to infinite volume is probably neglecting sizable
effects. In any case, the important aspect is that the two estimates
agree on the qualitative behavior of $\lambda_c$ as a function of
$\beta$.

The decrease of $\lambda_c$ with $\beta$ can be understood noticing
that the width of the spectral pseudogap, where the density of modes
is low, essentially coincides with the lowest lattice Matsubara
frequency, $\lambda_{(0)}$, which depends only on $N_t$ and so is
fixed here.  This makes the low end an effective edge of the spectrum,
where disorder localizes the outermost modes first. Increasing $\beta$
at fixed $N_t$, the disorder {\it decreases} as the system gets more
ordered, pushing the mobility edge towards the end of the spectrum,
which in this case means reducing $\lambda_c$. The decrease in
disorder is apparently compensated by the decrease of the lattice
spacing $1/\beta = e^2 a$, so that in physical units the mobility edge
stays constant. We return on this point below in
Sec.~\ref{sec:numrefsi}. The difference with other theories studied so
far is evident, and presumably due to the discrete nature of the gauge
group. In the other cases where the dependence of $\lambda_c$ on
$\beta$ was studied so far~\cite{Kovacs:2012zq,Holicki:2018sms,
  Giordano:2019pvc,Vig:2020pgq,Bonati:2020lal,Kovacs:2021fwq,
  Cardinali:2021fpu,Kehr:2023wrs,Baranka:2023ani}, the gauge group was
continuous, and in that case probably the spectral pseudogap does not
depend only on $N_t$ but on $\beta$ as well.

\subsection{Center cluster percolation}
\label{sec:cp}

\begin{figure}[t]
  \centering
    \includegraphics[width=0.48\textwidth]{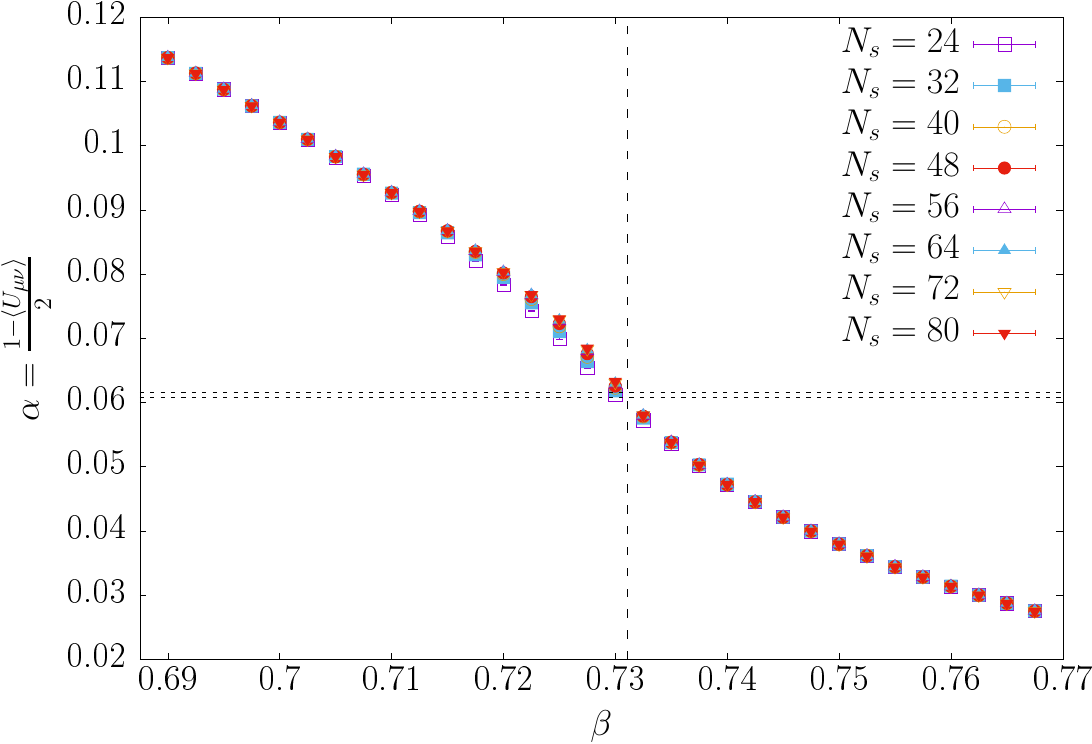} 
    \caption{Concentration of active bonds in the percolation problem
      associated with $\mathbb{Z}_2$ gauge theory as a function of the
      lattice coupling $\beta$, for $N_t=4$ and various $N_s$. The
      horizontal dashed lines enclose the confidence band for the
      critical concentration [see Eq.~\eqref{eq:perc_crit}], while the
      vertical dashed line marks the deconfinement temperature,
      $\beta_c$~\cite{Caselle:1995wn}.}
  \label{fig:conc}
\end{figure}

\begin{figure}[t!]
  \centering
    \includegraphics[width=0.48\textwidth]{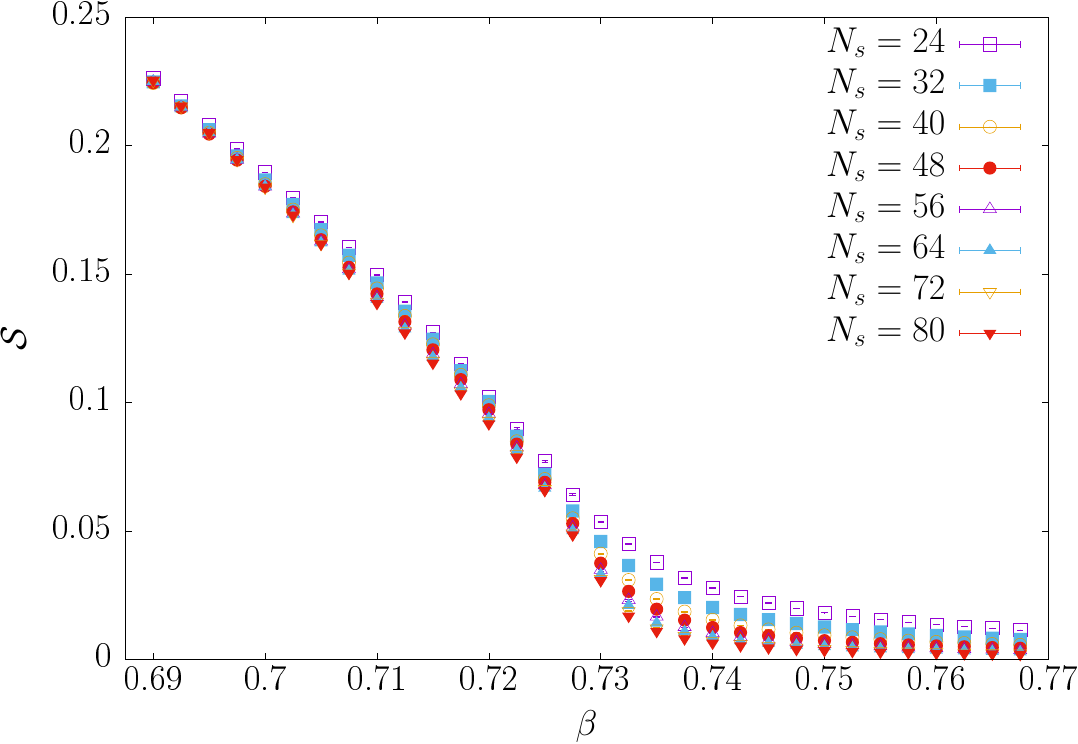}
    \caption{Size of the largest cluster divided by the lattice size,
      $\mathcal{S}$, Eq.~\eqref{eq:sclustsize}, as a function of the
      lattice coupling $\beta$ for $N_t=4$ and various $N_s$.}
  \label{fig:sclsust}
\end{figure}

To study the percolation properties of center clusters across the
transition we performed a second set of simulations on lattices with
$N_s=24,32,40,48,56,64,72,80$, accumulating between 60k and 135k
configurations for the various volumes. For each configuration, after
counting the number of active bonds, we identified all the center
clusters, finding then the largest cluster and measuring its size in
terms of the dual sites it touched. We then determined the
concentration of active bonds $\alpha$, Eq.~\eqref{eq:probnegplaq},
and largest cluster density $\mathcal{S}$, Eq.~\eqref{eq:sclustsize},
as functions of the lattice coupling $\beta$. Results for
$\alpha(\beta)$ for various lattice sizes are shown in
Fig.~\ref{fig:conc}, and results for $\mathcal{S}(\beta)$ are shown in
Fig.~\ref{fig:sclsust}. The function $\alpha(\beta)$ is monotonic, and
so invertible. At low $\beta$, corresponding to larger active-bond
concentrations, $\mathcal{S}(\beta)$ depends mildly on the volume and
tends to a finite value as the volume increases, signaling the
presence of an infinite cluster.


\begin{figure}[t]
  \centering
  \includegraphics[width=0.48\textwidth]{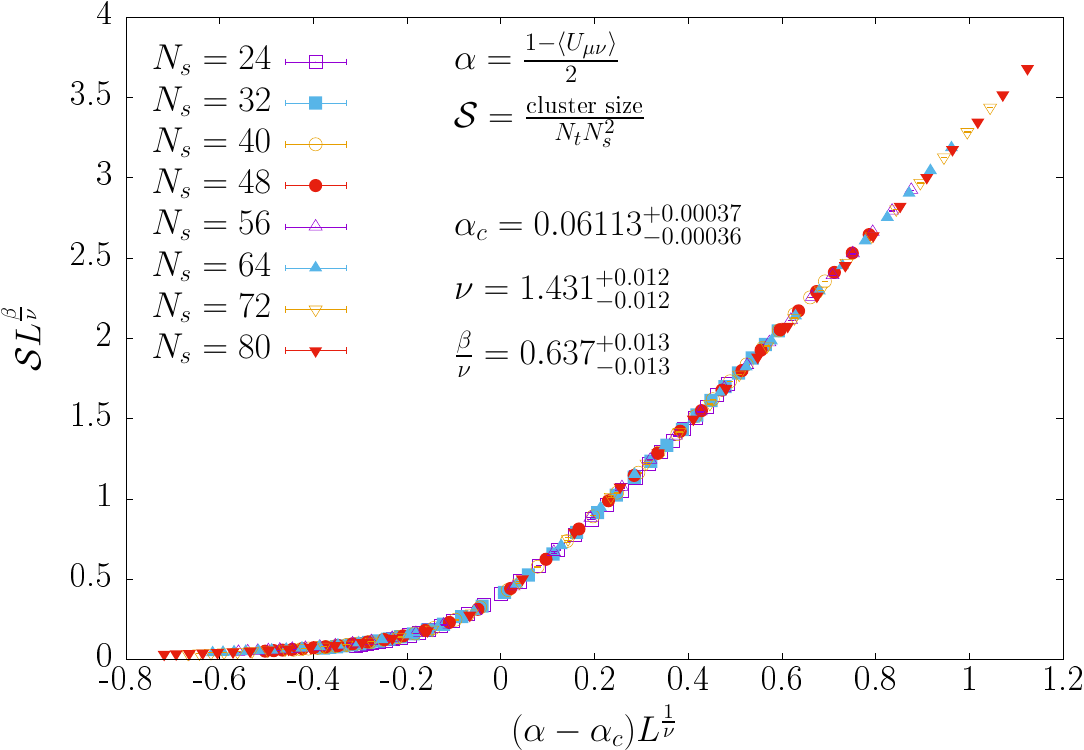} 
  \caption{Collapse plot for the rescaled largest-cluster density
    $\mathcal{S}(\alpha,L)L^{\f{\beta}{\nu}}$, for $N_t=4$ and various
    $N_s$.}
  \label{fig:clust_fss}
\end{figure}

We then studied quantitatively the volume scaling of the largest-cluster
density $\mathcal{S}$ as a function of the active-bond concentration
$\alpha$. To show that the system does indeed undergo a percolation
transition, we performed a finite-size scaling analysis of
$\mathcal{S}$ as a function of $\alpha$. We fitted our numerical data
with the functional form
\begin{equation}
  \label{eq:fss_clust}
  \mathcal{S}(\alpha,L)L^{\f{\beta}{\nu}} =
  s((\alpha-\alpha_c)L^{\f{1}{\nu}})\,,  
\end{equation}
truncating the analytic function $s(x)$ to a polynomial of varying
order. Here $\beta$ denotes one of the critical exponents, and should
not be confused with the lattice coupling.  We then used again the
constrained fitting techniques of Ref.~\cite{Lepage:2001ym}, with very
broad priors on Taylor coefficients starting only from the
fourth-order one, and using no priors for lower-order coefficients and
for the critical concentration $\alpha_c$ and critical exponents $\nu$
and $\beta$. Fits were again performed using the MINUIT
library~\cite{James:1975dr}. The results of our analysis are
summarized in the collapse plot in Fig.~\ref{fig:clust_fss}.  The
quality of the collapse is excellent. We find for the critical
parameters
\begin{equation}
  \label{eq:perc_crit}
  \begin{aligned}
    \alpha_c&=0.06113^{+0.00037}_{-0.00036}\,,\\
    \nu&=1.431^{+0.012}_{-0.012}\,,\\ \f{\beta}{\nu}&=
    0.637^{+0.013}_{-0.012}\,.
  \end{aligned}
\end{equation}
The critical exponents clearly deviate from those of two-dimensional
bond percolation, $\nu=4/3$ and $\beta/\nu=5/48$ (see
Refs.~\cite{RevModPhys.54.235,percolation,smirnov2001,10.1214/11-AOP740}),
in spite of their expected universality (see the proof in
Ref.~\cite{10.1214/11-AOP740} for square, triangular, and hexagonal
lattices). This is probably due to the topological constraint imposed
on clusters, mentioned above in Sec.~\ref{sec:cv}, and possibly to the
presence of an additional dimension (albeit of finite extent), that
change the universality class of the transition.

\begin{figure}[thb]
  \centering
    \includegraphics[width=0.48\textwidth]{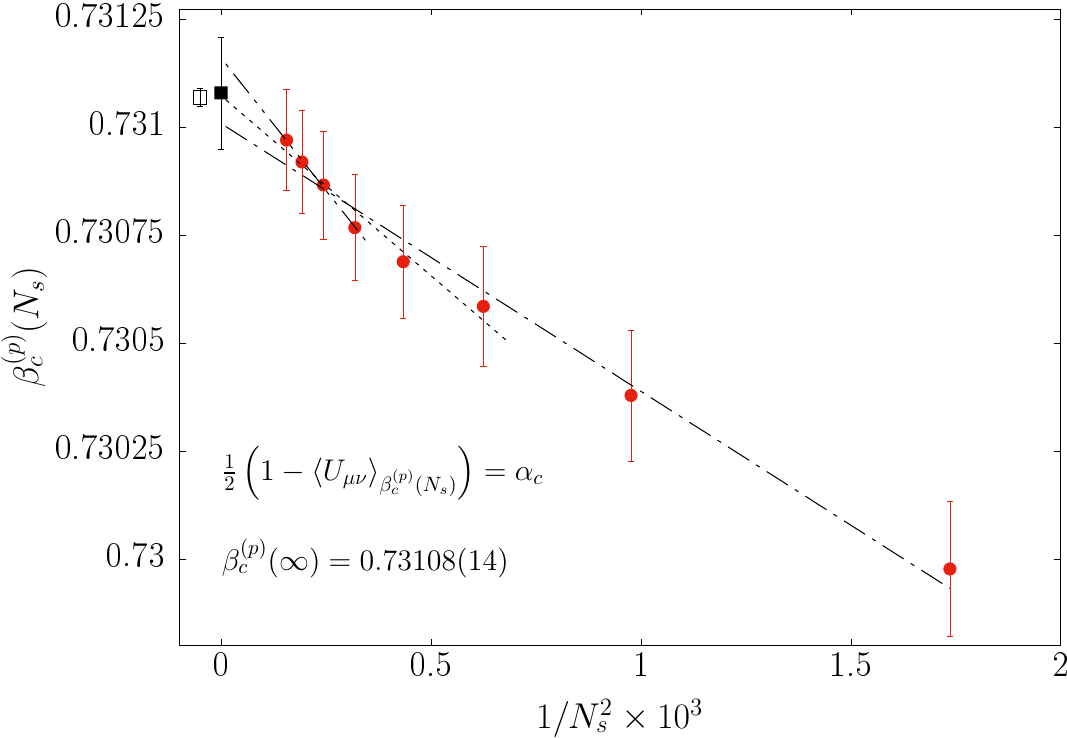}   

    \caption{Infinite-volume extrapolation of the critical coupling
      $\beta_c^{(p)}$ for the percolation transition, for $N_t=4$. The
      three dashed straight lines correspond to linear extrapolations
      in $1/N_s^2$, each ranging only over the volumes used in the
      extrapolation. The filled black square is our final result for
      the extrapolation, including statistical and finite-size
      systematic errors, while the empty black square is the critical
      lattice coupling for deconfinement obtained in
      Ref.~\cite{Caselle:1995wn}.}
  \label{fig:betac}
\end{figure}
\begin{figure}[t]
  \centering
    \includegraphics[width=0.48\textwidth]{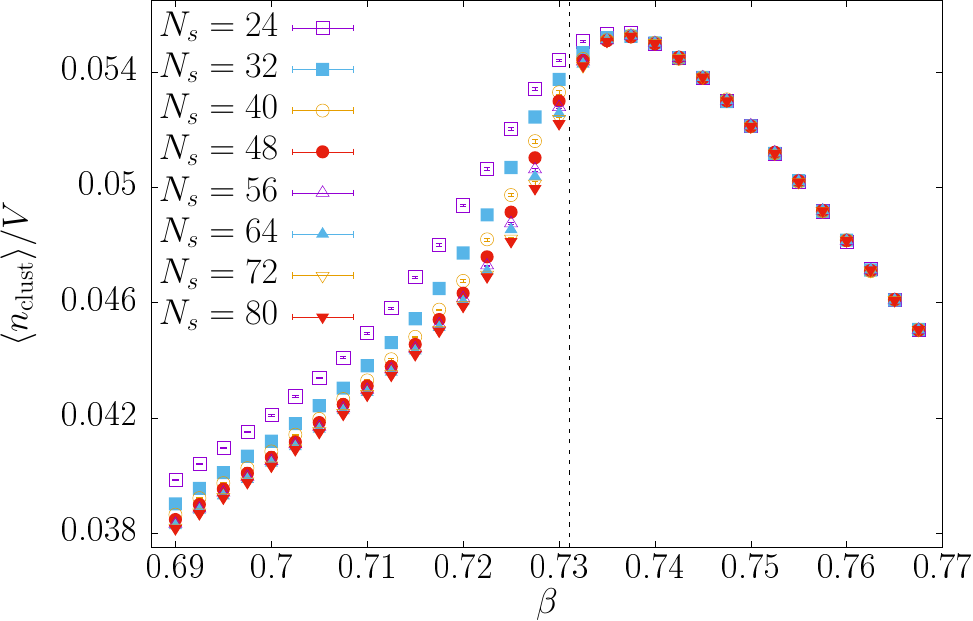} 
    \caption{Average number of center clusters
      $\la n_{\mathrm{clust}}\ra$ divided by the lattice spatial
      volume $V=N_s^2$ as a function of $\beta$ for various lattice
      sizes and $N_t=4$.}
  \label{fig:clust_no}
\end{figure}

Figure~\ref{fig:conc} shows that the critical concentration is
attained at the deconfinement transition, which corresponds then to
the percolation transition for center vortices. To make this statement
quantitative, we determined the critical percolation coupling
$\beta_c^{(p)}$ from the critical concentration $\alpha_c$ through the
following procedure. For each spatial size $N_s$, we define
$\beta_c^{(p)}(N_s)$ as the crossing point
$\alpha(\beta_c^{(p)}(N_s)) = \alpha_c$. This is obtained numerically
by means of spline interpolation of $\alpha(\beta)$ and a standard
bisection method. Next, we extrapolate $\beta_c^{(p)}(N_s)$ to
infinite volume. We performed a linear extrapolation in the inverse
volume, fitting $\beta_c^{(p)}(N_s)=a + b/N_s^2$ including all data,
only data for $N_s\ge 40$, and only data for $N_s\ge 56$, in order to
estimate finite-size systematic effects. Results are shown in
Fig.~\ref{fig:betac}. We took as central value the average of the
three determinations of $a$, as statistical error the average of the
statistical errors from the fits, and as systematic error the standard
deviation of $a$. The final result is
$\beta_c^{(p)} =
0.73108(12)_{\text{stat}}(06)_{\text{syst}}=0.73108(13)$, in good
agreement with the critical coupling $\beta_c = 0.73107(2)$ for the
deconfinement transition found in Ref.~\cite{Caselle:1995wn}.

To conclude this subsection, in Fig.~\ref{fig:clust_no} we show how
the average number of clusters in a configuration,
$\la n_{\mathrm{clust}}\ra $, scales with the volume. Deep in the
deconfined phase $\la n_{\mathrm{clust}}\ra/V$ is clearly constant,
showing that the number of clusters grows linearly with the volume.
Closer to the transition in the deconfined phase, as well as in the
confined phase, the volume scaling seems again to asymptotically
approach $n_{\mathrm{clust}}\propto V$, although more slowly. This
suggests that a finite density of finite, localized clusters is found
also in the confined phase, even in the presence of an infinite
cluster.

\begin{figure}[t]
  \centering
    \includegraphics[width=0.45\textwidth]{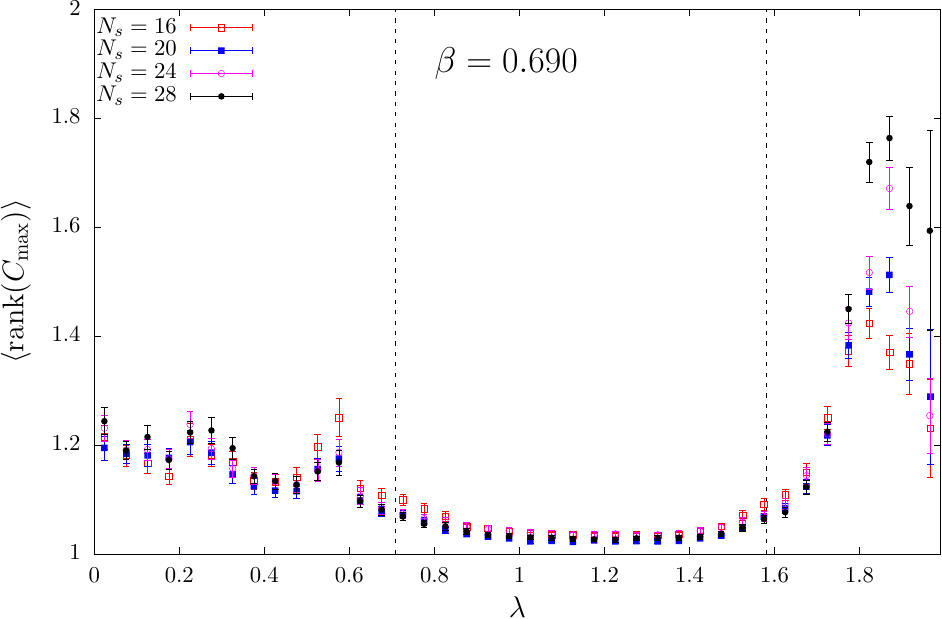}

    \includegraphics[width=0.45\textwidth]{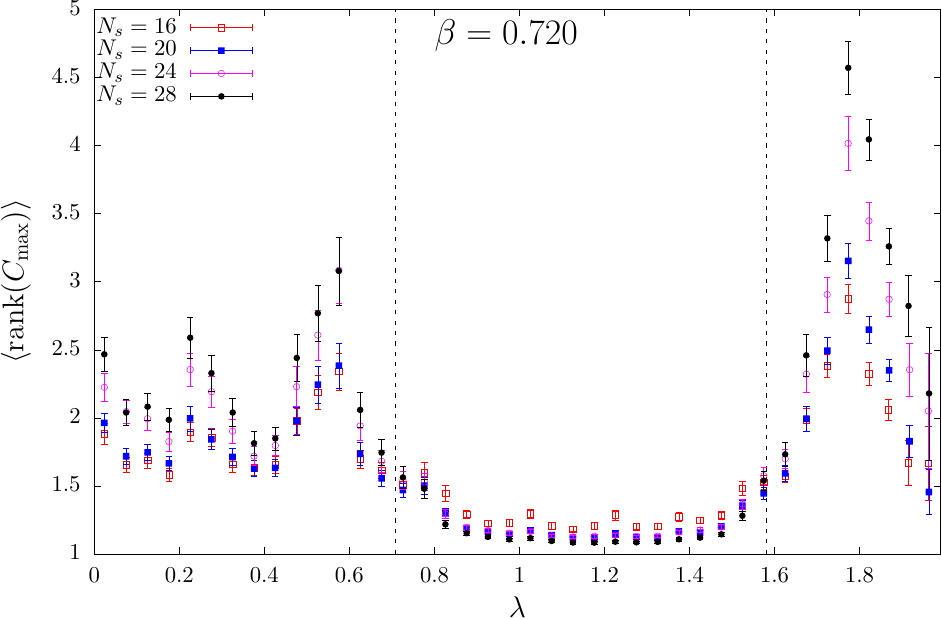}

    \includegraphics[width=0.45\textwidth]{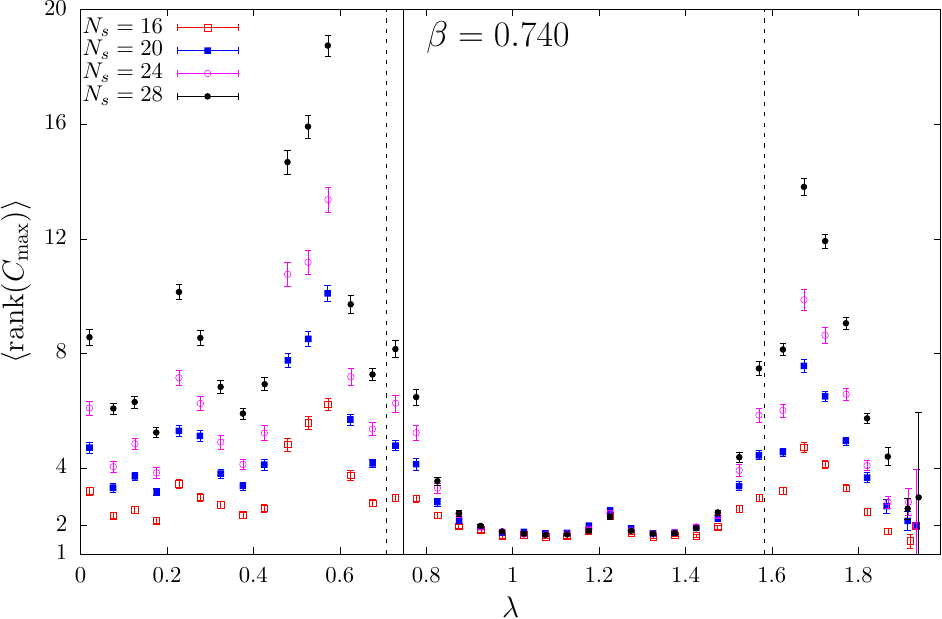}
    \caption{Average rank in magnitude of the cluster
      $C_{\mathrm{max}}$ on which modes have maximal weight at
      $\beta=0.69$ (top panel) and $\beta=0.72$ (center panel), in the
      confined phase, and $\beta=0.74$ (bottom panel), in the
      deconfined phase, for $N_t=4$ and various $N_s$. Dashed vertical
      lines mark the ends of the bulk, $\lambda_{(0)}$ and
      $\lambda_{(1)}$; the solid vertical line in the bottom panel
      marks the mobility edge $\lambda_c$ between low and bulk modes.}
  \label{fig:rank}
\end{figure}

\begin{figure}[thb]
  \centering
  \includegraphics[width=0.45\textwidth]{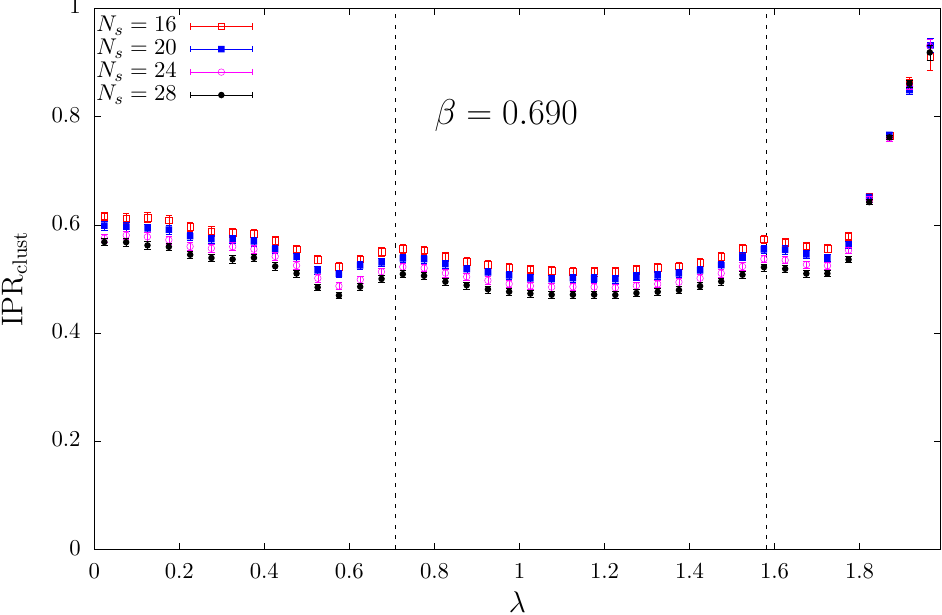}

  \vspace{0.2cm}
  \includegraphics[width=0.45\textwidth]{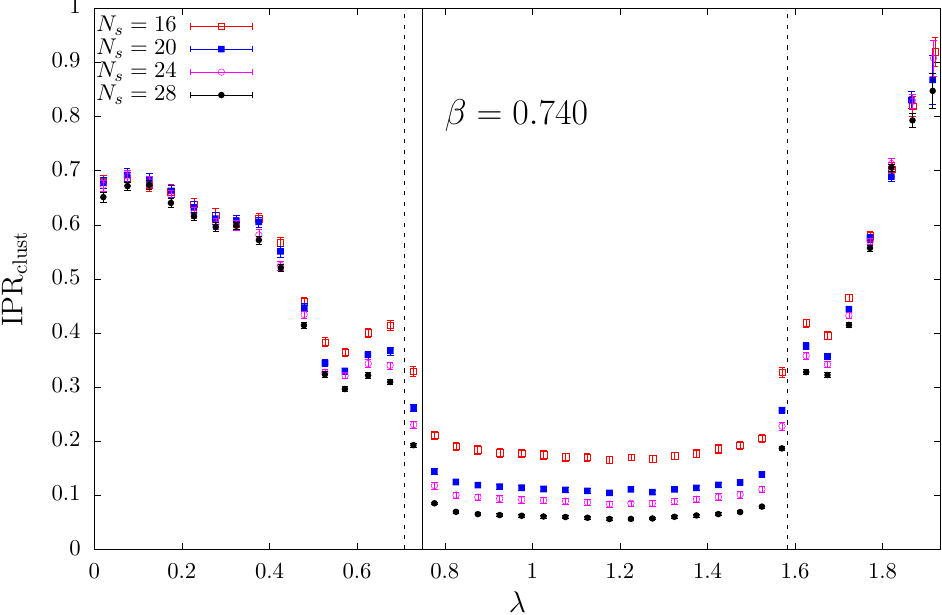}

  \vspace{0.2cm}
  \includegraphics[width=0.45\textwidth]{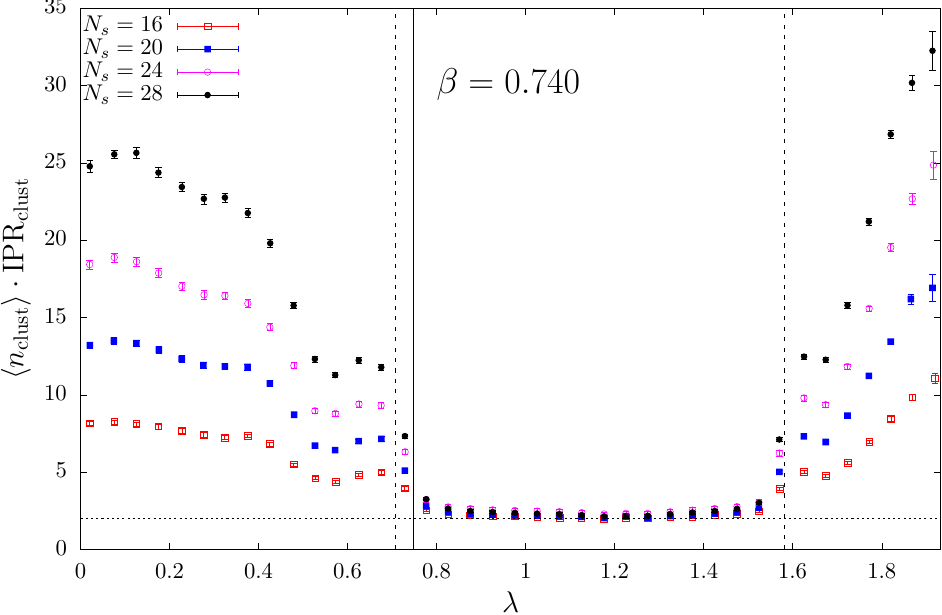}

  \caption{Average cluster IPR, $\mathrm{IPR}_{\mathrm{clust}}$, along
    the spectrum for $\beta=0.69$ (top panel) and $\beta=0.74$ (center
    panel), and average $\mathrm{IPR}_{\mathrm{clust}}$ times average
    number of clusters $\la n_{\mathrm{clust}}\ra$ along the spectrum
    for $\beta=0.74$ (bottom panel), for $N_t=4$ and various
    $N_s$. Vertical dashed lines correspond to the bulk ends
    $\lambda_{(0)}$ and $\lambda_{(1)}$. The solid line in the center
    and bottom panels corresponds to $\lambda_c$.  In the bottom panel,
    the dashed horizontal line corresponds to the value $2$.}
  \label{fig:clipr_vol}
\end{figure}

\begin{figure}[thb]
  \centering
  \includegraphics[width=0.45\textwidth]{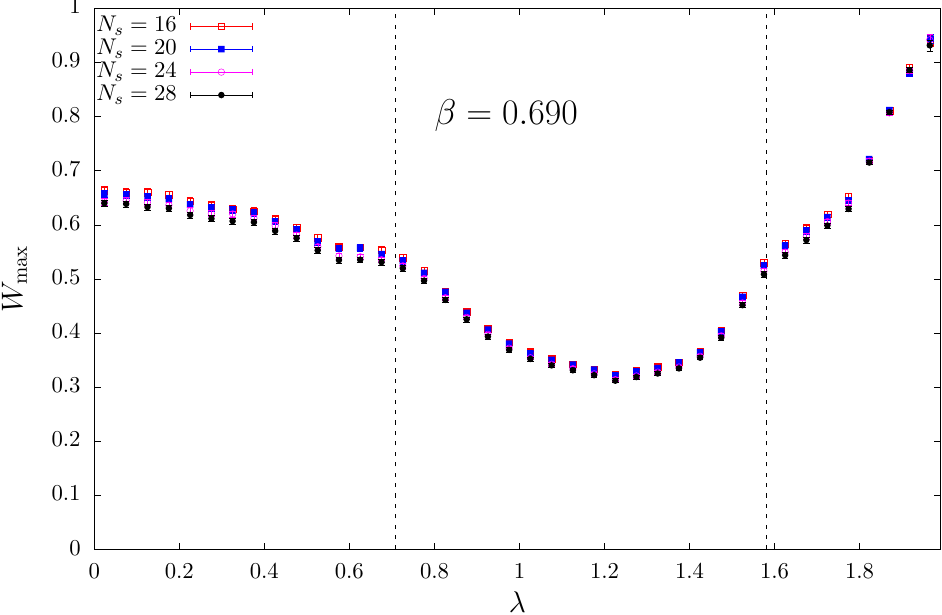}

  \vspace{0.2cm}
  \includegraphics[width=0.45\textwidth]{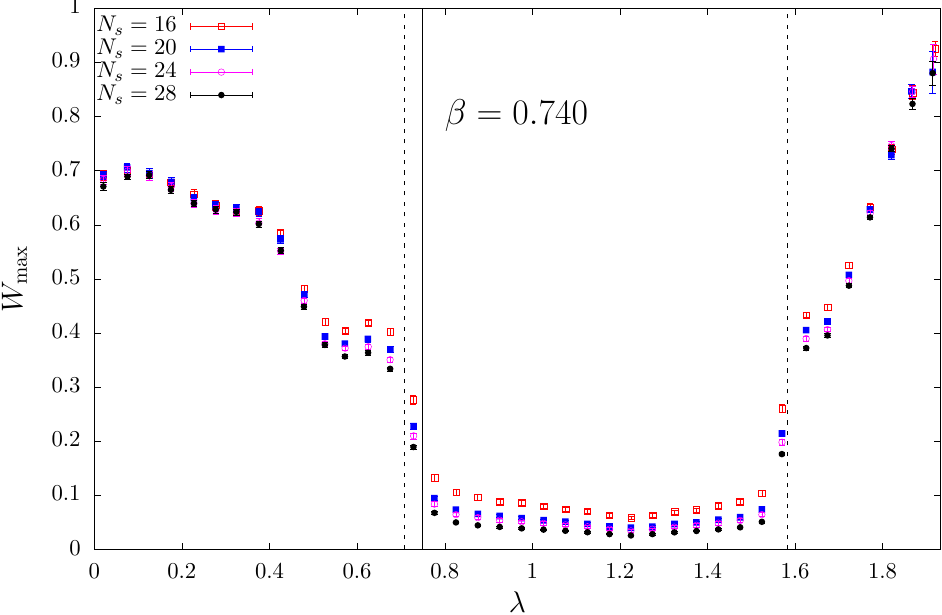}

  \vspace{0.2cm}
  \includegraphics[width=0.45\textwidth]{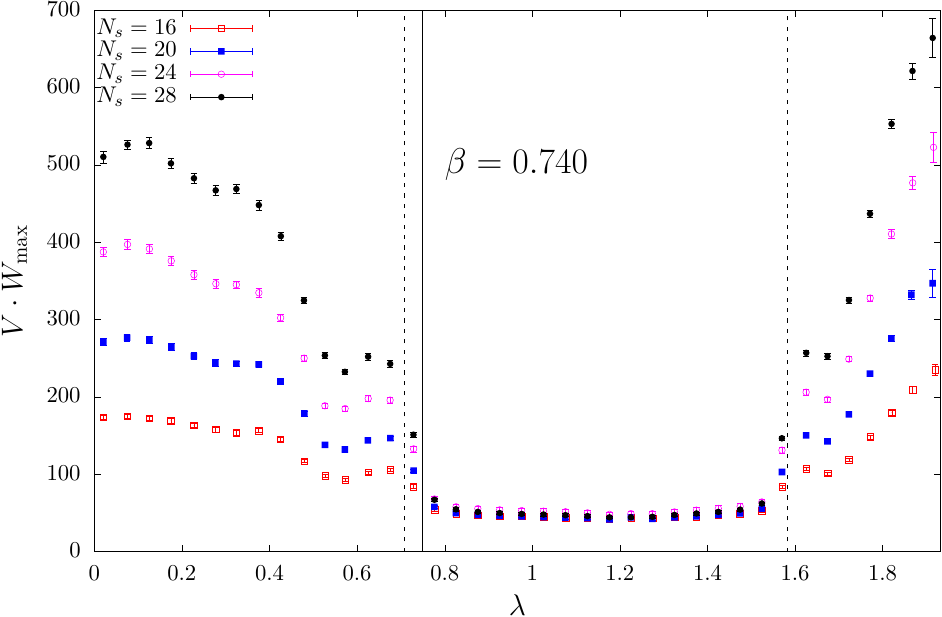}

  \caption{Average maximal weight on a cluster along the spectrum for
    $\beta=0.69$ (top panel) and $\beta=0.74$ (center panel), and
    average maximal weight on a cluster times the lattice volume along
    the spectrum for $\beta=0.74$ (bottom panel), for $N_t=4$ and
    various $N_s$. Vertical dashed lines correspond to the bulk ends
    $ \lambda_{(0)}$ and $\lambda_{(1)}$. In the center and bottom
    panel, the solid line corresponds to $\lambda_c$.}
  \label{fig:clmaxw_vol}
\end{figure}

\begin{figure}[thb]
  \centering
  \includegraphics[width=0.45\textwidth]{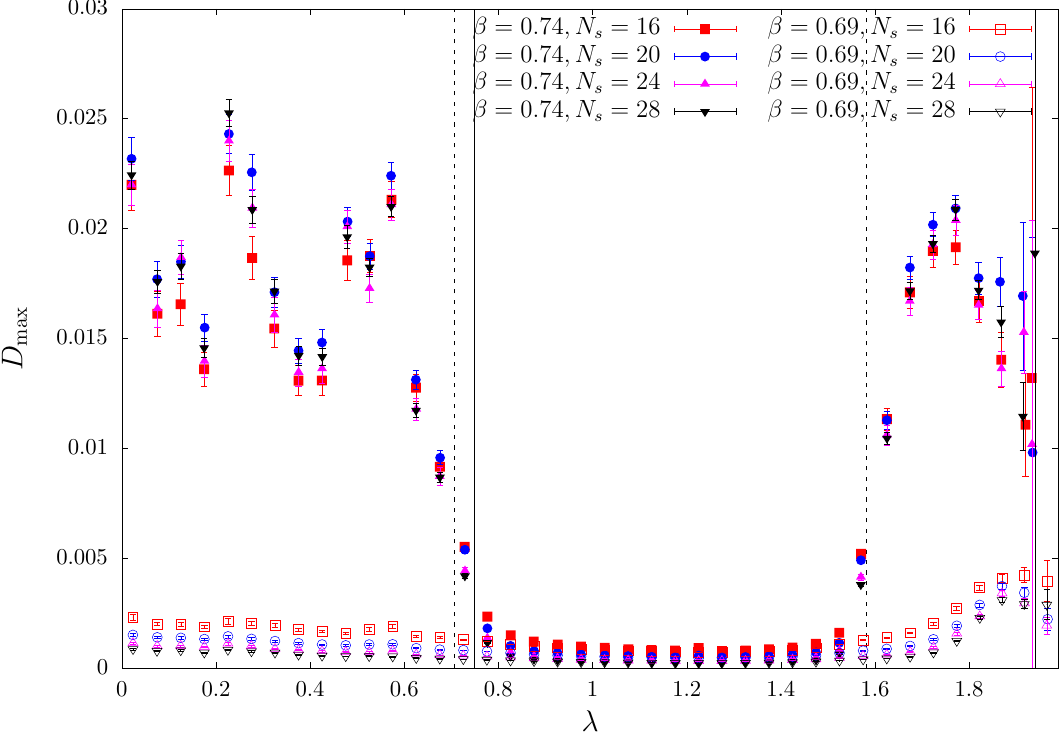}

  \includegraphics[width=0.44\textwidth]{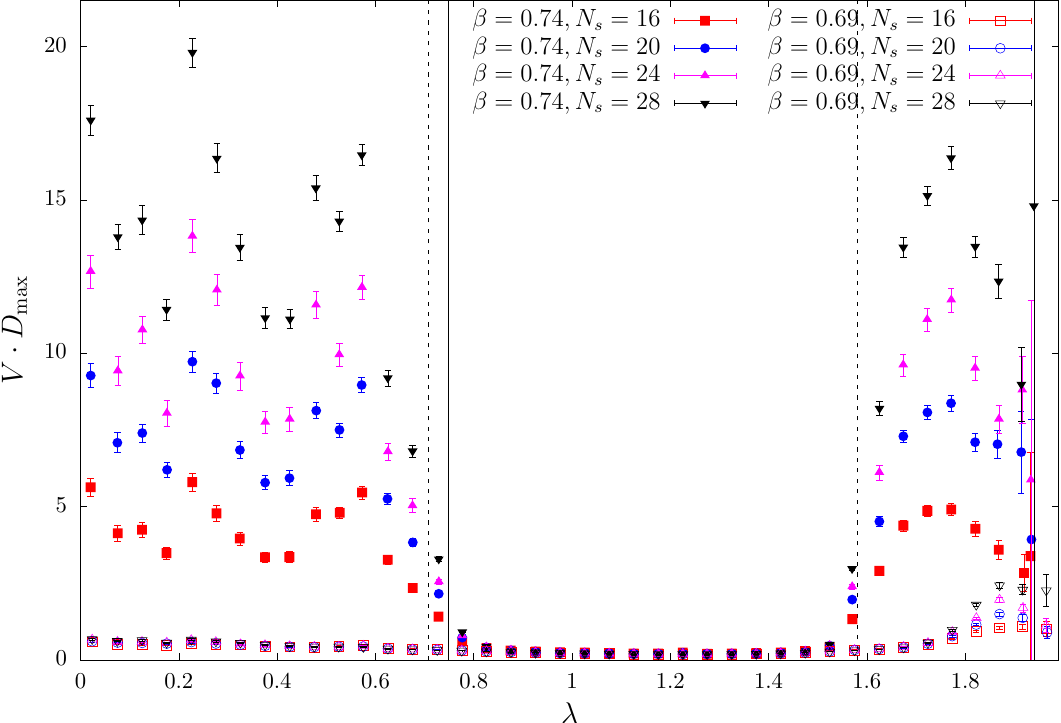}

  \includegraphics[width=0.45\textwidth]{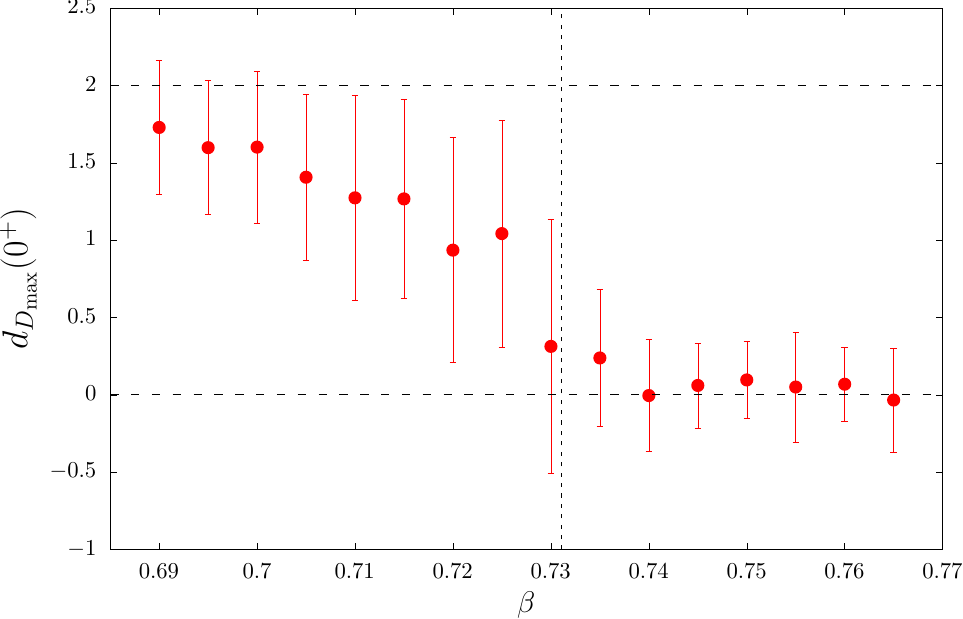}
      
  \caption{Top panel: Average maximal weight on a cluster divided by
    the cluster size, $D_{\mathrm{max}}$, computed locally in the
    spectrum, for $\beta=0.69$ and $\beta=0.74$, for $N_t=4$ and
    various $N_s$. Center panel: same quantity as in the top panel but
    rescaled by the lattice spatial volume $V$. In both panels,
    vertical dashed lines correspond to the bulk ends $ \lambda_{(0)}$
    and $\lambda_{(1)}$, and the solid line corresponds to
    $\lambda_c(\beta=0.74)$. Bottom panel: Scaling dimension of
    $D_{\mathrm{max}}$ [see Eq.~\eqref{eq:genscaldim}] in the lowest
    spectral bin, averaged over volume pairs.  The vertical dashed
    line corresponds to $\beta_c(N_t=4)$.}
  \label{fig:clmaxwpersite_vol0}
\end{figure}

\subsection{Correlation between eigenmodes and clusters}
\label{sec:eccorr}

We studied the correlation between center vortices and Dirac modes on
lattices with $N_s=16,20,24,28$, for several $\beta$s across the
transition, using 300 configurations for each setup. We identified all
center clusters $\{C\}$, and obtained all the staggered modes
$\{\psi_l\}$ using the LAPACK library~\cite{anderson1999lapack}.  We
then measured several types of cross-correlations between the two
objects.

In Fig.~\ref{fig:rank} we show the average rank in magnitude of the
cluster $C_{\mathrm{max}}$ on which modes have maximal weight (see
Sec.~\ref{sec:cv}), studied locally in spectrum. At $\beta=0.69$ (top
panel), well within the confined phase, this is close to 1 for low and
bulk modes, reaching slightly higher values for high modes. This is in
agreement with the modes' localization properties: the delocalized low
and bulk modes favor the largest cluster, which is the infinite
cluster extended over the whole lattice discussed in the previous
subsection, while the localized high modes tend to favor smaller
clusters, of lower rank, and more so as the system size increases.  At
$\beta=0.72$ (center panel), closer to the deconfined phase, low modes
start showing a tendency to favor smaller clusters. At $\beta=0.74$
(bottom panel), well within the deconfined phase, the infinite cluster
has disappeared, and the relation between the eigenvalue and the
average rank of the corresponding $C_{\mathrm{max}}$ is far from
straightforward, with the lowest modes not favoring the largest
cluster, as one would naively expect, and marked peaks in the
low-lying part of the spectrum (already visible, to a lower extent,
already in the confined phase).

The cluster IPR, $\mathrm{IPR}_{\mathrm{clust}}$,
Eq.~\eqref{eq:clipr}, averaged locally in the spectrum according to
Eq.~\eqref{eq:genav}, is shown for $\beta=0.69$ in the confined phase
and $\beta=0.74$ in the deconfined phase in the top and center panels
of Fig.~\ref{fig:clipr_vol}, respectively. At $\beta=0.69$, both low
and bulk modes spread on average on $1.5\,\text{--}\,2$ clusters
(including the infinite cluster, as shown above), with a mild increase
in the spreading as $N_s$ increases. Localized high modes instead
typically lie on fewer clusters ($1\,\text{--}\,1.5$), without any
apparent volume dependence of $\mathrm{IPR}_{\mathrm{clust}}$. This
situation changes in the deconfined phase, where the infinite cluster
is absent. For bulk modes,
$\mathrm{IPR}_{\mathrm{clust}}\simeq 2\la n_{\mathrm{clust}} \ra^{-1}$
(see Fig.~\ref{fig:clipr_vol}, bottom panel), i.e., they spread on
about half of the $n_{\mathrm{clust}}$ (finite) clusters in a
configuration; this also determines the marked volume dependence of
$\mathrm{IPR}_{\mathrm{clust}}$, since
$\la n_{\mathrm{clust}} \ra \propto V$, see Fig.~\ref{fig:clust_no}.
Low and high modes, instead, spread on about $1.5\,\text{--}\,3$ and
$1\,\text{--}\,3$ clusters, respectively, with a mild volume
dependence. This is again in agreement with the modes' localization
properties.

The maximal weight on a cluster $W_{\mathrm{max}}$,
Eq.~\eqref{eq:clmaxw}, again averaged locally in the spectrum, is
shown for the same values of $\beta$ in the top and center panels of
Fig.~\ref{fig:clmaxw_vol}. In the confined phase this quantity does
not show a big difference between low and bulk modes, and a mild
volume dependence for both, reflecting the fact that these modes are
delocalized on the infinite center cluster. High modes show a larger
value and very little volume dependence, reflecting their localized
nature. In the deconfined phase the difference between low and bulk
modes becomes marked, and only bulk modes show a noticeable, $1/V$
volume dependence due to their delocalizing over a finite fraction of
clusters (see Fig.~\ref{fig:clmaxw_vol}, bottom panel). For the lowest
and highest modes, most of the mode weight is found on a single finite
cluster even as the volume increases; closer to the mobility
edges\footnote{\label{foot:highlc}We did not determine precisely the
  position of the mobility edge separating delocalized bulk and
  localized high modes, but all observables point at it being found
  near the high end of the bulk.} the maximal fraction of weight on a
cluster goes below $0.5$, but it still seems to converge to a finite
value in the large-volume limit.

Finally, in Fig.~\ref{fig:clmaxwpersite_vol0}, top panel, we show the
local averages of the density
$D_{\mathrm{max}}=W_{\mathrm{max}}/S_{\mathrm{max}}$,
Eq.~\eqref{eq:clmdens}, i.e., the mode weight on the cluster carrying
the maximal amount of mode weight, divided by the cluster size.  This
reveals marked peaks at the same points in the spectrum where the
spectral density and the mode size also show structures. While these
peaks tend to flatten out with increasing volume in the confined
phase, they remain essentially constant in the deconfined phase. The
volume dependence for bulk modes in both the confined and the
deconfined phase is entirely explained by their delocalized nature: a
rescaling of $D$ by the spatial volume $V$ makes the curves
volume-independent in the bulk region, see
Fig.~\ref{fig:clmaxwpersite_vol0}, center panel. In the confined
phase, low modes still show a mild volume dependence after this
rescaling, due to their fractal dimension being smaller than
2~\cite{Baranka:2021san}. The scaling dimension $d_{D_{\mathrm{max}}}$
of $D_{\mathrm{max}}$ [see Eq.~\eqref{eq:genscaldim}] in the lowest
spectral bin is shown in Fig.~\ref{fig:clmaxwpersite_vol0}, bottom
panel. Our estimate is obtained averaging over the determinations
obtained using different pairs of volumes, and adding in quadrature
the average of the corresponding statistical errors and the standard
deviation of the determinations to estimate the error. Our results are
qualitatively consistent with the nontrivial behavior of the fractal
dimension of near-zero modes found in Ref.~\cite{Baranka:2021san},
although $d_{D_{\mathrm{max}}}$ as a function of $\beta$ approaches 2
faster than the fractal dimension as one goes deeper in the confined
phase.

The results of this subsection show that the localized low modes
appearing in the deconfined phase (and the localized high modes
present both in the confined and in the deconfined phase) have a clear
tendency to localize on one, or at most a few center clusters. This
shows that the gauge field structures related to the confining
properties of the theory coincide with the localization centers for
these modes.

\begin{figure}[t]
  \centering

    \includegraphics[width=0.45\textwidth]{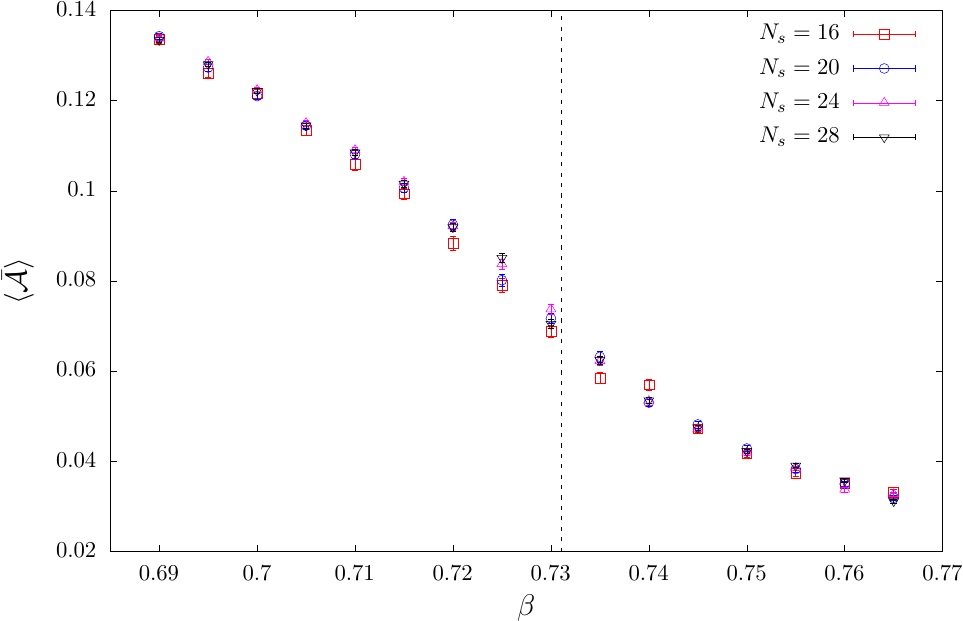}

    \includegraphics[width=0.45\textwidth]{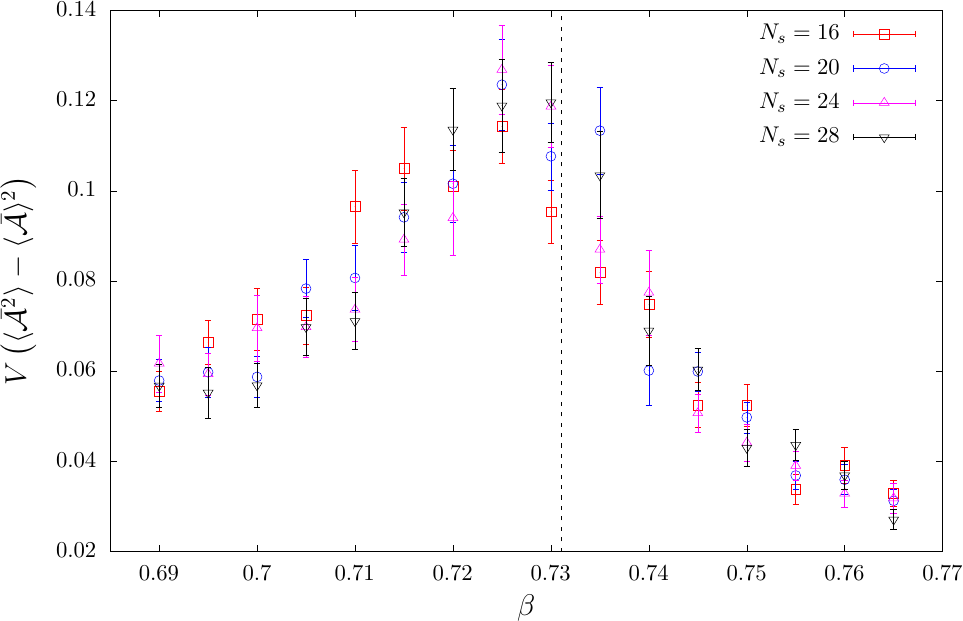}

    \caption{The average $\la \bar{\mathcal{A}}\ra$ (top panel) and
      the susceptibility
      $V\left(\la \bar{\mathcal{A}}^2\ra-\la
        \bar{\mathcal{A}}\ra^2\right)$ (bottom panel) of
      $\mathcal{A}(\vec{x})$, for several $\beta$s across the
      transition, $N_t=4$, and various $N_s$. The vertical dashed line
      corresponds to $\beta_c(N_t=4)$.}
  \label{fig:avA}
\end{figure}
\begin{figure}[t!]
  \centering

  \includegraphics[width=0.45\textwidth]{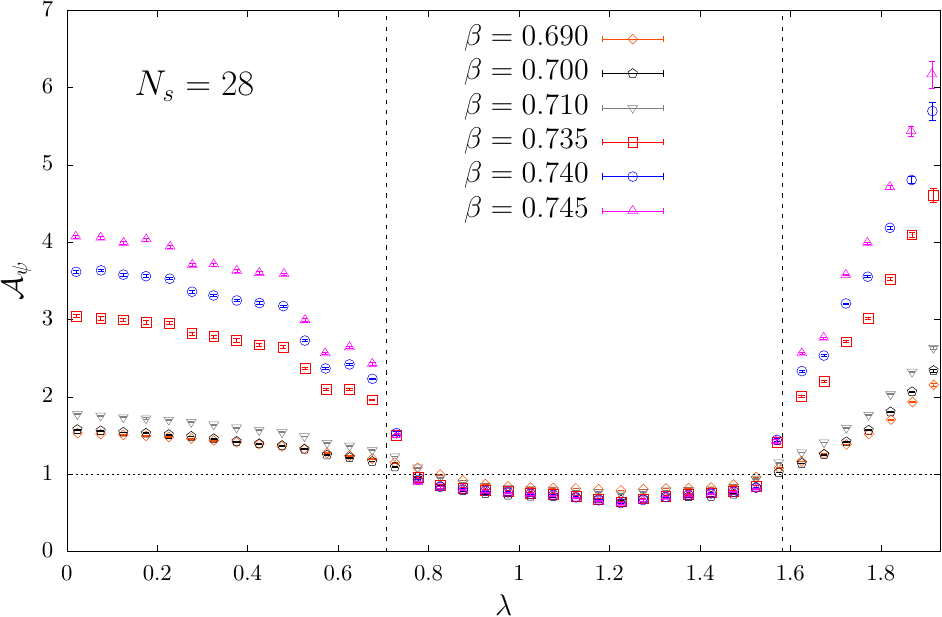}

  \caption{The quantity $\mathcal{A}_\psi$, measuring the correlation
    between islands and eigenmodes, for several values of $\beta$ both
    in the confined and deconfined phase, on a $4\times 28^2$ lattice.
    Vertical dashed lines correspond to the bulk ends $ \lambda_{(0)}$
    and $\lambda_{(1)}$.}
  \label{fig:si_simple}
\end{figure}

\begin{figure}[thb]
  \centering

  \includegraphics[width=0.45\textwidth]{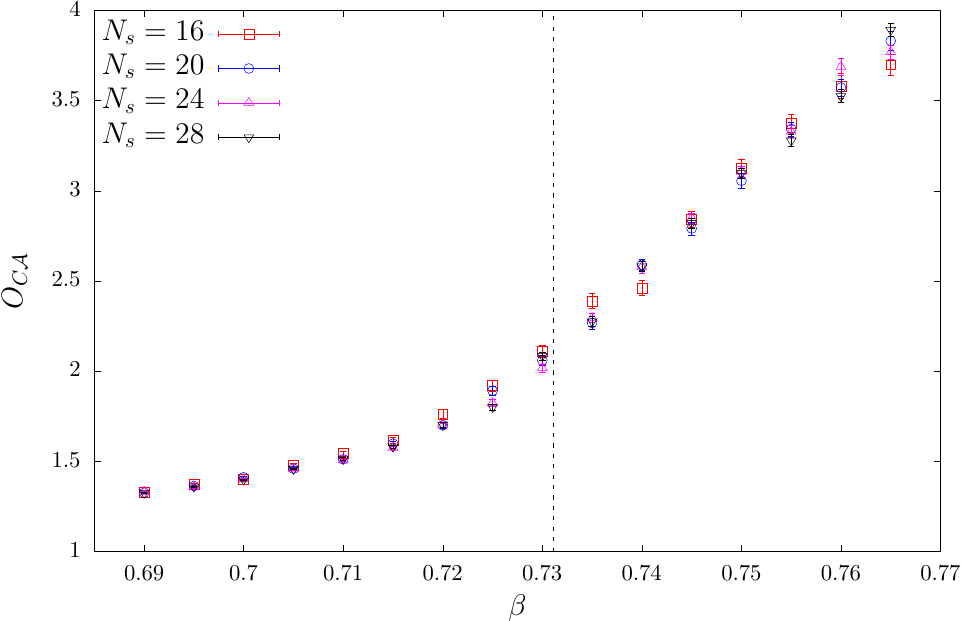}

  \caption{The quantity $O_{C\mathcal{A}}$, Eq.~\eqref{eq:cluA_alt3},
    measuring the correlation between center vortices and islands, for
    several values of $\beta$ both in the confined and deconfined
    phase, for $N_t=4$ and various volumes. The vertical dashed line
    corresponds to $\beta_c$.}
  \label{fig:si_clua}
\end{figure}

\begin{figure}[thb]
  \centering

   \includegraphics[width=0.45\textwidth]{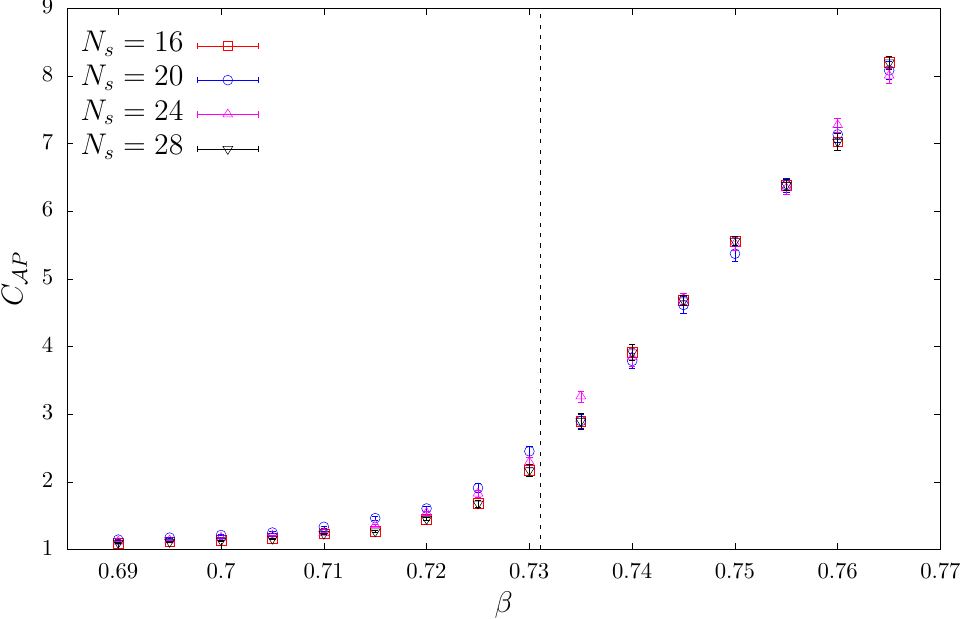}
  
   \caption{The quantity $C_{\mathcal{A}P}$, Eq.~\eqref{eq:AP_alt2},
     measuring the correlation between Polyakov loops and islands, for
     several values of $\beta$ both in the confined and deconfined
     phase, for various volumes. The vertical dashed line corresponds
     to $\beta_c$.}
  \label{fig:si_AP}
\end{figure}

\subsection{Refined sea/islands picture}
\label{sec:numrefsi}

To test the refined sea/islands picture of Ref.~\cite{Baranka:2022dib}
we measured the quantity ${\cal A}(\vec{x}) $ of
Eq.~\eqref{eq:newsi_meas} and its correlation with staggered
eigenmodes, using the same lattice setups and configurations as in the
previous subsection.

The average of
$\bar{\mathcal{A}}\equiv\f{1}{V}\sum_{\vec{x}}{\cal A}(\vec{x})$ is
shown in Fig.~\ref{fig:avA}, top panel. It steadily decreases as
$\beta$ increases, indicating that the gauge fluctuations that locally
increase the magnitude of $A_j$ become less and less frequent as one
moves into the deconfined phase, as expected. The susceptibility
$V \left(\la\bar{\mathcal{A}}^2\ra-\la\bar{\mathcal{A}}\ra^2\right)$
is shown in Fig.~\ref{fig:avA}, bottom panel: it displays a peak near
the critical coupling, but does not show a clear scaling with the
volume.

The correlation between islands and modes is studied by looking at
$ {\cal A}_\psi$, i.e., ${\cal A}(\vec{x})$ as seen by the modes
normalized by $ \la {\cal A}\ra$, Eq.~\eqref{eq:modwA}. This is shown
in Fig.~\ref{fig:si_simple}. For both low and high modes,
$ {\cal A}_\psi$ is larger than 1 in both phases. In the deconfined
phase the ratio shoots up for low modes, showing that they tend to
localize where ${\cal A}(\vec{x})$ is larger. High modes are localized
already in the confined phase, but for them the ratio shoots up in the
deconfined phase even more than for low modes. For bulk modes instead
$ {\cal A}_\psi$ is smaller than 1 in both phases, indicating that
these modes are repelled by the islands.

Finally, in Figs.~\ref{fig:si_clua} and \ref{fig:si_AP} we show the
correlation between islands and center vortices, and between islands
and Polyakov loops, respectively. As a measure of the correlation
between clusters and islands we used
\begin{equation}
  \label{eq:cluA_alt3}
  O_{C\mathcal{A}} =  \f{\left\la \sum_n
      \sum_{\vec{x}} \mathcal{A}(\vec{x}) \f{1}{N_t}\sum_{t}  \f{
        c_n(t,\vec{x})}{n_c(t,\vec{x})}\right\ra}{
    \left\la \mathcal{A}(\vec{x}) \right\ra
    \left\la \sum_n \sum_{\vec{x}}
      \f{1}{N_t}\sum_{t}  \f{
        c_n(t,\vec{x})}{n_c(t,\vec{x})}\right\ra}
  \,,  
\end{equation}
where $c_n(t,\vec{x})=1$ if the site $(t,\vec{x})$ on the direct
lattice is a corner of a plaquette whose corresponding dual link
belongs to cluster $n$, and $n_c(t,\vec{x})$ is the number of clusters
that touch site $(t,\vec{x})$. For the correlation between islands and
Polyakov loops we used $C_{\mathcal{A}P}$ defined in
Eq.~\eqref{eq:AP_alt2}, measured on 700 configurations. Both
correlations become very strong in the deconfined phase.\footnote{For
  a quantitative comparison between the magnitudes of the two types of
  correlation one should normalize by the fluctuations of the various
  quantities instead of their averages. On the other hand, an accurate
  quantitative comparison is not really possible: while for
  $C_{\mathcal{A}P}$ both quantities are defined on the direct lattice
  sites, $O_{C\mathcal{A}}$ involves a quantity defined on the direct
  lattice and one defined on the dual lattice.} This strongly suggests
that the islands relevant to localization are mainly found at the
location of those center vortices associated with Polyakov-loop
fluctuations away from order. Notice, however, that only about half of
the mode weight is found on the lattice sites corresponding to these
fluctuations for the lowest modes, a fraction that decreases as one
approaches the mobility edge~\cite{Baranka:2021san}. This should be
compared with the larger fraction found on center vortices, that
therefore identify more accurately the localization region of low
modes. Moreover and more importantly, as we show in the next
subsection, low modes localize mostly where the gauge configuration
shows a more complex structure on the dual lattice than for a simple
Polyakov-loop fluctuation, indicating that center vortices play a more
fundamental role.

These results suggest a qualitative explanation for the peculiar
dependence of the mobility edge on $\beta$.  As already pointed out,
the size of the pseudogap is almost equal to $\lambda_{(0)}$,
independently of $\beta$, probably due to the discrete nature of the
gauge group. Indeed, the diagonal matrix $E$,
Eq.~\eqref{eq:DA_general}, has only two, sharply different possible
sets of entries, corresponding to sites where $P(\vec{x})=1$ or
$P(\vec{x})=-1$. Looking at the hopping, off-diagonal term in the
Dirac-Anderson Hamiltonian $H^{\rm DA}$ as a perturbation of the
diagonal term, one sees then that unperturbed eigenvectors localized
on sites where $P(\vec{x})=-1$ (i.e., on islands) do not mix easily
with those localized on sites where $P(\vec{x})=1$ (i.e., in the sea),
and so the delocalized region of the spectrum should start from around
$\lambda_{(0)}$. Moreover, since avoiding islands likely comes at a
cost, one expects the bulk eigenvalues to be pushed further up,
increasing the mobility edge above $\lambda_{(0)}$. As the density of
islands decreases with increasing $\beta$, their avoidance becomes
easier and the mobility edge decreases towards $\lambda_{(0)}$.

\subsection{Center vortex structures and localization}
\label{sec:spconf}

As we have already pointed out, the spectral density $\rho$,
Fig.~\ref{fig:spd}, shows clear peak structures at specific points in
the low part of the spectrum. At the same points, the mode size
$\mathrm{IPR}^{-1}$, Fig.~\ref{fig:PR}, shows clear dip structures,
indicating correspondingly a stronger localization; and the maximal
mode weight on a cluster divided by the cluster size,
$D_{\mathrm{max}}$, Fig.~\ref{fig:clmaxwpersite_vol0}, also shows
clear peaks, indicating a higher concentration of the mode on a single
cluster. Structure is visible, although to a much lesser extent, also
in other observables, such as $I_{s_0}$, Fig.~\ref{fig:9}, the cluster
IPR, $\mathrm{IPR}_{\mathrm{clust}}$, Fig.~\ref{fig:clipr_vol}, and
the maximal mode weight on a cluster, $W_{\mathrm{max}}$,
Fig.~\ref{fig:clmaxw_vol}.  These peaks and dips survive, or even
become sharper, as the volume increases. As we now show, the presence
of these structures in the observables can be traced back to the
presence of specific structures in the gauge field, best described in
terms of the center vortices they form in the dual lattice.

\begin{table}[t]
  \centering
  \begin{tabular}{c|c|c|c|c}
    & object $O$ & $\lambda_O$ & $\tilde{\lambda}_{O}$
    & $\mathcal{D}_O$ \\    \hline
    \raisebox{-0.4cm}{\includegraphics[width=0.07\textwidth]{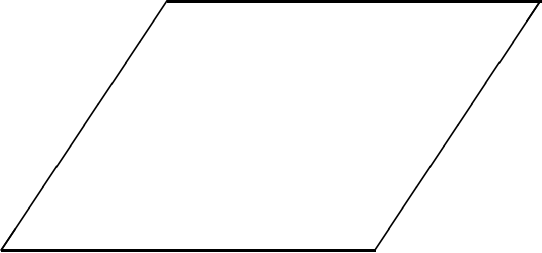}}
    & \parbox[t]{1.5cm}{horizontal\\ square}
    & \parbox[t]{1.5cm}{$0.4932906$ \\ $0.5896407$ } &
                                                       \parbox[t]{1.5cm}{$1.709904$\\     $1.628596$}
    & \parbox[t]{1.5cm}{$1$\\$4$}
    \\                                    
    \raisebox{-0.6cm}{\includegraphics[width=0.05\textwidth]{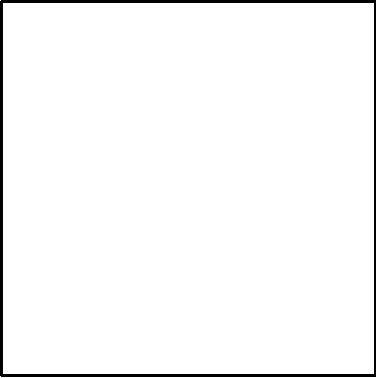}}  &
                                                                              \parbox[t]{1.5cm}{vertical\\
    square}\rule{0pt}{16pt}
    &    \parbox[t]{1.5cm}{ $0.5511360$\\ $0.6697194$} & 
                                                         \parbox[t]{1.5cm}{$1.672366$ \\  $1.597334$ }
    & \parbox[t]{1.5cm}{$1$\\$4$}
    \\    
    \raisebox{-0.2cm}{\includegraphics[width=0.11\textwidth]{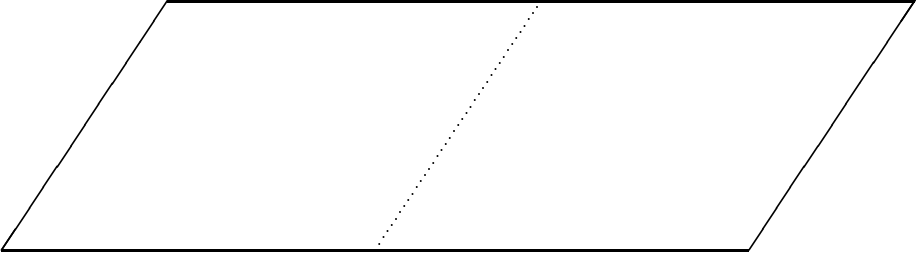}}
    & flat brick\rule{0pt}{22pt}
                 & $0.5462419$ & $1.673471$ & $2$\\
    \raisebox{-0.4cm}{\includegraphics[width=0.1\textwidth]{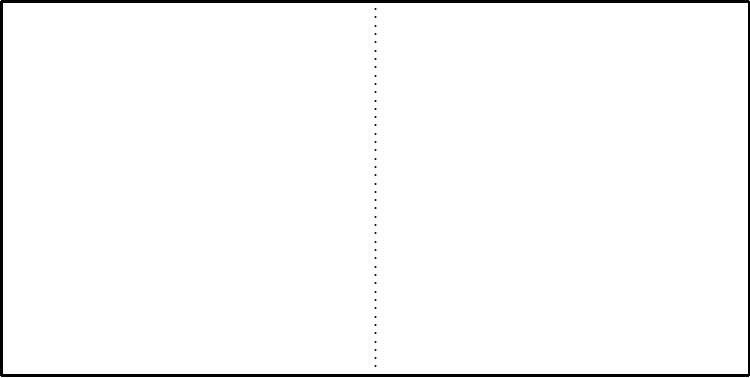}}
    & lying brick\rule{0pt}{22pt} & $0.5943152$ & $1.626896$  & $2$
    \\
    \raisebox{-1.4cm}{
    \includegraphics[width=0.02\textwidth]{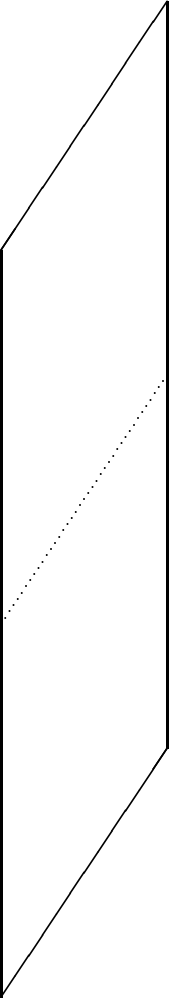}} &
                                                            \parbox[t]{1.5cm}{standing\\
    brick}\rule{0pt}{28pt}
    &
      $0.6420283$ & $1.646262$
                               & $2$
    \\
    \raisebox{-0.8cm}{
    \includegraphics[width=0.07\textwidth]{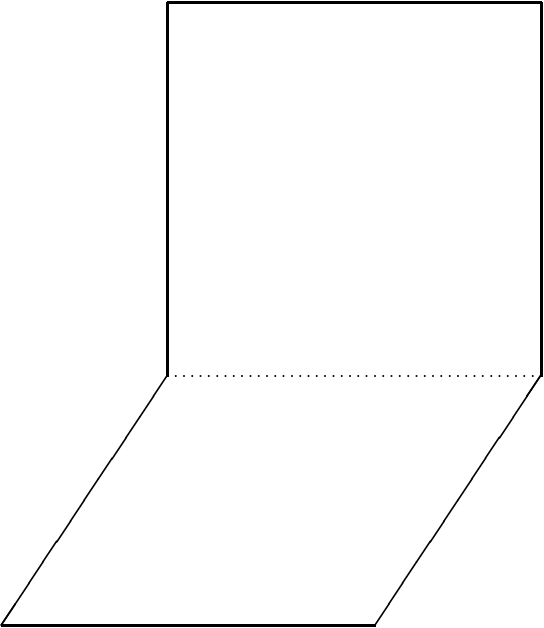}} &
                                                         chair \rule{0pt}{24pt}
                 & $0.2390408$ & $1.752455$ & $1$ \\
    \raisebox{-0.8cm}{
    \includegraphics[width=0.07\textwidth]{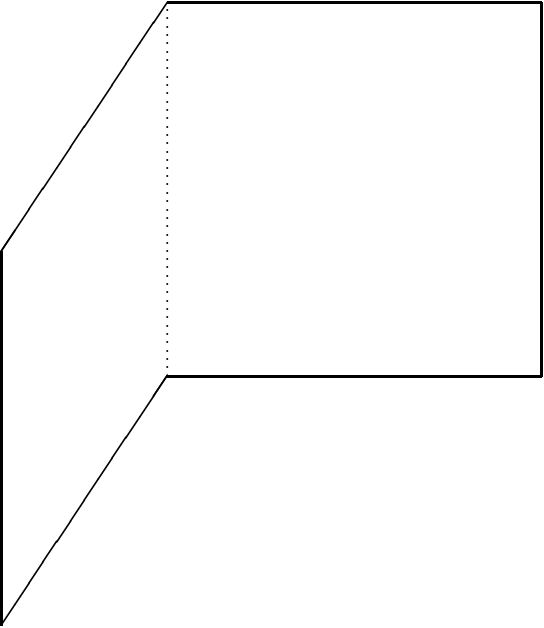}} & open
                                                        book \rule{0pt}{24pt}
                 & $0.2928932$ & $1.707107$ & $1$\\
    \raisebox{-0.8cm}{\includegraphics[width=0.07\textwidth]{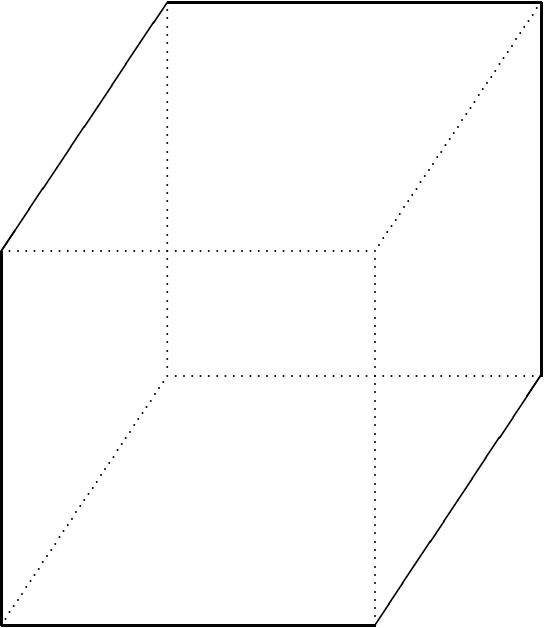}}
    & twister\rule{0pt}{24pt} & $0$ & $\sqrt{3}$ & $1$
      \end{tabular}
      \caption{Simplest dual-lattice center-vortex configurations $O$,
        eigenvalues of the corresponding low ($\lambda_O$) and high
        ($\tilde{\lambda}_O$) localized mode (computed on a
        $4\times 20^2$ lattice) and their degeneracy $\mathcal{D}_O$.
        The temporal direction is along the vertical. For the
        horizontal and vertical square, also the eigenvalues
        corresponding to their ``big'', $2\times 2$ versions are
        reported on a subsequent line. For the ``twister''
        configuration we found eigenvalues consistent with $0$ and
        $\sqrt{3}$ within numerical precision.}
  \label{tab:obj}
\end{table}

\begin{figure}[t]
  \centering
  \includegraphics[width=0.48\textwidth]{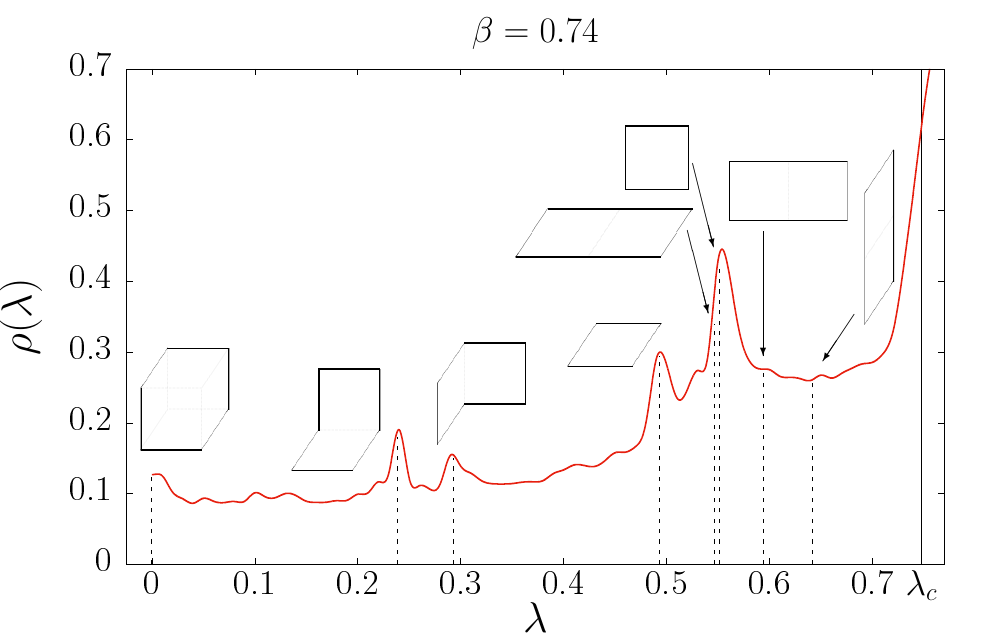}
  \caption{Comparison of the normalized spectral density
    $\rho(\lambda)$, Eq.~\eqref{eq:spdens_def}, at $\beta=0.74$ on a
    $4\times 72^2$ lattice with the localized low modes of the
    simplest vortex configurations of Tab.~\ref{tab:obj}.}
  \label{fig:specdensobj}
\end{figure}

\begin{figure}[t!]
  \centering
  \includegraphics[width=0.48\textwidth]{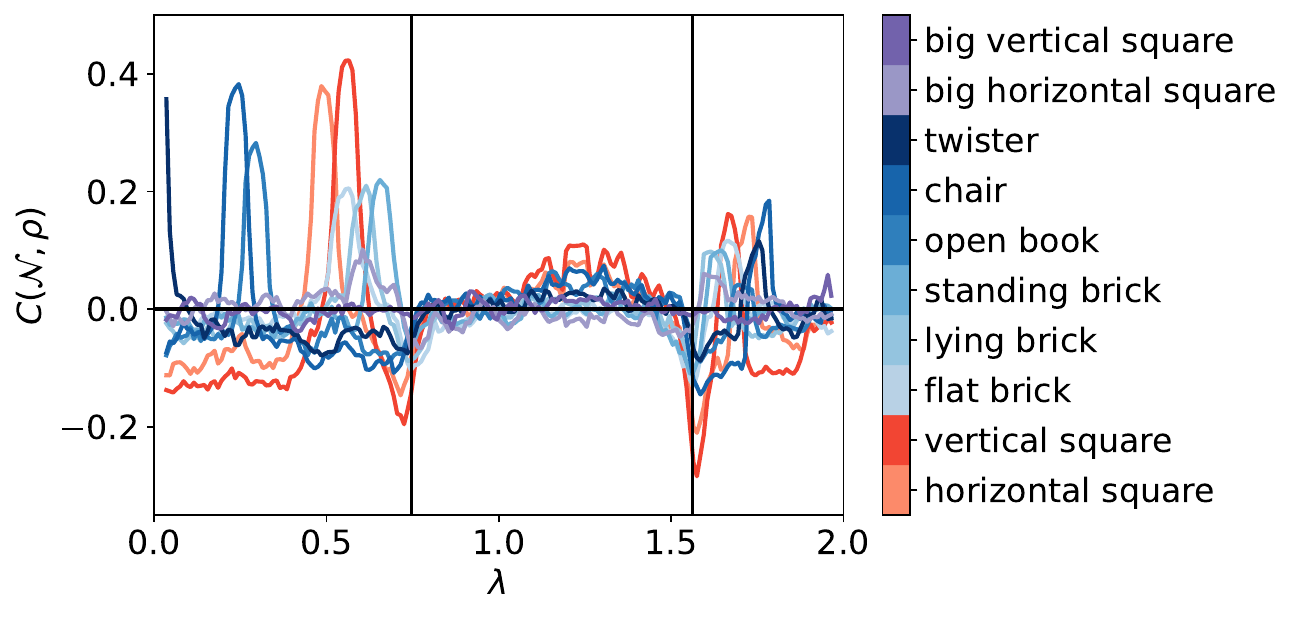}

  \includegraphics[width=0.34\textwidth]{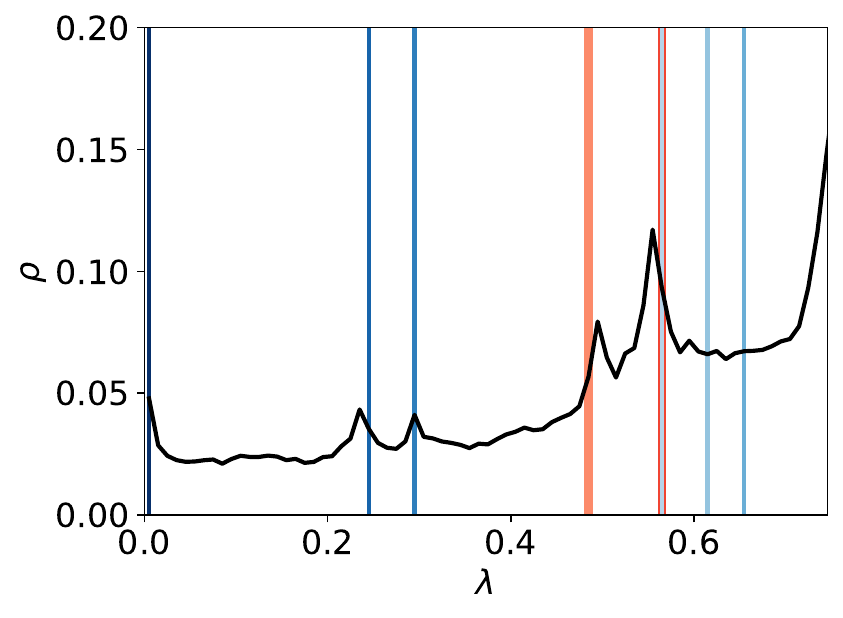}\hfil

  \includegraphics[width=0.34\textwidth]{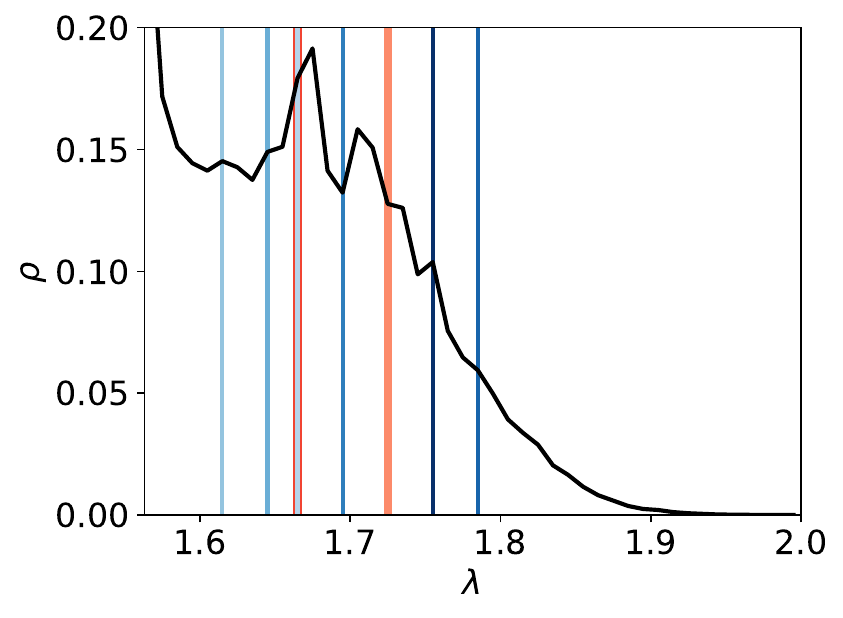}

  \includegraphics[width=0.34\textwidth]{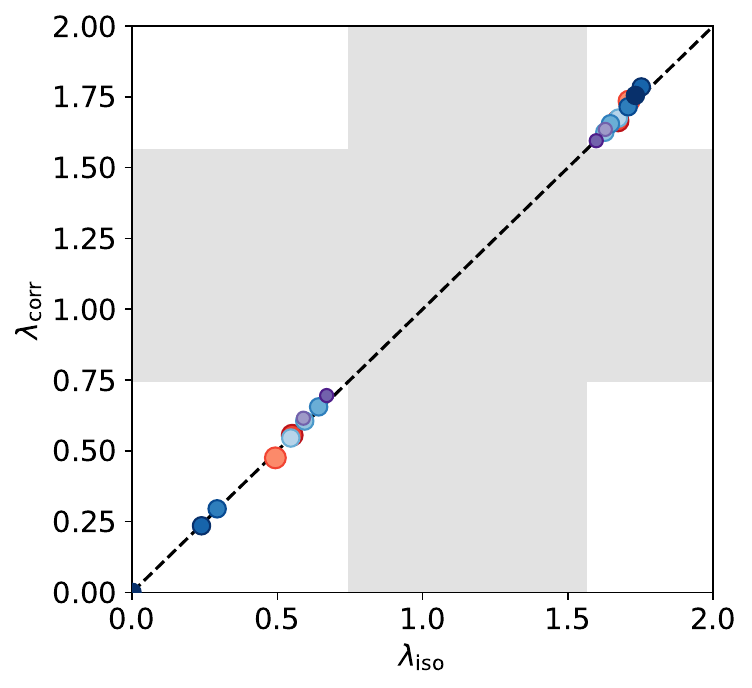}
  \caption{From top to bottom: Correlation between the number of
    center-vortex objects listed in Tab.~\ref{tab:obj} and the
    spectral density, Eq.~\eqref{eq:corrNrho} (first panel);
    correlation peaks of the first panel superimposed on the
    normalized spectral density for the low (second panel) and high
    modes (third panel); Position of the correlation peaks
    $\lambda_{\mathrm{corr}}$ against the eigenvalues
    $\lambda_{\mathrm{iso}}=\{\lambda_O,\tilde{\lambda}_O\}$ of the
    localized modes found in the presence of isolated objects of the
    type listed in Tab.~\ref{tab:obj} (fourth panel).  Here
    $\beta=0.74$, $N_s=20$, and $N_t=4$.}
  \label{fig:corrobj}
\end{figure}
The simplest center vortices are those of length 4 and 6, depicted in
Tab.~\ref{tab:obj}.  Diagonalizing gauge configurations that contain
one object of type $O$,\footnote{Both the vortex structure and the
  Dirac spectrum are gauge invariant, so the association of a
  staggered spectrum with $O$ is unambiguous.}  one obtains a spectrum
mostly comprising delocalized modes in the bulk region of the
spectrum, and one (possibly degenerate) localized mode $\lambda_O$ at
the low end of the spectrum, and one, $\tilde{\lambda}_O$, at the high
end of the spectrum (again possibly degenerate, with the same
degeneracy $\mathcal{D}_O$ as its low-mode counterpart). The
eigenvalues of the localized modes found outside the bulk in these
configurations match almost exactly the main peaks and dips in
Figs.~\ref{fig:spd}, \ref{fig:PR}, \ref{fig:9}, \ref{fig:clipr_vol},
\ref{fig:clmaxw_vol}, and \ref{fig:clmaxwpersite_vol0}. We show this
by comparing the spectral density of low modes with the position of
the localized eigenmodes corresponding to the various objects in
Fig.~\ref{fig:specdensobj}. This is no coincidence. In
Fig.~\ref{fig:corrobj} (first panel) we show the correlation between
the number $\mathcal{N}_O$ of objects of type $O$ and the spectral
density,
\begin{equation}
  \label{eq:corrNrho}
  C(\mathcal{N}_O,\rho(\lambda)) \equiv \f{\la \mathcal{N}_O
    \sum_n\delta(\lambda-\lambda_n)\ra}{
    \la \mathcal{N}_O\ra \la \sum_n\delta(\lambda-\lambda_n)\ra} -1\,.
\end{equation}
Here objects are counted by identifying exact matches between
disconnected components of the vortex configuration on the dual
lattice and the configurations shown in Tab.~\ref{tab:obj}, possibly
rotated or reflected in space, or reflected in the temporal
direction, or both.  For a given type of object, a strong correlation is found
precisely at the position of the corresponding localized low and high
eigenmodes, $\lambda_O$ and $\tilde{\lambda}_O$, found in the presence
of a single, isolated object; a clear anticorrelation is also seen
near the mobility edges (see footnote \ref{foot:highlc}). The
position of the peaks in the correlation match almost perfectly those
in the spectral density of low modes, as shown in
Fig.~\ref{fig:corrobj} (second panel), while the matching seems less
nice at the high end of the spectrum (Fig.~\ref{fig:corrobj}, third
panel). However, although relatively rare, the objects are not
isolated from each other, and so the spectrum is not simply the union
of the individual spectra of the configurations corresponding to the
objects in Tab.~\ref{tab:obj}, and the positions of the peaks in
$\rho$ are expected to deviate from the isolated-object case. On the
other hand, a comparison between the positions of the correlation
peaks and the eigenvalues of the localized modes for isolated objects
(fourth panel) shows that even in a heterogeneous environment the
spectrum responds the most to the center-vortex configurations of
Tab.~\ref{tab:obj} almost exactly where the ``free'' eigenvalue would
be. The bigger deviation of the spectral density peaks from the
correlation peaks for the high modes is then likely due to the fact
that the localized high modes associated with isolated objects have
eigenvalues closer to each other than the corresponding localized low
modes, and therefore are more likely to mix when several objects are
present, distorting the peak structure from the isolated-object case
more than at the low end of the spectrum.

It is worth noting that the number of objects of a certain type,
divided by the number of its different possible orientations, appears
to be distributed according to a Poisson distribution with parameter
$\Lambda_O = N_0 V e^{-k_O s}$, where $k_O$ is the ``order'' of the
object, i.e., the number of dual lattice plaquettes in the minimal
surface having as boundary the closed vortex line that defines the
object. At $\beta=0.74$ the numerical coefficients $N_0$ and $s$ are
$N_0=1.82(3)\times 10^{-3}$ and $s=3.17(3)$.

\begin{figure}[t]
  \centering
  \includegraphics[width=0.46\textwidth]{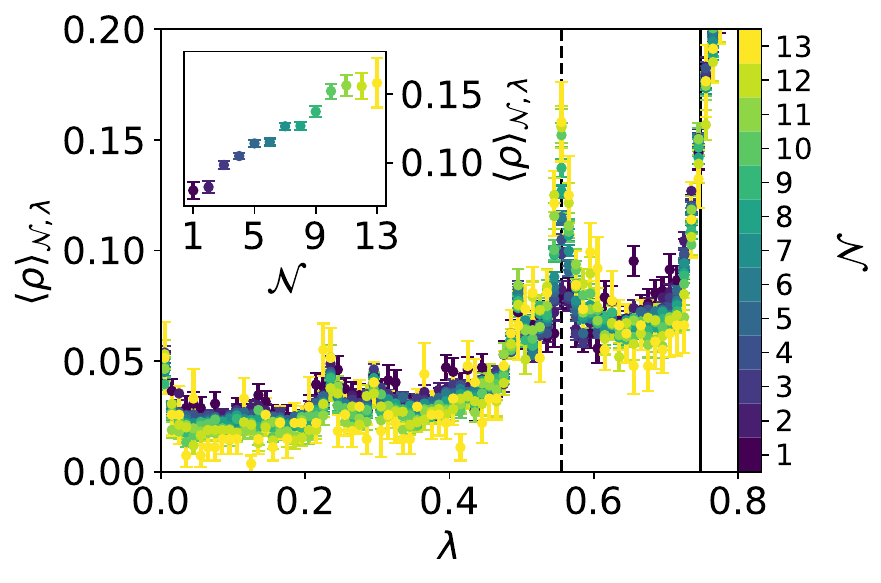}

  \includegraphics[width=0.46\textwidth]{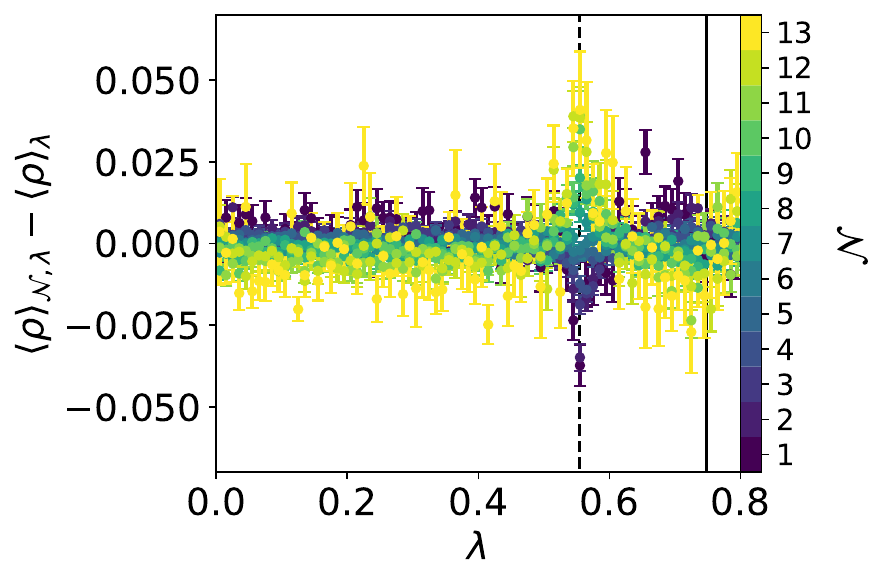}
  
  \caption{Normalized spectral density at fixed number $\mathcal{N}$
    of vertical squares (top panel), and its difference with the
    full normalized spectral density (bottom panel). The dashed line
    is at $\lambda\simeq 0.55$, which is the eigenvalue of the
    localized low mode corresponding to an isolated vertical square;
    the solid line indicates the position of the mobility edge. Here
    $\beta=0.74$, $N_s=20$, and $N_t=4$.}
  \label{fig:linresobj}
\end{figure}
We mention in passing that sharp peaks ending in a cusp were noticed
in the spectral density in the deconfined phase in
Ref.~\cite{Baranka:2021san}, at $\sqrt{3/2}$ in the physical sector
($\bar{P}>0$), and at $1$ and $\sqrt{2}$ in the unphysical sector
($\bar{P}<0$). These peaks are of an entirely different origin, and
are related to the Van Hove singularities of the free spectrum. We
discuss this point in Appendix~\ref{sec:vhsing}.

\begin{figure}[t]
  \centering
    \includegraphics[width=0.49\textwidth]{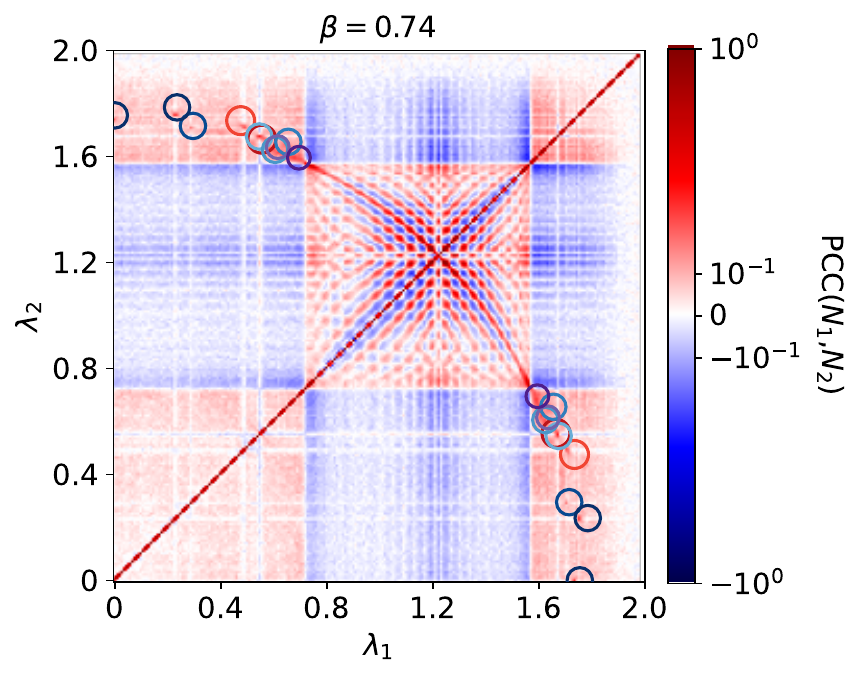}
  
    \caption{Pearson correlation coefficient for the mode number,
      Eq.~\eqref{eq:modenumbcorr}.  Circles correspond to the pairs of
      localized low and high modes associated with the objects listed
      in Tab.~\ref{tab:obj}.  Here $\beta=0.74$, $N_s=20$, and
      $N_t=4$.}
  \label{fig:eigcorr}
\end{figure}
\begin{figure}[t!]
  \centering
   \includegraphics[width=0.49\textwidth]{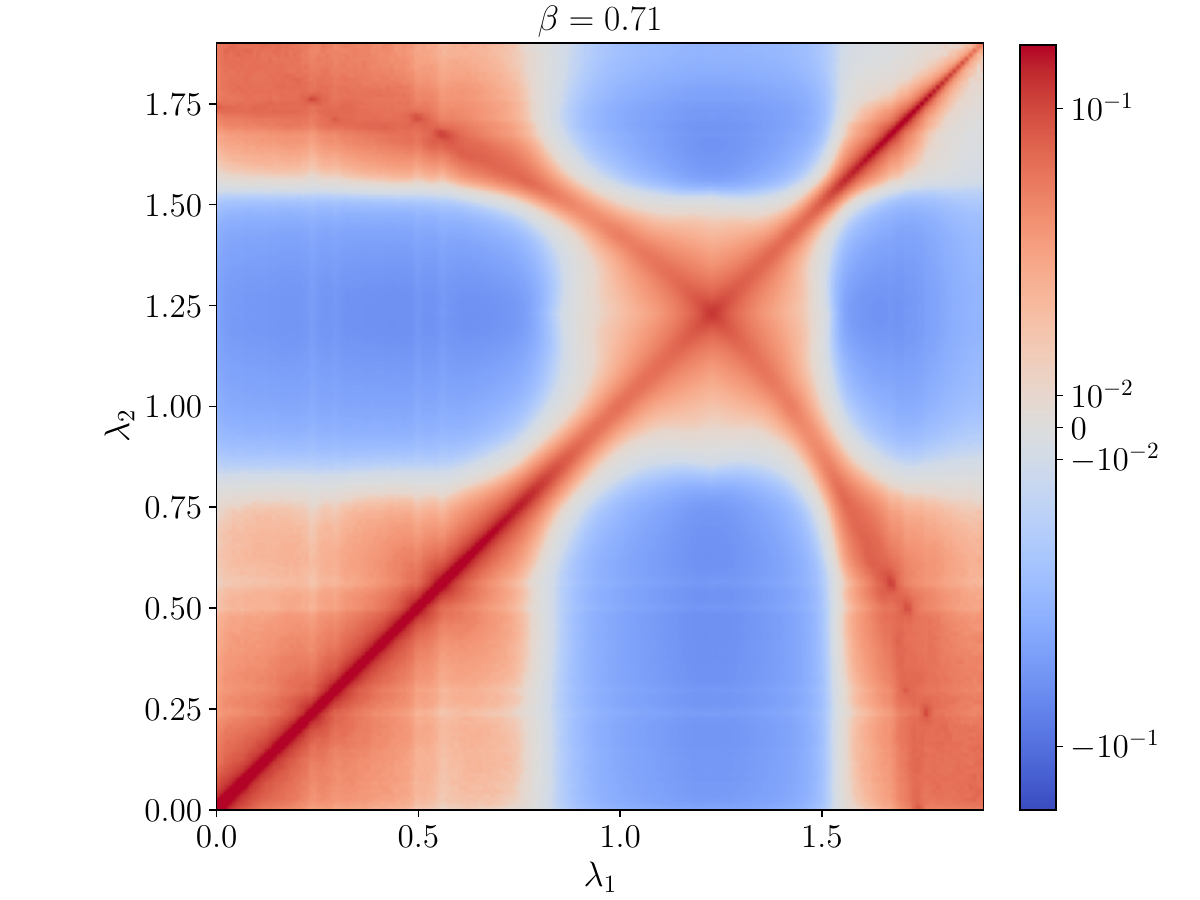}
  
   \includegraphics[width=0.49\textwidth]{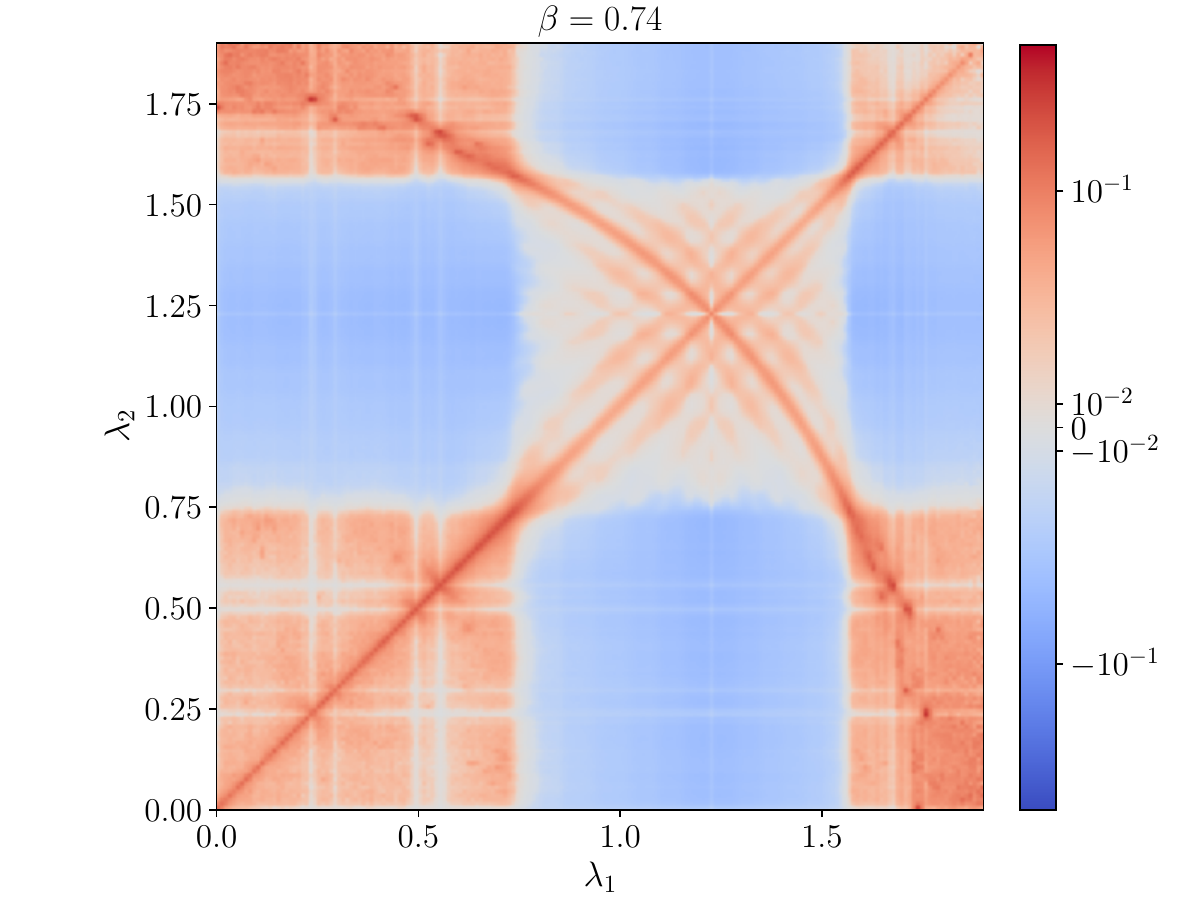}
   \caption{Two-point eigenvector correlator $C$,
     Eq.~\eqref{eq:eigenveccorr}, in the confined phase at
     $\beta=0.71$ (top panel) and in the deconfined phase at
     $\beta=0.74$ (bottom panel).  Here $N_s=20$ and $N_t=4$. For
     better visualization, a linear scale is used in the range
     $|C|\le 10^{-2}$ and a logarithmic scale outside of it.}
  \label{fig:eigcorr2}
\end{figure}

Further evidence of the relation between the peak structures and the
simple center vortex structures is provided by the nearly linear
response of the spectral density to the number of objects of type $O$
near $\lambda_O$.  In Fig.~\ref{fig:linresobj} we show the spectral
density of low modes obtained using those configurations from one of
our ensembles that contain a fixed number of ``vertical squares'', for
several such numbers. The strongest, and almost linear response is
found near $\lambda\simeq 0.55$, which is the eigenvalue of the
corresponding localized low mode. For each object type we then counted
the number of modes $N_O^{(0)}$ within the two peaks of positive
correlation seen in Fig.~\ref{fig:corrobj}, and checked how it related
with the number $\mathcal{N}_O$ of objects of the given type. We found
an approximately linear response, with
\begin{equation}
  \label{eq:linresp}
  \left \la \f{N_O^{(0)} - \la N_O^{(0)}\ra}{\mathcal{N}_O-\la
      \mathcal{N}_O\ra}\right\ra \simeq c \mathcal{D}_O \,,
\end{equation}
with $\mathcal{D}_O$ the degeneracy of the localized modes given in
Tab.~\ref{tab:obj}, and $c\simeq 0.79(3)$. The distortive effects on
the spectrum due to several objects being present at the same time in
a typical gauge configuration are likely the reason of the reduction
of $c$ from the naive expectation $c=2$.

Since localized modes originate from the same gauge-field structures,
being basically the result of mixing of the localized modes associated
with isolated center vortices, an increase in the number of center
vortices in the configuration should drive up the number of localized
modes independently of where they are found in the spectrum. One then
expects a positive correlation in the local density of modes at
different points in the localized regime of the spectrum, including
between low and high modes. This is clearly visible in
Fig.~\ref{fig:eigcorr}, where we show the Pearson correlation
coefficient
\begin{equation}
  \label{eq:modenumbcorr}
  \mathrm{PCC}(N_1,N_2)\equiv \f{\la (N_1-\la N_1\ra) (N_2-\la
    N_2\ra)\ra}{\sqrt{(\la N_1^2\ra-\la N_1\ra^2) (\la N_2^2\ra-\la N_2\ra^2)}}
  \,,
\end{equation}
between the number of modes $N_{1,2}$ found in disjoint spectral bins
of equal size $\Delta\lambda=0.01$. Particularly strong correlations
are observed for the pair of bins corresponding to the low and high
localized modes associated with one of the simplest objects (i.e.,
center vortices), $\lambda_O$ and $\tilde{\lambda}_O$.

One expects to find a similarly strong correlation also between the
eigenvectors corresponding to low and high modes near $\lambda_O$ and
$\tilde{\lambda}_O$, as they should be localized in the same region
near one of the simplest center-vortex structures. This is confirmed
by Fig.~\ref{fig:eigcorr2}, where we show the following quantity,
\begin{equation}
  \label{eq:eigenveccorr}
  \begin{aligned}
    C(\lambda_1,\lambda_2) & \equiv \f{\la \sum_{l_1\neq l_2}
      \delta(\lambda_1-\lambda_{l_1})\delta(\lambda_2-\lambda_{l_2})
      X_{l_1 l_2} \ra}{\left[ \Delta(\lambda_1)\Delta(\lambda_2)\right]^{\f{1}{2}}}\,,\\
    X_{l_1 l_2} &\equiv
    \left( \textstyle\sum_n|\psi_{l_1}(n)|^2 |\psi_{l_2}(n)|^2\right)- \tf{1}{V}\,, \\
    \Delta(\lambda) &\equiv \la
    \textstyle\sum_{l}\delta(\lambda-\lambda_{l})\left(\mathrm{IPR}_{l}
      -\tf{1}{V}\right)\ra\,.
  \end{aligned}
\end{equation}
The anticorrelation between localized and bulk modes can be explained
by taking into account the sum rule $\sum_l |\psi_l(n)|^2=1$, that
follows from the completeness relation for the eigenvectors.  In fact,
localized modes give large contributions to the sum if $n$ is in the
regions where they are localized, i.e., those occupied by center
vortices. This leads to almost saturating the sum rule, so that little
is left for the delocalized modes to share, leading in practice to
them avoiding the vortices.

The positive correlation between localized modes seen in
Fig.~\ref{fig:eigcorr2} can be understood as a consequence of the
mixing of modes associated with isolated objects, mentioned above. If
a configuration contains $N_O$ objects of size $V_O$, and localized
modes typically spread over $n_O$ of them, then their typical size is
$n_OV_O$ and the chance that two modes share one and the same object
is $(n_O/N_O)^2$, and so one finds
\begin{equation}
  \label{eq:eigcorr_loc1}
  \begin{aligned}
    \sum_n|\psi_{l_1}(n)|^2 |\psi_{l_2}(n)|^2 &\sim N_OV_O
    \left(\f{n_O}{N_O}\right)^2\left(\f{1}{n_OV_O}\right)^2 \\ &=
    \f{1}{N_OV_O}\,,
  \end{aligned}
\end{equation}
leading to
\begin{equation}
  \label{eq:eigcorr_loc2}
  C(\lambda_1,\lambda_2) \sim
  \left(\f{V}{N_O V_O}-1\right)[\rho(\lambda_1)\rho(\lambda_2)]^{\f{1}{2}}n_OV_O\,.
\end{equation}
Since $N_OV_O/V<1$ is the fraction of the lattice occupied by objects,
the correlation is positive.

Interestingly, in Fig.~\ref{fig:eigcorr2} a positive correlation
between low and high modes, particularly strong near
$(\lambda_1,\lambda_2)=(\lambda_O,\tilde{\lambda}_O)$, is observed
also in the confined phase. As a nonzero density of finite clusters is
present also in the confined phase (see Sec.~\ref{sec:cp}), it is not
surprising that they support the nonzero spectral density of localized
high modes, as this result further confirms. It is somewhat
surprising, though, that a strong correlation with the spectral
regions where the low-mode counterparts would be found survives also
in the confined phase, where low modes are delocalized. This indicates
that low modes still carry some local structure highly correlated with
the center-vortex configuration. This also explains their
anticorrelation with the bulk modes.

The quarter-circular structure visible in the bulk in
Figs.~\ref{fig:eigcorr} and \ref{fig:eigcorr2} reflects the symmetry
properties of the free staggered spectrum, Eq.~\eqref{eq:spec_free}.
In fact, for $N_{t}$ and $N_{s}$ multiples of 4, there are pairs of
free eigenvalues obeying
$ \lambda(\omega,\vec{p}\,)^2 + \lambda(\omega',\vec{p}^{\,\prime})^2
= 3$, where $\omega'-\omega=p_1'-p_1=p_2'-p_2=\f{\pi}{2}$
corresponding to $k_0'-k_0=k_1'-k_1=k_2'-k_2 =\f{N_t}{4}$.  The
correlation between modes lying on the line
$\lambda_1^2 + \lambda_2^2=3$ remains even in the interacting case,
although it is partially smeared out in the confined phase (see
Fig.~\ref{fig:eigcorr2}, top panel).  This can be understood by
noticing that if one could neglect the effect of nontrivial
plaquettes, then in $d$ spatial dimensions and for generic (compact)
gauge group one would find
\begin{equation}
  \label{eq:approx_stag}
  \begin{aligned}
    (-iD^{\mathrm{stag}})^2&\approx \f{d+1}{2} -
    \f{1}{4}\sum_{\mu=1}^{d+1} \left[(U_\mu T_\mu)^2 + (T_\mu^\dag
      U_\mu^\dag)^2\right] \\ &\equiv \f{d+1}{2} - \tilde{H}\,.
  \end{aligned}
\end{equation}
Setting $\zeta(n)\equiv i^{\sum_\mu n_\mu}$, one has
$\{\zeta,\tilde{H}\}=0$, meaning that $\tilde{H}$ has a symmetric
spectrum,
$\tilde{H}\tilde{\psi}_{\pm l}=\pm \tilde{\lambda}_l\tilde{\psi}_{\pm
  l}$, with $\tilde{\lambda}_l>0$ and
$\tilde{\psi}_{-l}=\zeta\tilde{\psi}_l$.  One then finds for
$(-iD^{\mathrm{stag}})^2$ the approximate relations
\begin{equation}
  \label{eq:approx_stag2}
  (-iD^{\mathrm{stag}})^2\tilde{\psi}_{\pm l} \approx \left(\f{d+1}{2}
    \mp \tilde{\lambda}_l\right)\tilde{\psi}_{\pm l}\,,
\end{equation}
where $\tilde{\psi}_l$ and $\tilde{\psi}_{-l}$ have the same local
amplitude squared,
$\tilde{\psi}_l(n)^\dag\tilde{\psi}_l(n)=
\tilde{\psi}_{-l}(n)^\dag\tilde{\psi}_{-l}(n)$. Since also
$[\varepsilon,\tilde{H}]=0$, each eigenvalue is doubly degenerate, and
one finds the usual pair of positive and negative eigenvalues of equal
magnitude for $-iD^{\mathrm{stag}}$.  Consistency with the
(anti)periodic boundary conditions requires that
$\zeta(n+N_t\hat{1}) = \zeta(n+N_s\hat{\jmath}) = \zeta(n)$ for
$j=2,\ldots, d+1$, so $i^{N_t}=i^{N_s}=1$, restricting the validity of
this result to the case when $N_{t}$ and $N_s$ are both multiples of
4. Negative plaquettes are indeed rare in the deconfined phase, so
that the analysis above applies, explaining the strong correlation
found in the bulk at $\lambda_1^2 + \lambda_2^2=3$. In the confined
phase negative plaquettes are certainly not rare, but it is reasonable
that their effect on the bulk modes partially cancels out, making the
explanation above viable.

Although much less well defined, the quarter-circular structure seems
to extend also in the (low, high) and (high, low) corners of the
plot. Here the argument above certainly does not apply, since it is
precisely the nontrivial plaquettes that give rise to localized modes,
and so they cannot be neglected. A possible explanation can be
obtained combining the sum rule
$\sum_{l} \lambda_l^2 |\psi_l(n)|^2=\f{d+1}{2}$ (with $d$ again the
spatial dimensionality), specific to staggered fermions, with the
completeness sum rule $\sum_l |\psi_l(n)|^2=1$ to get
$\sum_{l} [\lambda_l^2-\f{d+1}{2}] |\psi_l(n)|^2=0$. If bulk modes in
the presence of interactions do not differ much from the free bulk
modes, then their total contribution to the sum rule is small, with
positive and negative contributions nearly cancelling out.  This seems
to be the case at least in the deconfined phase, where the argument
above shows the existence of pairs of eigenmodes $\psi_l$,
$\psi'_l\approx \zeta\psi_l$ of $-iD^{\mathrm{stag}}$ with
approximately equal local weight,
$|\psi'_l(n)|^2\approx|\zeta(n)\psi_l(n)|^2 = |\psi_l(n)|^2$, and
eigenvalues $\lambda_l$, $\lambda'_l$ approximately related as
$\lambda_l^2+\lambda^{\prime\,2}_l \approx d+1$. On the other hand, if
a localized low mode exists at some $\lambda^2<\f{d+1}{2}$, then a
similarly localized high mode should exist near $d+1- \lambda^2$ to
balance the sum rule.

The results of this subsection show in further detail how the
localization of low Dirac modes in the deconfined phase of
$\mathbb{Z}_2$ gauge theory is explained by the opening of the
spectral pseudogap due to Polyakov loop ordering, and by the presence
of disconnected finite clusters, i.e., center vortices not spanning
the whole lattice, that are the main localization centers. As both
Polyakov-loop ordering and the fragmentation of the infinite center
cluster are direct manifestations of deconfinement, the strong
connection between deconfinement and localization is evident. Notice
that center vortices being the main localization centers is not at all
in contradiction with the sea/islands picture, as Polyakov loop,
islands, and center vortices are very strongly correlated with each
other, as shown in the previous subsection. Note, however, that the
vortex structures associated with the lowest modes (see
Fig.~\ref{fig:specdensobj}) correspond to more complex structures in
the gauge configuration than a simple Polyakov-loop fluctuation. The
simplest center vortex associated with such a fluctuation is the
``horizontal square'', that is more relevant higher up in the
spectrum. This shows that center vortices play a more fundamental role
than Polyakov-loop fluctuations in setting up favorable localization
centers for the low Dirac modes.

\section{Conclusions}
\label{sec:concl}

In this paper we have investigated different aspects of the
deconfinement transition in finite-temperature (2+1)-dimensional
$\mathbb{Z}_2$ gauge theory on the lattice, and of its relation with
the localization of low Dirac modes.

As a first step, we confirmed the presence of an Anderson transition
in the spectrum of the staggered Dirac operator, separating localized
low-lying modes from delocalized bulk modes, in the deconfined phase
of the theory. The presence of localized low modes, appearing at the
deconfinement transition, was shown in Ref.~\cite{Baranka:2021san},
short only of a finite-size scaling analysis. Here we performed this
analysis, fully validating the claims of Ref.~\cite{Baranka:2021san}.

Our results are interesting also from the point of view of disordered
systems, independently of their connection with the confinement of
quarks. Our finite-size-scaling analysis shows that (2+1)-dimensional
$\mathbb{Z}_2$ gauge theory is another (spatially) two-dimensional
disordered system in the orthogonal class displaying a BKT-type
Anderson transition~\cite{Wen-ShengLiu_1999,zhang2009localization}.
Furthermore, the dependence of the position of the mobility edge on
the temperature is rather counterintuitive, and different from what
has been observed so far in lattice gauge theories. One surprising
aspect is that our results are consistent with the sudden appearance
of the mobility edge at a finite distance from the origin of the
spectrum at the critical temperature. Such a behavior seems natural
for a first-order deconfinement
transition~\cite{Bonati:2020lal,Kovacs:2021fwq}, but quite less so for
the second-order one found in the present model.

We then further established the characterization of the deconfinement
transition as a percolation transition of center
vortices~\cite{Langfeld:1998cz,Engelhardt:1999fd,Gliozzi:2002ht} (see
also Ref.~\cite{Agrawal:2024mra}), determining the percolation
temperature and finding it in agreement with the deconfinement
temperature~\cite{Caselle:1995wn}. An interesting finding is that the
critical properties of the vortex percolation transition are different
from those of ordinary two-dimensional
percolation~\cite{RevModPhys.54.235,percolation,
  smirnov2001,10.1214/11-AOP740}. This is probably due to the
long-range nature of the constraint on the percolating objects, i.e.,
the requirement that active bonds form closed loops on the dual
lattice.

We then studied the correlation between center vortices and localized
Dirac modes. Not only did we find a strong correlation between the
two, but we could also precisely identify the gauge field fluctuations
mostly responsible for the localization of both low and high modes,
and for the peculiar peak and dip features found in the spectral
density and in the mode size.

In (2+1)-dimensional $\mathbb{Z}_2$ gauge theory the very close
connection between confinement and specific topological objects
(center vortices, in this case) is almost trivially expected, due to
the simplicity of the dynamics (it is nonetheless nice to be able to
confirm it explicitly). This allowed us to investigate in an
inequivocal way (albeit, perhaps, in too simplified an environment)
the connection between deconfinement and localization of Dirac modes,
strongly suggested by a wealth of results~\cite{GarciaGarcia:2005vj,
  GarciaGarcia:2006gr,Kovacs:2009zj,Kovacs:2010wx,Bruckmann:2011cc,
  Kovacs:2012zq,Giordano:2013taa,Nishigaki:2013uya,Ujfalusi:2015nha,
  Cossu:2016scb,Giordano:2016nuu,Kovacs:2017uiz,Holicki:2018sms,
  Giordano:2019pvc,Vig:2020pgq,Bonati:2020lal,Kovacs:2021fwq,
  Cardinali:2021fpu,Baranka:2021san,Baranka:2022dib,
  Kehr:2023wrs,Baranka:2023ani,Bonanno:2023mzj,Giordano:2021qav}.
Specifically, we looked directly at the correlation between localized
modes and the center vortices whose percolation signals
confinement. Such a correlation is demonstrated here at the
microscopic level, identifying the structures in the gauge
configurations mainly responsible for localizing the Dirac modes:
these are precisely the most elementary center vortices.  This is not
trivial: localization could have been more directly related to the
local fluctuations of the (temporal) Polyakov loops, as expected in
the original sea/islands
picture~\cite{Bruckmann:2011cc,Giordano:2015vla,Giordano:2016cjs,
  Giordano:2016vhx,Giordano:2021qav}, which are correlated with center
vortices but do not coincide with them. The resulting picture is then
that as the temperature is lowered and the confinement transition is
approached, such vortices approach a percolation transition, and the
localized modes they support tend to mix more strongly, therefore
tending to delocalize. The connection between the
confinement-deconfinement transition and delocalization-localization
transition is then fully elucidated by identifying the gauge
structures that are responsible for both phenomena. This supports our
hope that identifying the structures responsible for low-mode
localization in more realistic gauge theories, including the
physically relevant case of QCD, can lead to a better understanding of
the microscopic mechanism behind confinement.

\begin{acknowledgments}
  We thank Z.~R\'acz for useful discussions, and Z.~Tulip\'ant for
  collaboration in the early stages of this work. This work was
  partially supported by NKFIH grants K-147396 and KKP-126769, and by
  the NKFIH excellence grant TKP2021-NKTA-64.
\end{acknowledgments}

  \mbox{}

\appendix

\section{Van Hove singularities}
\label{sec:vhsing}

In Ref.~\cite{Baranka:2021san}, sharp peaks ending in a cusp were
noticed in the spectral density in the deconfined phase, both in the
physical and in the unphysical sector, but no explanation was offered
for them. This is provided here.

Due to the ordering of the Polyakov loop, in the deconfined phase the
bulk of the spectrum should resemble that of the free theory in the
physical sector, and that of the free theory supplemented by a uniform
nontrivial Polyakov loop in the unphysical sector. This includes
reproducing the Van Hove singularities~\cite{ashcroft_mermin1976} of
the spectral density where the ``band energies''
$\mathcal{E}_n(\vec{p}\,) = \sqrt{(\sin\omega_n)^2 + (\sin p_1)^2 +
  (\sin p_2)^2}$ have a vanishing gradient,
$\vec{\nabla}\mathcal{E}_n(\vec{p}\,) =
[2\mathcal{E}_n(\vec{p}\,)]^{-1}(\sin 2p_1,\sin 2p_2)=0$.  Here the
various bands are characterized by
$\omega_n^{\mathrm{phys}} = \f{(2n+1)\pi}{N_t}$ for the physical
sector, and $\omega_n^{\mathrm{unphys}} = \f{2n\pi}{N_t}$ for the
unphysical sector, with $n=0,\ldots,N_t-1$ in both cases. The spatial
momentum $\vec{p}=(p_1,p_2)$ is restricted to the first Brillouin
zone; in the large-volume limit, $p_i\in [-\pi,\pi]$.

For the $N_t=4$ case considered here and in
Ref.~\cite{Baranka:2021san}, one has in the physical sector
$(\sin\omega_n^{\mathrm{phys}} )^2=1/2$, for $n=0,\ldots,3$, and the
gradient vanishes for $p_{1,2}=0,\pm\f{\pi}{2},\pm\pi$, corresponding
to $\mathcal{E}_n = \sqrt{1/2}, \sqrt{3/2}, \sqrt{5/2}$.  In the
interacting case at $\beta>\beta_c$, the smallest and largest values
are found in the localized regions of the spectrum , while a Van Hove
singularity is indeed seen at $\sqrt{3/2}$. In the unphysical sector
one has instead $(\sin\omega_n^{\mathrm{unphys}} )^2=0$, for $n=0,2$,
and $(\sin\omega_n^{\mathrm{unphys}} )^2=1$, for $n=1,3$.  The
gradient again vanishes for $p_{1,2}=0,\pm\f{\pi}{2},\pm\pi$,
corresponding to $\mathcal{E}_n = 0,1,\sqrt{2},\sqrt{3}$. (More
precisely, if $n=0,2$ and all the $p_i$ approach $0$ or $\pm \pi$,
then $\vec{\nabla}\mathcal{E}_n$ tends to zero.) The smallest and
largest values are exactly at the edge of the positive spectrum, where
deviations from the free behavior are expected and observed in the
interacting case; Van Hove singularities are indeed seen at $1$ and
$\sqrt{2}$.

\mbox{}
\bibliographystyle{apsrev4-2}
\bibliography{references_gt}

\begin{thebibliography}{80}%
\makeatletter
\providecommand \@ifxundefined [1]{%
 \@ifx{#1\undefined}
}%
\providecommand \@ifnum [1]{%
 \ifnum #1\expandafter \@firstoftwo
 \else \expandafter \@secondoftwo
 \fi
}%
\providecommand \@ifx [1]{%
 \ifx #1\expandafter \@firstoftwo
 \else \expandafter \@secondoftwo
 \fi
}%
\providecommand \natexlab [1]{#1}%
\providecommand \enquote  [1]{``#1''}%
\providecommand \bibnamefont  [1]{#1}%
\providecommand \bibfnamefont [1]{#1}%
\providecommand \citenamefont [1]{#1}%
\providecommand \href@noop [0]{\@secondoftwo}%
\providecommand \href [0]{\begingroup \@sanitize@url \@href}%
\providecommand \@href[1]{\@@startlink{#1}\@@href}%
\providecommand \@@href[1]{\endgroup#1\@@endlink}%
\providecommand \@sanitize@url [0]{\catcode `\\12\catcode `\$12\catcode
  `\&12\catcode `\#12\catcode `\^12\catcode `\_12\catcode `\%12\relax}%
\providecommand \@@startlink[1]{}%
\providecommand \@@endlink[0]{}%
\providecommand \url  [0]{\begingroup\@sanitize@url \@url }%
\providecommand \@url [1]{\endgroup\@href {#1}{\urlprefix }}%
\providecommand \urlprefix  [0]{URL }%
\providecommand \Eprint [0]{\href }%
\providecommand \doibase [0]{https://doi.org/}%
\providecommand \selectlanguage [0]{\@gobble}%
\providecommand \bibinfo  [0]{\@secondoftwo}%
\providecommand \bibfield  [0]{\@secondoftwo}%
\providecommand \translation [1]{[#1]}%
\providecommand \BibitemOpen [0]{}%
\providecommand \bibitemStop [0]{}%
\providecommand \bibitemNoStop [0]{.\EOS\space}%
\providecommand \EOS [0]{\spacefactor3000\relax}%
\providecommand \BibitemShut  [1]{\csname bibitem#1\endcsname}%
\let\auto@bib@innerbib\@empty
%</preamble>
\bibitem [{\citenamefont {Garc\'ia-Garc\'ia}\ and\ \citenamefont
  {Osborn}(2006)}]{GarciaGarcia:2005vj}%
  \BibitemOpen
  \bibfield  {author} {\bibinfo {author} {\bibfnamefont {A.~M.}\ \bibnamefont
  {Garc\'ia-Garc\'ia}}\ and\ \bibinfo {author} {\bibfnamefont {J.~C.}\
  \bibnamefont {Osborn}},\ }\href
  {https://doi.org/10.1016/j.nuclphysa.2006.02.011} {\bibfield  {journal}
  {\bibinfo  {journal} {Nucl. Phys.}\ }\textbf {\bibinfo {volume} {A770}},\
  \bibinfo {pages} {141} (\bibinfo {year} {2006})},\ \Eprint
  {https://arxiv.org/abs/hep-lat/0512025} {arXiv:hep-lat/0512025 [hep-lat]}
  \BibitemShut {NoStop}%
%%CITATION = HEP-LAT/0512025;%%
\bibitem [{\citenamefont {Garc\'ia-Garc\'ia}\ and\ \citenamefont
  {Osborn}(2007)}]{GarciaGarcia:2006gr}%
  \BibitemOpen
  \bibfield  {author} {\bibinfo {author} {\bibfnamefont {A.~M.}\ \bibnamefont
  {Garc\'ia-Garc\'ia}}\ and\ \bibinfo {author} {\bibfnamefont {J.~C.}\
  \bibnamefont {Osborn}},\ }\href {https://doi.org/10.1103/PhysRevD.75.034503}
  {\bibfield  {journal} {\bibinfo  {journal} {Phys. Rev. D}\ }\textbf {\bibinfo
  {volume} {75}},\ \bibinfo {pages} {034503} (\bibinfo {year} {2007})},\
  \Eprint {https://arxiv.org/abs/hep-lat/0611019} {arXiv:hep-lat/0611019
  [hep-lat]} \BibitemShut {NoStop}%
%%CITATION = HEP-LAT/0611019;%%
\bibitem [{\citenamefont {Kov\'acs}(2010)}]{Kovacs:2009zj}%
  \BibitemOpen
  \bibfield  {author} {\bibinfo {author} {\bibfnamefont {T.~G.}\ \bibnamefont
  {Kov\'acs}},\ }\href {https://doi.org/10.1103/PhysRevLett.104.031601}
  {\bibfield  {journal} {\bibinfo  {journal} {Phys. Rev. Lett.}\ }\textbf
  {\bibinfo {volume} {104}},\ \bibinfo {pages} {031601} (\bibinfo {year}
  {2010})},\ \Eprint {https://arxiv.org/abs/0906.5373} {arXiv:0906.5373
  [hep-lat]} \BibitemShut {NoStop}%
%%CITATION = ARXIV:0906.5373;%%
\bibitem [{\citenamefont {Kov\'acs}\ and\ \citenamefont
  {Pittler}(2010)}]{Kovacs:2010wx}%
  \BibitemOpen
  \bibfield  {author} {\bibinfo {author} {\bibfnamefont {T.~G.}\ \bibnamefont
  {Kov\'acs}}\ and\ \bibinfo {author} {\bibfnamefont {F.}~\bibnamefont
  {Pittler}},\ }\href {https://doi.org/10.1103/PhysRevLett.105.192001}
  {\bibfield  {journal} {\bibinfo  {journal} {Phys. Rev. Lett.}\ }\textbf
  {\bibinfo {volume} {105}},\ \bibinfo {pages} {192001} (\bibinfo {year}
  {2010})},\ \Eprint {https://arxiv.org/abs/1006.1205} {arXiv:1006.1205
  [hep-lat]} \BibitemShut {NoStop}%
%%CITATION = ARXIV:1006.1205;%%
\bibitem [{\citenamefont {Bruckmann}\ \emph {et~al.}(2011)\citenamefont
  {Bruckmann}, \citenamefont {Kov\'acs},\ and\ \citenamefont
  {Schierenberg}}]{Bruckmann:2011cc}%
  \BibitemOpen
  \bibfield  {author} {\bibinfo {author} {\bibfnamefont {F.}~\bibnamefont
  {Bruckmann}}, \bibinfo {author} {\bibfnamefont {T.~G.}\ \bibnamefont
  {Kov\'acs}},\ and\ \bibinfo {author} {\bibfnamefont {S.}~\bibnamefont
  {Schierenberg}},\ }\href {https://doi.org/10.1103/PhysRevD.84.034505}
  {\bibfield  {journal} {\bibinfo  {journal} {Phys. Rev. D}\ }\textbf {\bibinfo
  {volume} {84}},\ \bibinfo {pages} {034505} (\bibinfo {year} {2011})},\
  \Eprint {https://arxiv.org/abs/1105.5336} {arXiv:1105.5336 [hep-lat]}
  \BibitemShut {NoStop}%
%%CITATION = ARXIV:1105.5336;%%
\bibitem [{\citenamefont {Kov{\'a}cs}\ and\ \citenamefont
  {Pittler}(2012)}]{Kovacs:2012zq}%
  \BibitemOpen
  \bibfield  {author} {\bibinfo {author} {\bibfnamefont {T.~G.}\ \bibnamefont
  {Kov{\'a}cs}}\ and\ \bibinfo {author} {\bibfnamefont {F.}~\bibnamefont
  {Pittler}},\ }\href {https://doi.org/10.1103/PhysRevD.86.114515} {\bibfield
  {journal} {\bibinfo  {journal} {Phys. Rev. D}\ }\textbf {\bibinfo {volume}
  {86}},\ \bibinfo {pages} {114515} (\bibinfo {year} {2012})},\ \Eprint
  {https://arxiv.org/abs/1208.3475} {arXiv:1208.3475 [hep-lat]} \BibitemShut
  {NoStop}%
\bibitem [{\citenamefont {Giordano}\ \emph {et~al.}(2014)\citenamefont
  {Giordano}, \citenamefont {Kov\'acs},\ and\ \citenamefont
  {Pittler}}]{Giordano:2013taa}%
  \BibitemOpen
  \bibfield  {author} {\bibinfo {author} {\bibfnamefont {M.}~\bibnamefont
  {Giordano}}, \bibinfo {author} {\bibfnamefont {T.~G.}\ \bibnamefont
  {Kov\'acs}},\ and\ \bibinfo {author} {\bibfnamefont {F.}~\bibnamefont
  {Pittler}},\ }\href {https://doi.org/10.1103/PhysRevLett.112.102002}
  {\bibfield  {journal} {\bibinfo  {journal} {Phys. Rev. Lett.}\ }\textbf
  {\bibinfo {volume} {112}},\ \bibinfo {pages} {102002} (\bibinfo {year}
  {2014})},\ \Eprint {https://arxiv.org/abs/1312.1179} {arXiv:1312.1179
  [hep-lat]} \BibitemShut {NoStop}%
%%CITATION = ARXIV:1312.1179;%%
\bibitem [{\citenamefont {Nishigaki}\ \emph {et~al.}(2014)\citenamefont
  {Nishigaki}, \citenamefont {Giordano}, \citenamefont {Kov\'acs},\ and\
  \citenamefont {Pittler}}]{Nishigaki:2013uya}%
  \BibitemOpen
  \bibfield  {author} {\bibinfo {author} {\bibfnamefont {S.~M.}\ \bibnamefont
  {Nishigaki}}, \bibinfo {author} {\bibfnamefont {M.}~\bibnamefont {Giordano}},
  \bibinfo {author} {\bibfnamefont {T.~G.}\ \bibnamefont {Kov\'acs}},\ and\
  \bibinfo {author} {\bibfnamefont {F.}~\bibnamefont {Pittler}},\ }\href
  {https://doi.org/10.22323/1.187.0018} {\bibfield  {journal} {\bibinfo
  {journal} {PoS}\ }\textbf {\bibinfo {volume} {LATTICE2013}},\ \bibinfo
  {pages} {018} (\bibinfo {year} {2014})},\ \Eprint
  {https://arxiv.org/abs/1312.3286} {arXiv:1312.3286 [hep-lat]} \BibitemShut
  {NoStop}%
%%CITATION = ARXIV:1312.3286;%%
\bibitem [{\citenamefont {Ujfalusi}\ \emph {et~al.}(2015)\citenamefont
  {Ujfalusi}, \citenamefont {Giordano}, \citenamefont {Pittler}, \citenamefont
  {Kov\'acs},\ and\ \citenamefont {Varga}}]{Ujfalusi:2015nha}%
  \BibitemOpen
  \bibfield  {author} {\bibinfo {author} {\bibfnamefont {L.}~\bibnamefont
  {Ujfalusi}}, \bibinfo {author} {\bibfnamefont {M.}~\bibnamefont {Giordano}},
  \bibinfo {author} {\bibfnamefont {F.}~\bibnamefont {Pittler}}, \bibinfo
  {author} {\bibfnamefont {T.~G.}\ \bibnamefont {Kov\'acs}},\ and\ \bibinfo
  {author} {\bibfnamefont {I.}~\bibnamefont {Varga}},\ }\href
  {https://doi.org/10.1103/PhysRevD.92.094513} {\bibfield  {journal} {\bibinfo
  {journal} {Phys. Rev. D}\ }\textbf {\bibinfo {volume} {92}},\ \bibinfo
  {pages} {094513} (\bibinfo {year} {2015})},\ \Eprint
  {https://arxiv.org/abs/1507.02162} {arXiv:1507.02162 [cond-mat.dis-nn]}
  \BibitemShut {NoStop}%
%%CITATION = ARXIV:1507.02162;%%
\bibitem [{\citenamefont {Cossu}\ and\ \citenamefont
  {Hashimoto}(2016)}]{Cossu:2016scb}%
  \BibitemOpen
  \bibfield  {author} {\bibinfo {author} {\bibfnamefont {G.}~\bibnamefont
  {Cossu}}\ and\ \bibinfo {author} {\bibfnamefont {S.}~\bibnamefont
  {Hashimoto}},\ }\href {https://doi.org/10.1007/JHEP06(2016)056} {\bibfield
  {journal} {\bibinfo  {journal} {{J. High Energy Phys.}}\ }\textbf {\bibinfo
  {volume} {06}},\ \bibinfo {pages} {056} (\bibinfo {year} {2016})},\ \Eprint
  {https://arxiv.org/abs/1604.00768} {arXiv:1604.00768 [hep-lat]} \BibitemShut
  {NoStop}%
%%CITATION = ARXIV:1604.00768;%%
\bibitem [{\citenamefont {Giordano}\ \emph
  {et~al.}(2017{\natexlab{a}})\citenamefont {Giordano}, \citenamefont {Katz},
  \citenamefont {Kov\'acs},\ and\ \citenamefont {Pittler}}]{Giordano:2016nuu}%
  \BibitemOpen
  \bibfield  {author} {\bibinfo {author} {\bibfnamefont {M.}~\bibnamefont
  {Giordano}}, \bibinfo {author} {\bibfnamefont {S.~D.}\ \bibnamefont {Katz}},
  \bibinfo {author} {\bibfnamefont {T.~G.}\ \bibnamefont {Kov\'acs}},\ and\
  \bibinfo {author} {\bibfnamefont {F.}~\bibnamefont {Pittler}},\ }\href
  {https://doi.org/10.1007/JHEP02(2017)055} {\bibfield  {journal} {\bibinfo
  {journal} {{J. High Energy Phys.}}\ }\textbf {\bibinfo {volume} {02}},\
  \bibinfo {pages} {055} (\bibinfo {year} {2017}{\natexlab{a}})},\ \Eprint
  {https://arxiv.org/abs/1611.03284} {arXiv:1611.03284 [hep-lat]} \BibitemShut
  {NoStop}%
%%CITATION = ARXIV:1611.03284;%%
\bibitem [{\citenamefont {Kov\'acs}\ and\ \citenamefont
  {Vig}(2018)}]{Kovacs:2017uiz}%
  \BibitemOpen
  \bibfield  {author} {\bibinfo {author} {\bibfnamefont {T.~G.}\ \bibnamefont
  {Kov\'acs}}\ and\ \bibinfo {author} {\bibfnamefont {R.~{\'A}.}\ \bibnamefont
  {Vig}},\ }\href {https://doi.org/10.1103/PhysRevD.97.014502} {\bibfield
  {journal} {\bibinfo  {journal} {Phys. Rev. D}\ }\textbf {\bibinfo {volume}
  {97}},\ \bibinfo {pages} {014502} (\bibinfo {year} {2018})},\ \Eprint
  {https://arxiv.org/abs/1706.03562} {arXiv:1706.03562 [hep-lat]} \BibitemShut
  {NoStop}%
%%CITATION = ARXIV:1706.03562;%%
\bibitem [{\citenamefont {Holicki}\ \emph {et~al.}(2018)\citenamefont
  {Holicki}, \citenamefont {Ilgenfritz},\ and\ \citenamefont {von
  Smekal}}]{Holicki:2018sms}%
  \BibitemOpen
  \bibfield  {author} {\bibinfo {author} {\bibfnamefont {L.}~\bibnamefont
  {Holicki}}, \bibinfo {author} {\bibfnamefont {E.-M.}\ \bibnamefont
  {Ilgenfritz}},\ and\ \bibinfo {author} {\bibfnamefont {L.}~\bibnamefont {von
  Smekal}},\ }\href {https://doi.org/10.22323/1.334.0180} {\bibfield  {journal}
  {\bibinfo  {journal} {PoS}\ }\textbf {\bibinfo {volume} {LATTICE2018}},\
  \bibinfo {pages} {180} (\bibinfo {year} {2018})},\ \Eprint
  {https://arxiv.org/abs/1810.01130} {arXiv:1810.01130 [hep-lat]} \BibitemShut
  {NoStop}%
\bibitem [{\citenamefont {Giordano}(2019)}]{Giordano:2019pvc}%
  \BibitemOpen
  \bibfield  {author} {\bibinfo {author} {\bibfnamefont {M.}~\bibnamefont
  {Giordano}},\ }\href {https://doi.org/10.1007/JHEP05(2019)204} {\bibfield
  {journal} {\bibinfo  {journal} {{J. High Energy Phys.}}\ }\textbf {\bibinfo
  {volume} {05}},\ \bibinfo {pages} {204} (\bibinfo {year} {2019})},\ \Eprint
  {https://arxiv.org/abs/1903.04983} {arXiv:1903.04983 [hep-lat]} \BibitemShut
  {NoStop}%
\bibitem [{\citenamefont {Vig}\ and\ \citenamefont
  {Kov{\'a}cs}(2020)}]{Vig:2020pgq}%
  \BibitemOpen
  \bibfield  {author} {\bibinfo {author} {\bibfnamefont {R.~{\'A}.}\
  \bibnamefont {Vig}}\ and\ \bibinfo {author} {\bibfnamefont {T.~G.}\
  \bibnamefont {Kov{\'a}cs}},\ }\href
  {https://doi.org/10.1103/PhysRevD.101.094511} {\bibfield  {journal} {\bibinfo
   {journal} {Phys. Rev. D}\ }\textbf {\bibinfo {volume} {101}},\ \bibinfo
  {pages} {094511} (\bibinfo {year} {2020})},\ \Eprint
  {https://arxiv.org/abs/2001.06872} {arXiv:2001.06872 [hep-lat]} \BibitemShut
  {NoStop}%
\bibitem [{\citenamefont {Bonati}\ \emph {et~al.}(2021)\citenamefont {Bonati},
  \citenamefont {Cardinali}, \citenamefont {D'Elia}, \citenamefont {Giordano},\
  and\ \citenamefont {Mazziotti}}]{Bonati:2020lal}%
  \BibitemOpen
  \bibfield  {author} {\bibinfo {author} {\bibfnamefont {C.}~\bibnamefont
  {Bonati}}, \bibinfo {author} {\bibfnamefont {M.}~\bibnamefont {Cardinali}},
  \bibinfo {author} {\bibfnamefont {M.}~\bibnamefont {D'Elia}}, \bibinfo
  {author} {\bibfnamefont {M.}~\bibnamefont {Giordano}},\ and\ \bibinfo
  {author} {\bibfnamefont {F.}~\bibnamefont {Mazziotti}},\ }\href
  {https://doi.org/10.1103/PhysRevD.103.034506} {\bibfield  {journal} {\bibinfo
   {journal} {Phys. Rev. D}\ }\textbf {\bibinfo {volume} {103}},\ \bibinfo
  {pages} {034506} (\bibinfo {year} {2021})},\ \Eprint
  {https://arxiv.org/abs/2012.13246} {arXiv:2012.13246 [hep-lat]} \BibitemShut
  {NoStop}%
\bibitem [{\citenamefont {Kov\'acs}(2022)}]{Kovacs:2021fwq}%
  \BibitemOpen
  \bibfield  {author} {\bibinfo {author} {\bibfnamefont {T.~G.}\ \bibnamefont
  {Kov\'acs}},\ }\href {https://doi.org/10.22323/1.396.0238} {\bibfield
  {journal} {\bibinfo  {journal} {PoS}\ }\textbf {\bibinfo {volume}
  {LATTICE2021}},\ \bibinfo {pages} {238} (\bibinfo {year} {2022})},\ \Eprint
  {https://arxiv.org/abs/2112.05454} {arXiv:2112.05454 [hep-lat]} \BibitemShut
  {NoStop}%
\bibitem [{\citenamefont {Cardinali}\ \emph {et~al.}(2022)\citenamefont
  {Cardinali}, \citenamefont {D'Elia}, \citenamefont {Garosi},\ and\
  \citenamefont {Giordano}}]{Cardinali:2021fpu}%
  \BibitemOpen
  \bibfield  {author} {\bibinfo {author} {\bibfnamefont {M.}~\bibnamefont
  {Cardinali}}, \bibinfo {author} {\bibfnamefont {M.}~\bibnamefont {D'Elia}},
  \bibinfo {author} {\bibfnamefont {F.}~\bibnamefont {Garosi}},\ and\ \bibinfo
  {author} {\bibfnamefont {M.}~\bibnamefont {Giordano}},\ }\href
  {https://doi.org/10.1103/PhysRevD.105.014506} {\bibfield  {journal} {\bibinfo
   {journal} {Phys. Rev. D}\ }\textbf {\bibinfo {volume} {105}},\ \bibinfo
  {pages} {014506} (\bibinfo {year} {2022})},\ \Eprint
  {https://arxiv.org/abs/2110.10029} {arXiv:2110.10029 [hep-lat]} \BibitemShut
  {NoStop}%
\bibitem [{\citenamefont {Baranka}\ and\ \citenamefont
  {Giordano}(2021)}]{Baranka:2021san}%
  \BibitemOpen
  \bibfield  {author} {\bibinfo {author} {\bibfnamefont {G.}~\bibnamefont
  {Baranka}}\ and\ \bibinfo {author} {\bibfnamefont {M.}~\bibnamefont
  {Giordano}},\ }\href {https://doi.org/10.1103/PhysRevD.104.054513} {\bibfield
   {journal} {\bibinfo  {journal} {Phys. Rev. D}\ }\textbf {\bibinfo {volume}
  {104}},\ \bibinfo {pages} {054513} (\bibinfo {year} {2021})},\ \Eprint
  {https://arxiv.org/abs/2104.03779} {arXiv:2104.03779 [hep-lat]} \BibitemShut
  {NoStop}%
\bibitem [{\citenamefont {Baranka}\ and\ \citenamefont
  {Giordano}(2022)}]{Baranka:2022dib}%
  \BibitemOpen
  \bibfield  {author} {\bibinfo {author} {\bibfnamefont {G.}~\bibnamefont
  {Baranka}}\ and\ \bibinfo {author} {\bibfnamefont {M.}~\bibnamefont
  {Giordano}},\ }\href {https://doi.org/10.1103/PhysRevD.106.094508} {\bibfield
   {journal} {\bibinfo  {journal} {Phys. Rev. D}\ }\textbf {\bibinfo {volume}
  {106}},\ \bibinfo {pages} {094508} (\bibinfo {year} {2022})},\ \Eprint
  {https://arxiv.org/abs/2210.00840} {arXiv:2210.00840 [hep-lat]} \BibitemShut
  {NoStop}%
\bibitem [{\citenamefont {Kehr}\ \emph {et~al.}(2024)\citenamefont {Kehr},
  \citenamefont {Smith},\ and\ \citenamefont {von Smekal}}]{Kehr:2023wrs}%
  \BibitemOpen
  \bibfield  {author} {\bibinfo {author} {\bibfnamefont {R.}~\bibnamefont
  {Kehr}}, \bibinfo {author} {\bibfnamefont {D.}~\bibnamefont {Smith}},\ and\
  \bibinfo {author} {\bibfnamefont {L.}~\bibnamefont {von Smekal}},\ }\href
  {https://doi.org/10.1103/PhysRevD.109.074512} {\bibfield  {journal} {\bibinfo
   {journal} {Phys. Rev. D}\ }\textbf {\bibinfo {volume} {109}},\ \bibinfo
  {pages} {074512} (\bibinfo {year} {2024})},\ \Eprint
  {https://arxiv.org/abs/2304.13617} {arXiv:2304.13617 [hep-lat]} \BibitemShut
  {NoStop}%
\bibitem [{\citenamefont {Baranka}\ and\ \citenamefont
  {Giordano}(2023)}]{Baranka:2023ani}%
  \BibitemOpen
  \bibfield  {author} {\bibinfo {author} {\bibfnamefont {G.}~\bibnamefont
  {Baranka}}\ and\ \bibinfo {author} {\bibfnamefont {M.}~\bibnamefont
  {Giordano}},\ }\href {https://doi.org/10.1103/PhysRevD.108.114508} {\bibfield
   {journal} {\bibinfo  {journal} {Phys. Rev. D}\ }\textbf {\bibinfo {volume}
  {108}},\ \bibinfo {pages} {114508} (\bibinfo {year} {2023})},\ \Eprint
  {https://arxiv.org/abs/2310.03542} {arXiv:2310.03542 [hep-lat]} \BibitemShut
  {NoStop}%
\bibitem [{\citenamefont {Bonanno}\ and\ \citenamefont
  {Giordano}(2024)}]{Bonanno:2023mzj}%
  \BibitemOpen
  \bibfield  {author} {\bibinfo {author} {\bibfnamefont {C.}~\bibnamefont
  {Bonanno}}\ and\ \bibinfo {author} {\bibfnamefont {M.}~\bibnamefont
  {Giordano}},\ }\href {https://doi.org/10.1103/PhysRevD.109.054510} {\bibfield
   {journal} {\bibinfo  {journal} {Phys. Rev. D}\ }\textbf {\bibinfo {volume}
  {109}},\ \bibinfo {pages} {054510} (\bibinfo {year} {2024})},\ \Eprint
  {https://arxiv.org/abs/2312.02857} {arXiv:2312.02857 [hep-lat]} \BibitemShut
  {NoStop}%
\bibitem [{\citenamefont {Giordano}\ and\ \citenamefont
  {Kov{\'a}cs}(2021)}]{Giordano:2021qav}%
  \BibitemOpen
  \bibfield  {author} {\bibinfo {author} {\bibfnamefont {M.}~\bibnamefont
  {Giordano}}\ and\ \bibinfo {author} {\bibfnamefont {T.~G.}\ \bibnamefont
  {Kov{\'a}cs}},\ }\href {https://doi.org/10.3390/universe7060194} {\bibfield
  {journal} {\bibinfo  {journal} {Universe}\ }\textbf {\bibinfo {volume} {7}},\
  \bibinfo {pages} {194} (\bibinfo {year} {2021})},\ \Eprint
  {https://arxiv.org/abs/2104.14388} {arXiv:2104.14388 [hep-lat]} \BibitemShut
  {NoStop}%
\bibitem [{\citenamefont {Anderson}(1958)}]{Anderson:1958vr}%
  \BibitemOpen
  \bibfield  {author} {\bibinfo {author} {\bibfnamefont {P.~W.}\ \bibnamefont
  {Anderson}},\ }\href {https://doi.org/10.1103/PhysRev.109.1492} {\bibfield
  {journal} {\bibinfo  {journal} {Phys. Rev.}\ }\textbf {\bibinfo {volume}
  {109}},\ \bibinfo {pages} {1492} (\bibinfo {year} {1958})}\BibitemShut
  {NoStop}%
\bibitem [{\citenamefont {Thouless}(1974)}]{thouless1974electrons}%
  \BibitemOpen
  \bibfield  {author} {\bibinfo {author} {\bibfnamefont {D.~J.}\ \bibnamefont
  {Thouless}},\ }\href {https://doi.org/10.1016/0370-1573(74)90029-5}
  {\bibfield  {journal} {\bibinfo  {journal} {Phys. Rep.}\ }\textbf {\bibinfo
  {volume} {13}},\ \bibinfo {pages} {93} (\bibinfo {year} {1974})}\BibitemShut
  {NoStop}%
\bibitem [{\citenamefont {Lee}\ and\ \citenamefont
  {Ramakrishnan}(1985)}]{lee1985disordered}%
  \BibitemOpen
  \bibfield  {author} {\bibinfo {author} {\bibfnamefont {P.~A.}\ \bibnamefont
  {Lee}}\ and\ \bibinfo {author} {\bibfnamefont {T.~V.}\ \bibnamefont
  {Ramakrishnan}},\ }\href {https://doi.org/10.1103/RevModPhys.57.287}
  {\bibfield  {journal} {\bibinfo  {journal} {Rev. Mod. Phys.}\ }\textbf
  {\bibinfo {volume} {57}},\ \bibinfo {pages} {287} (\bibinfo {year}
  {1985})}\BibitemShut {NoStop}%
\bibitem [{\citenamefont {Kramer}\ and\ \citenamefont
  {MacKinnon}(1993)}]{kramer1993localization}%
  \BibitemOpen
  \bibfield  {author} {\bibinfo {author} {\bibfnamefont {B.}~\bibnamefont
  {Kramer}}\ and\ \bibinfo {author} {\bibfnamefont {A.}~\bibnamefont
  {MacKinnon}},\ }\href {https://doi.org/10.1088/0034-4885/56/12/001}
  {\bibfield  {journal} {\bibinfo  {journal} {Rep. Prog. Phys.}\ }\textbf
  {\bibinfo {volume} {56}},\ \bibinfo {pages} {1469} (\bibinfo {year}
  {1993})}\BibitemShut {NoStop}%
\bibitem [{\citenamefont {Evers}\ and\ \citenamefont
  {Mirlin}(2008)}]{Evers:2008zz}%
  \BibitemOpen
  \bibfield  {author} {\bibinfo {author} {\bibfnamefont {F.}~\bibnamefont
  {Evers}}\ and\ \bibinfo {author} {\bibfnamefont {A.~D.}\ \bibnamefont
  {Mirlin}},\ }\href {https://doi.org/10.1103/RevModPhys.80.1355} {\bibfield
  {journal} {\bibinfo  {journal} {Rev. Mod. Phys.}\ }\textbf {\bibinfo {volume}
  {80}},\ \bibinfo {pages} {1355} (\bibinfo {year} {2008})},\ \Eprint
  {https://arxiv.org/abs/0707.4378} {arXiv:0707.4378 [cond-mat.mes-hall]}
  \BibitemShut {NoStop}%
\bibitem [{\citenamefont {Giordano}\ \emph {et~al.}(2015)\citenamefont
  {Giordano}, \citenamefont {Kov\'acs},\ and\ \citenamefont
  {Pittler}}]{Giordano:2015vla}%
  \BibitemOpen
  \bibfield  {author} {\bibinfo {author} {\bibfnamefont {M.}~\bibnamefont
  {Giordano}}, \bibinfo {author} {\bibfnamefont {T.~G.}\ \bibnamefont
  {Kov\'acs}},\ and\ \bibinfo {author} {\bibfnamefont {F.}~\bibnamefont
  {Pittler}},\ }\href {https://doi.org/10.1007/JHEP04(2015)112} {\bibfield
  {journal} {\bibinfo  {journal} {{J. High Energy Phys.}}\ }\textbf {\bibinfo
  {volume} {04}},\ \bibinfo {pages} {112} (\bibinfo {year} {2015})},\ \Eprint
  {https://arxiv.org/abs/1502.02532} {arXiv:1502.02532 [hep-lat]} \BibitemShut
  {NoStop}%
%%CITATION = ARXIV:1502.02532;%%
\bibitem [{\citenamefont {Giordano}\ \emph {et~al.}(2016)\citenamefont
  {Giordano}, \citenamefont {Kov\'acs},\ and\ \citenamefont
  {Pittler}}]{Giordano:2016cjs}%
  \BibitemOpen
  \bibfield  {author} {\bibinfo {author} {\bibfnamefont {M.}~\bibnamefont
  {Giordano}}, \bibinfo {author} {\bibfnamefont {T.~G.}\ \bibnamefont
  {Kov\'acs}},\ and\ \bibinfo {author} {\bibfnamefont {F.}~\bibnamefont
  {Pittler}},\ }\href {https://doi.org/10.1007/JHEP06(2016)007} {\bibfield
  {journal} {\bibinfo  {journal} {{J. High Energy Phys.}}\ }\textbf {\bibinfo
  {volume} {06}},\ \bibinfo {pages} {007} (\bibinfo {year} {2016})},\ \Eprint
  {https://arxiv.org/abs/1603.09548} {arXiv:1603.09548 [hep-lat]} \BibitemShut
  {NoStop}%
%%CITATION = ARXIV:1603.09548;%%
\bibitem [{\citenamefont {Giordano}\ \emph
  {et~al.}(2017{\natexlab{b}})\citenamefont {Giordano}, \citenamefont
  {Kov\'acs},\ and\ \citenamefont {Pittler}}]{Giordano:2016vhx}%
  \BibitemOpen
  \bibfield  {author} {\bibinfo {author} {\bibfnamefont {M.}~\bibnamefont
  {Giordano}}, \bibinfo {author} {\bibfnamefont {T.~G.}\ \bibnamefont
  {Kov\'acs}},\ and\ \bibinfo {author} {\bibfnamefont {F.}~\bibnamefont
  {Pittler}},\ }\href {https://doi.org/10.1103/PhysRevD.95.074503} {\bibfield
  {journal} {\bibinfo  {journal} {Phys. Rev. D}\ }\textbf {\bibinfo {volume}
  {95}},\ \bibinfo {pages} {074503} (\bibinfo {year} {2017}{\natexlab{b}})},\
  \Eprint {https://arxiv.org/abs/1612.05059} {arXiv:1612.05059 [hep-lat]}
  \BibitemShut {NoStop}%
%%CITATION = ARXIV:1612.05059;%%
\bibitem [{\citenamefont {Wipf}(2013)}]{Wipf:2013vp}%
  \BibitemOpen
  \bibfield  {author} {\bibinfo {author} {\bibfnamefont {A.}~\bibnamefont
  {Wipf}},\ }\href {https://doi.org/10.1007/978-3-642-33105-3} {\emph {\bibinfo
  {title} {{Statistical Approach to Quantum Field Theory}: {An
  Introduction}}}},\ Vol.\ \bibinfo {volume} {864}\ (\bibinfo  {publisher}
  {Springer-Verlag},\ \bibinfo {address} {Berlin Heidelberg},\ \bibinfo {year}
  {2013})\BibitemShut {NoStop}%
\bibitem [{\citenamefont {Caselle}\ and\ \citenamefont
  {Hasenbusch}(1996)}]{Caselle:1995wn}%
  \BibitemOpen
  \bibfield  {author} {\bibinfo {author} {\bibfnamefont {M.}~\bibnamefont
  {Caselle}}\ and\ \bibinfo {author} {\bibfnamefont {M.}~\bibnamefont
  {Hasenbusch}},\ }\href {https://doi.org/10.1016/0550-3213(96)00161-7}
  {\bibfield  {journal} {\bibinfo  {journal} {Nucl. Phys. B}\ }\textbf
  {\bibinfo {volume} {470}},\ \bibinfo {pages} {435} (\bibinfo {year}
  {1996})},\ \Eprint {https://arxiv.org/abs/hep-lat/9511015}
  {arXiv:hep-lat/9511015} \BibitemShut {NoStop}%
\bibitem [{\citenamefont {von Smekal}(2012)}]{vonSmekal:2012vx}%
  \BibitemOpen
  \bibfield  {author} {\bibinfo {author} {\bibfnamefont {L.}~\bibnamefont {von
  Smekal}},\ }\href {https://doi.org/10.1016/j.nuclphysbps.2012.06.006}
  {\bibfield  {journal} {\bibinfo  {journal} {Nucl. Phys. B Proc. Suppl.}\
  }\textbf {\bibinfo {volume} {228}},\ \bibinfo {pages} {179} (\bibinfo {year}
  {2012})},\ \Eprint {https://arxiv.org/abs/1205.4205} {arXiv:1205.4205
  [hep-ph]} \BibitemShut {NoStop}%
\bibitem [{\citenamefont {Swendsen}\ and\ \citenamefont
  {Wang}(1987)}]{Swendsen:1987ce}%
  \BibitemOpen
  \bibfield  {author} {\bibinfo {author} {\bibfnamefont {R.~H.}\ \bibnamefont
  {Swendsen}}\ and\ \bibinfo {author} {\bibfnamefont {J.-S.}\ \bibnamefont
  {Wang}},\ }\href {https://doi.org/10.1103/PhysRevLett.58.86} {\bibfield
  {journal} {\bibinfo  {journal} {Phys. Rev. Lett.}\ }\textbf {\bibinfo
  {volume} {58}},\ \bibinfo {pages} {86} (\bibinfo {year} {1987})}\BibitemShut
  {NoStop}%
\bibitem [{\citenamefont {Wolff}(1989)}]{Wolff:1988uh}%
  \BibitemOpen
  \bibfield  {author} {\bibinfo {author} {\bibfnamefont {U.}~\bibnamefont
  {Wolff}},\ }\href {https://doi.org/10.1103/PhysRevLett.62.361} {\bibfield
  {journal} {\bibinfo  {journal} {Phys. Rev. Lett.}\ }\textbf {\bibinfo
  {volume} {62}},\ \bibinfo {pages} {361} (\bibinfo {year} {1989})}\BibitemShut
  {NoStop}%
\bibitem [{\citenamefont {Greensite}(2003)}]{Greensite:2003bk}%
  \BibitemOpen
  \bibfield  {author} {\bibinfo {author} {\bibfnamefont {J.}~\bibnamefont
  {Greensite}},\ }\href {https://doi.org/10.1016/S0146-6410(03)90012-3}
  {\bibfield  {journal} {\bibinfo  {journal} {Prog. Part. Nucl. Phys.}\
  }\textbf {\bibinfo {volume} {51}},\ \bibinfo {pages} {1} (\bibinfo {year}
  {2003})},\ \Eprint {https://arxiv.org/abs/hep-lat/0301023}
  {arXiv:hep-lat/0301023} \BibitemShut {NoStop}%
\bibitem [{\citenamefont {'t~Hooft}(1978)}]{tHooft:1977nqb}%
  \BibitemOpen
  \bibfield  {author} {\bibinfo {author} {\bibfnamefont {G.}~\bibnamefont
  {'t~Hooft}},\ }\href {https://doi.org/10.1016/0550-3213(78)90153-0}
  {\bibfield  {journal} {\bibinfo  {journal} {Nucl. Phys. B}\ }\textbf
  {\bibinfo {volume} {138}},\ \bibinfo {pages} {1} (\bibinfo {year}
  {1978})}\BibitemShut {NoStop}%
\bibitem [{\citenamefont {Del~Debbio}\ \emph {et~al.}(1997)\citenamefont
  {Del~Debbio}, \citenamefont {Faber}, \citenamefont {Greensite},\ and\
  \citenamefont {Olejn{\'i}k}}]{DelDebbio:1996lih}%
  \BibitemOpen
  \bibfield  {author} {\bibinfo {author} {\bibfnamefont {L.}~\bibnamefont
  {Del~Debbio}}, \bibinfo {author} {\bibfnamefont {M.}~\bibnamefont {Faber}},
  \bibinfo {author} {\bibfnamefont {J.}~\bibnamefont {Greensite}},\ and\
  \bibinfo {author} {\bibfnamefont {{\v S}.}~\bibnamefont {Olejn{\'i}k}},\
  }\href {https://doi.org/10.1103/PhysRevD.55.2298} {\bibfield  {journal}
  {\bibinfo  {journal} {Phys. Rev. D}\ }\textbf {\bibinfo {volume} {55}},\
  \bibinfo {pages} {2298} (\bibinfo {year} {1997})},\ \Eprint
  {https://arxiv.org/abs/hep-lat/9610005} {arXiv:hep-lat/9610005} \BibitemShut
  {NoStop}%
\bibitem [{\citenamefont {Engelhardt}\ \emph {et~al.}(1998)\citenamefont
  {Engelhardt}, \citenamefont {Langfeld}, \citenamefont {Reinhardt},\ and\
  \citenamefont {Tennert}}]{Engelhardt:1998wu}%
  \BibitemOpen
  \bibfield  {author} {\bibinfo {author} {\bibfnamefont {M.}~\bibnamefont
  {Engelhardt}}, \bibinfo {author} {\bibfnamefont {K.}~\bibnamefont
  {Langfeld}}, \bibinfo {author} {\bibfnamefont {H.}~\bibnamefont
  {Reinhardt}},\ and\ \bibinfo {author} {\bibfnamefont {O.}~\bibnamefont
  {Tennert}},\ }\href {https://doi.org/10.1016/S0370-2693(98)00583-8}
  {\bibfield  {journal} {\bibinfo  {journal} {Phys. Lett. B}\ }\textbf
  {\bibinfo {volume} {431}},\ \bibinfo {pages} {141} (\bibinfo {year}
  {1998})},\ \Eprint {https://arxiv.org/abs/hep-lat/9801030}
  {arXiv:hep-lat/9801030} \BibitemShut {NoStop}%
\bibitem [{\citenamefont {Langfeld}\ \emph {et~al.}(1999)\citenamefont
  {Langfeld}, \citenamefont {Tennert}, \citenamefont {Engelhardt},\ and\
  \citenamefont {Reinhardt}}]{Langfeld:1998cz}%
  \BibitemOpen
  \bibfield  {author} {\bibinfo {author} {\bibfnamefont {K.}~\bibnamefont
  {Langfeld}}, \bibinfo {author} {\bibfnamefont {O.}~\bibnamefont {Tennert}},
  \bibinfo {author} {\bibfnamefont {M.}~\bibnamefont {Engelhardt}},\ and\
  \bibinfo {author} {\bibfnamefont {H.}~\bibnamefont {Reinhardt}},\ }\href
  {https://doi.org/10.1016/S0370-2693(99)00252-X} {\bibfield  {journal}
  {\bibinfo  {journal} {Phys. Lett. B}\ }\textbf {\bibinfo {volume} {452}},\
  \bibinfo {pages} {301} (\bibinfo {year} {1999})},\ \Eprint
  {https://arxiv.org/abs/hep-lat/9805002} {arXiv:hep-lat/9805002} \BibitemShut
  {NoStop}%
\bibitem [{\citenamefont {Engelhardt}\ \emph {et~al.}(2000)\citenamefont
  {Engelhardt}, \citenamefont {Langfeld}, \citenamefont {Reinhardt},\ and\
  \citenamefont {Tennert}}]{Engelhardt:1999fd}%
  \BibitemOpen
  \bibfield  {author} {\bibinfo {author} {\bibfnamefont {M.}~\bibnamefont
  {Engelhardt}}, \bibinfo {author} {\bibfnamefont {K.}~\bibnamefont
  {Langfeld}}, \bibinfo {author} {\bibfnamefont {H.}~\bibnamefont
  {Reinhardt}},\ and\ \bibinfo {author} {\bibfnamefont {O.}~\bibnamefont
  {Tennert}},\ }\href {https://doi.org/10.1103/PhysRevD.61.054504} {\bibfield
  {journal} {\bibinfo  {journal} {Phys. Rev. D}\ }\textbf {\bibinfo {volume}
  {61}},\ \bibinfo {pages} {054504} (\bibinfo {year} {2000})},\ \Eprint
  {https://arxiv.org/abs/hep-lat/9904004} {arXiv:hep-lat/9904004} \BibitemShut
  {NoStop}%
\bibitem [{\citenamefont {Engelhardt}\ \emph {et~al.}(2004)\citenamefont
  {Engelhardt}, \citenamefont {Quandt},\ and\ \citenamefont
  {Reinhardt}}]{Engelhardt:2003wm}%
  \BibitemOpen
  \bibfield  {author} {\bibinfo {author} {\bibfnamefont {M.}~\bibnamefont
  {Engelhardt}}, \bibinfo {author} {\bibfnamefont {M.}~\bibnamefont {Quandt}},\
  and\ \bibinfo {author} {\bibfnamefont {H.}~\bibnamefont {Reinhardt}},\ }\href
  {https://doi.org/10.1016/j.nuclphysb.2004.02.036} {\bibfield  {journal}
  {\bibinfo  {journal} {Nucl. Phys. B}\ }\textbf {\bibinfo {volume} {685}},\
  \bibinfo {pages} {227} (\bibinfo {year} {2004})},\ \Eprint
  {https://arxiv.org/abs/hep-lat/0311029} {arXiv:hep-lat/0311029} \BibitemShut
  {NoStop}%
\bibitem [{\citenamefont {Gattnar}\ \emph {et~al.}(2005)\citenamefont
  {Gattnar}, \citenamefont {Gattringer}, \citenamefont {Langfeld},
  \citenamefont {Reinhardt}, \citenamefont {Sch{\"a}fer}, \citenamefont
  {Solbrig},\ and\ \citenamefont {Tok}}]{Gattnar:2004gx}%
  \BibitemOpen
  \bibfield  {author} {\bibinfo {author} {\bibfnamefont {J.}~\bibnamefont
  {Gattnar}}, \bibinfo {author} {\bibfnamefont {C.}~\bibnamefont {Gattringer}},
  \bibinfo {author} {\bibfnamefont {K.}~\bibnamefont {Langfeld}}, \bibinfo
  {author} {\bibfnamefont {H.}~\bibnamefont {Reinhardt}}, \bibinfo {author}
  {\bibfnamefont {A.}~\bibnamefont {Sch{\"a}fer}}, \bibinfo {author}
  {\bibfnamefont {S.}~\bibnamefont {Solbrig}},\ and\ \bibinfo {author}
  {\bibfnamefont {T.}~\bibnamefont {Tok}},\ }\href
  {https://doi.org/10.1016/j.nuclphysb.2005.03.027} {\bibfield  {journal}
  {\bibinfo  {journal} {Nucl. Phys. B}\ }\textbf {\bibinfo {volume} {716}},\
  \bibinfo {pages} {105} (\bibinfo {year} {2005})},\ \Eprint
  {https://arxiv.org/abs/hep-lat/0412032} {arXiv:hep-lat/0412032} \BibitemShut
  {NoStop}%
\bibitem [{\citenamefont {Bornyakov}\ \emph {et~al.}(2008)\citenamefont
  {Bornyakov}, \citenamefont {Ilgenfritz}, \citenamefont {Martemyanov},
  \citenamefont {Morozov}, \citenamefont {M{\"u}ller-Preussker},\ and\
  \citenamefont {Veselov}}]{Bornyakov:2007fz}%
  \BibitemOpen
  \bibfield  {author} {\bibinfo {author} {\bibfnamefont {V.~G.}\ \bibnamefont
  {Bornyakov}}, \bibinfo {author} {\bibfnamefont {E.~M.}\ \bibnamefont
  {Ilgenfritz}}, \bibinfo {author} {\bibfnamefont {B.~V.}\ \bibnamefont
  {Martemyanov}}, \bibinfo {author} {\bibfnamefont {S.~M.}\ \bibnamefont
  {Morozov}}, \bibinfo {author} {\bibfnamefont {M.}~\bibnamefont
  {M{\"u}ller-Preussker}},\ and\ \bibinfo {author} {\bibfnamefont {A.~I.}\
  \bibnamefont {Veselov}},\ }\href {https://doi.org/10.1103/PhysRevD.77.074507}
  {\bibfield  {journal} {\bibinfo  {journal} {Phys. Rev. D}\ }\textbf {\bibinfo
  {volume} {77}},\ \bibinfo {pages} {074507} (\bibinfo {year} {2008})},\
  \Eprint {https://arxiv.org/abs/0708.3335} {arXiv:0708.3335 [hep-lat]}
  \BibitemShut {NoStop}%
\bibitem [{\citenamefont {H{\"o}llwieser}\ \emph {et~al.}(2008)\citenamefont
  {H{\"o}llwieser}, \citenamefont {Faber}, \citenamefont {Greensite},
  \citenamefont {Heller},\ and\ \citenamefont
  {Olejn{\'i}k}}]{Hollwieser:2008tq}%
  \BibitemOpen
  \bibfield  {author} {\bibinfo {author} {\bibfnamefont {R.}~\bibnamefont
  {H{\"o}llwieser}}, \bibinfo {author} {\bibfnamefont {M.}~\bibnamefont
  {Faber}}, \bibinfo {author} {\bibfnamefont {J.}~\bibnamefont {Greensite}},
  \bibinfo {author} {\bibfnamefont {U.~M.}\ \bibnamefont {Heller}},\ and\
  \bibinfo {author} {\bibfnamefont {{\v S}.}~\bibnamefont {Olejn{\'i}k}},\
  }\href {https://doi.org/10.1103/PhysRevD.78.054508} {\bibfield  {journal}
  {\bibinfo  {journal} {Phys. Rev. D}\ }\textbf {\bibinfo {volume} {78}},\
  \bibinfo {pages} {054508} (\bibinfo {year} {2008})},\ \Eprint
  {https://arxiv.org/abs/0805.1846} {arXiv:0805.1846 [hep-lat]} \BibitemShut
  {NoStop}%
\bibitem [{\citenamefont {Bornyakov}\ \emph {et~al.}(2009)\citenamefont
  {Bornyakov}, \citenamefont {Luschevskaya}, \citenamefont {Morozov},
  \citenamefont {Polikarpov}, \citenamefont {Ilgenfritz},\ and\ \citenamefont
  {M{\"u}ller-Preussker}}]{Bornyakov:2008bg}%
  \BibitemOpen
  \bibfield  {author} {\bibinfo {author} {\bibfnamefont {V.~G.}\ \bibnamefont
  {Bornyakov}}, \bibinfo {author} {\bibfnamefont {E.~V.}\ \bibnamefont
  {Luschevskaya}}, \bibinfo {author} {\bibfnamefont {S.~M.}\ \bibnamefont
  {Morozov}}, \bibinfo {author} {\bibfnamefont {M.~I.}\ \bibnamefont
  {Polikarpov}}, \bibinfo {author} {\bibfnamefont {E.~M.}\ \bibnamefont
  {Ilgenfritz}},\ and\ \bibinfo {author} {\bibfnamefont {M.}~\bibnamefont
  {M{\"u}ller-Preussker}},\ }\href {https://doi.org/10.1103/PhysRevD.79.054505}
  {\bibfield  {journal} {\bibinfo  {journal} {Phys. Rev. D}\ }\textbf {\bibinfo
  {volume} {79}},\ \bibinfo {pages} {054505} (\bibinfo {year} {2009})},\
  \Eprint {https://arxiv.org/abs/0807.1980} {arXiv:0807.1980 [hep-lat]}
  \BibitemShut {NoStop}%
\bibitem [{\citenamefont {Junior}\ \emph {et~al.}(2022)\citenamefont {Junior},
  \citenamefont {Oxman},\ and\ \citenamefont {Reinhardt}}]{Junior:2022bol}%
  \BibitemOpen
  \bibfield  {author} {\bibinfo {author} {\bibfnamefont {D.~R.}\ \bibnamefont
  {Junior}}, \bibinfo {author} {\bibfnamefont {L.~E.}\ \bibnamefont {Oxman}},\
  and\ \bibinfo {author} {\bibfnamefont {H.}~\bibnamefont {Reinhardt}},\ }\href
  {https://doi.org/10.1103/PhysRevD.106.114021} {\bibfield  {journal} {\bibinfo
   {journal} {Phys. Rev. D}\ }\textbf {\bibinfo {volume} {106}},\ \bibinfo
  {pages} {114021} (\bibinfo {year} {2022})},\ \Eprint
  {https://arxiv.org/abs/2211.03006} {arXiv:2211.03006 [hep-th]} \BibitemShut
  {NoStop}%
\bibitem [{\citenamefont {Sale}\ \emph {et~al.}(2023)\citenamefont {Sale},
  \citenamefont {Lucini},\ and\ \citenamefont {Giansiracusa}}]{Sale:2022qfn}%
  \BibitemOpen
  \bibfield  {author} {\bibinfo {author} {\bibfnamefont {N.}~\bibnamefont
  {Sale}}, \bibinfo {author} {\bibfnamefont {B.}~\bibnamefont {Lucini}},\ and\
  \bibinfo {author} {\bibfnamefont {J.}~\bibnamefont {Giansiracusa}},\ }\href
  {https://doi.org/10.1103/PhysRevD.107.034501} {\bibfield  {journal} {\bibinfo
   {journal} {Phys. Rev. D}\ }\textbf {\bibinfo {volume} {107}},\ \bibinfo
  {pages} {034501} (\bibinfo {year} {2023})},\ \Eprint
  {https://arxiv.org/abs/2207.13392} {arXiv:2207.13392 [hep-lat]} \BibitemShut
  {NoStop}%
\bibitem [{\citenamefont {Biddle}\ \emph {et~al.}(2023)\citenamefont {Biddle},
  \citenamefont {Kamleh},\ and\ \citenamefont {Leinweber}}]{Biddle:2023lod}%
  \BibitemOpen
  \bibfield  {author} {\bibinfo {author} {\bibfnamefont {J.~C.}\ \bibnamefont
  {Biddle}}, \bibinfo {author} {\bibfnamefont {W.}~\bibnamefont {Kamleh}},\
  and\ \bibinfo {author} {\bibfnamefont {D.~B.}\ \bibnamefont {Leinweber}},\
  }\href {https://doi.org/10.1103/PhysRevD.107.094507} {\bibfield  {journal}
  {\bibinfo  {journal} {Phys. Rev. D}\ }\textbf {\bibinfo {volume} {107}},\
  \bibinfo {pages} {094507} (\bibinfo {year} {2023})},\ \Eprint
  {https://arxiv.org/abs/2302.05897} {arXiv:2302.05897 [hep-lat]} \BibitemShut
  {NoStop}%
\bibitem [{\citenamefont {Kamleh}\ \emph {et~al.}(2024)\citenamefont {Kamleh},
  \citenamefont {Leinweber},\ and\ \citenamefont {Virgili}}]{Kamleh:2023gho}%
  \BibitemOpen
  \bibfield  {author} {\bibinfo {author} {\bibfnamefont {W.}~\bibnamefont
  {Kamleh}}, \bibinfo {author} {\bibfnamefont {D.~B.}\ \bibnamefont
  {Leinweber}},\ and\ \bibinfo {author} {\bibfnamefont {A.}~\bibnamefont
  {Virgili}},\ }\href {https://doi.org/10.1103/PhysRevD.110.L051502} {\bibfield
   {journal} {\bibinfo  {journal} {Phys. Rev. D}\ }\textbf {\bibinfo {volume}
  {110}},\ \bibinfo {pages} {L051502} (\bibinfo {year} {2024})},\ \Eprint
  {https://arxiv.org/abs/2305.18690} {arXiv:2305.18690 [hep-lat]} \BibitemShut
  {NoStop}%
\bibitem [{\citenamefont {Dehghan}\ \emph {et~al.}(2024)\citenamefont
  {Dehghan}, \citenamefont {Golubich}, \citenamefont {H\"ollwieser},\ and\
  \citenamefont {Faber}}]{Dehghan:2024rly}%
  \BibitemOpen
  \bibfield  {author} {\bibinfo {author} {\bibfnamefont {Z.}~\bibnamefont
  {Dehghan}}, \bibinfo {author} {\bibfnamefont {R.}~\bibnamefont {Golubich}},
  \bibinfo {author} {\bibfnamefont {R.}~\bibnamefont {H\"ollwieser}},\ and\
  \bibinfo {author} {\bibfnamefont {M.}~\bibnamefont {Faber}},\ }\href
  {https://doi.org/10.1103/PhysRevD.110.014501} {\bibfield  {journal} {\bibinfo
   {journal} {Phys. Rev. D}\ }\textbf {\bibinfo {volume} {110}},\ \bibinfo
  {pages} {014501} (\bibinfo {year} {2024})},\ \Eprint
  {https://arxiv.org/abs/2404.13304} {arXiv:2404.13304 [hep-lat]} \BibitemShut
  {NoStop}%
\bibitem [{\citenamefont {Gliozzi}\ \emph {et~al.}(2002)\citenamefont
  {Gliozzi}, \citenamefont {Panero},\ and\ \citenamefont
  {Provero}}]{Gliozzi:2002ht}%
  \BibitemOpen
  \bibfield  {author} {\bibinfo {author} {\bibfnamefont {F.}~\bibnamefont
  {Gliozzi}}, \bibinfo {author} {\bibfnamefont {M.}~\bibnamefont {Panero}},\
  and\ \bibinfo {author} {\bibfnamefont {P.}~\bibnamefont {Provero}},\ }\href
  {https://doi.org/10.1103/PhysRevD.66.017501} {\bibfield  {journal} {\bibinfo
  {journal} {Phys. Rev. D}\ }\textbf {\bibinfo {volume} {66}},\ \bibinfo
  {pages} {017501} (\bibinfo {year} {2002})},\ \Eprint
  {https://arxiv.org/abs/hep-lat/0204030} {arXiv:hep-lat/0204030} \BibitemShut
  {NoStop}%
\bibitem [{\citenamefont {Balian}\ \emph {et~al.}(1975)\citenamefont {Balian},
  \citenamefont {Drouffe},\ and\ \citenamefont {Itzykson}}]{Balian:1974ir}%
  \BibitemOpen
  \bibfield  {author} {\bibinfo {author} {\bibfnamefont {R.}~\bibnamefont
  {Balian}}, \bibinfo {author} {\bibfnamefont {J.~M.}\ \bibnamefont
  {Drouffe}},\ and\ \bibinfo {author} {\bibfnamefont {C.}~\bibnamefont
  {Itzykson}},\ }\href {https://doi.org/10.1103/PhysRevD.11.2098} {\bibfield
  {journal} {\bibinfo  {journal} {Phys. Rev. D}\ }\textbf {\bibinfo {volume}
  {11}},\ \bibinfo {pages} {2098} (\bibinfo {year} {1975})}\BibitemShut
  {NoStop}%
\bibitem [{\citenamefont {Bhanot}\ and\ \citenamefont
  {Creutz}(1980)}]{Bhanot:1980pc}%
  \BibitemOpen
  \bibfield  {author} {\bibinfo {author} {\bibfnamefont {G.}~\bibnamefont
  {Bhanot}}\ and\ \bibinfo {author} {\bibfnamefont {M.}~\bibnamefont
  {Creutz}},\ }\href {https://doi.org/10.1103/PhysRevD.21.2892} {\bibfield
  {journal} {\bibinfo  {journal} {Phys. Rev. D}\ }\textbf {\bibinfo {volume}
  {21}},\ \bibinfo {pages} {2892} (\bibinfo {year} {1980})}\BibitemShut
  {NoStop}%
\bibitem [{\citenamefont {Agrawal}\ \emph {et~al.}(2024)\citenamefont
  {Agrawal}, \citenamefont {Cugliandolo}, \citenamefont {Faoro}, \citenamefont
  {Ioffe},\ and\ \citenamefont {Picco}}]{Agrawal:2024mra}%
  \BibitemOpen
  \bibfield  {author} {\bibinfo {author} {\bibfnamefont {R.}~\bibnamefont
  {Agrawal}}, \bibinfo {author} {\bibfnamefont {L.~F.}\ \bibnamefont
  {Cugliandolo}}, \bibinfo {author} {\bibfnamefont {L.}~\bibnamefont {Faoro}},
  \bibinfo {author} {\bibfnamefont {L.~B.}\ \bibnamefont {Ioffe}},\ and\
  \bibinfo {author} {\bibfnamefont {M.}~\bibnamefont {Picco}},\ }\Eprint
  {https://arxiv.org/abs/2409.15123} {arXiv:2409.15123 [cond-mat.stat-mech]}
  (\bibinfo {year} {2024}),\ \bibinfo {note} {unpublished}\BibitemShut
  {NoStop}%
\bibitem [{\citenamefont {Neuberger}(1998)}]{Neuberger:1997fp}%
  \BibitemOpen
  \bibfield  {author} {\bibinfo {author} {\bibfnamefont {H.}~\bibnamefont
  {Neuberger}},\ }\href {https://doi.org/10.1016/S0370-2693(97)01368-3}
  {\bibfield  {journal} {\bibinfo  {journal} {Phys. Lett. B}\ }\textbf
  {\bibinfo {volume} {417}},\ \bibinfo {pages} {141} (\bibinfo {year}
  {1998})},\ \Eprint {https://arxiv.org/abs/hep-lat/9707022}
  {arXiv:hep-lat/9707022} \BibitemShut {NoStop}%
\bibitem [{\citenamefont {Bazavov}\ \emph {et~al.}(2010)\citenamefont {Bazavov}
  \emph {et~al.}}]{MILC:2009mpl}%
  \BibitemOpen
  \bibfield  {author} {\bibinfo {author} {\bibfnamefont {A.}~\bibnamefont
  {Bazavov}} \emph {et~al.} (\bibinfo {collaboration} {MILC}),\ }\href
  {https://doi.org/10.1103/RevModPhys.82.1349} {\bibfield  {journal} {\bibinfo
  {journal} {Rev. Mod. Phys.}\ }\textbf {\bibinfo {volume} {82}},\ \bibinfo
  {pages} {1349} (\bibinfo {year} {2010})},\ \Eprint
  {https://arxiv.org/abs/0903.3598} {arXiv:0903.3598 [hep-lat]} \BibitemShut
  {NoStop}%
\bibitem [{\citenamefont {Wu}(1982)}]{RevModPhys.54.235}%
  \BibitemOpen
  \bibfield  {author} {\bibinfo {author} {\bibfnamefont {F.~Y.}\ \bibnamefont
  {Wu}},\ }\href {https://doi.org/10.1103/RevModPhys.54.235} {\bibfield
  {journal} {\bibinfo  {journal} {Rev. Mod. Phys.}\ }\textbf {\bibinfo {volume}
  {54}},\ \bibinfo {pages} {235} (\bibinfo {year} {1982})}\BibitemShut
  {NoStop}%
\bibitem [{\citenamefont {Stauffer}\ and\ \citenamefont
  {Aharony}(1992)}]{percolation}%
  \BibitemOpen
  \bibfield  {author} {\bibinfo {author} {\bibfnamefont {D.}~\bibnamefont
  {Stauffer}}\ and\ \bibinfo {author} {\bibfnamefont {A.}~\bibnamefont
  {Aharony}},\ }\href {https://doi.org/10.1201/9781315274386} {\emph {\bibinfo
  {title} {{Introduction To Percolation Theory: Second Edition}}}}\ (\bibinfo
  {publisher} {Taylor \& Francis},\ \bibinfo {address} {London},\ \bibinfo
  {year} {1992})\BibitemShut {NoStop}%
\bibitem [{\citenamefont {Smirnov}\ and\ \citenamefont
  {Werner}(2001)}]{smirnov2001}%
  \BibitemOpen
  \bibfield  {author} {\bibinfo {author} {\bibfnamefont {S.}~\bibnamefont
  {Smirnov}}\ and\ \bibinfo {author} {\bibfnamefont {W.}~\bibnamefont
  {Werner}},\ }\href {https://doi.org/10.4310/MRL.2001.v8.n6.a4} {\bibfield
  {journal} {\bibinfo  {journal} {Math. Res. Lett.}\ }\textbf {\bibinfo
  {volume} {8}},\ \bibinfo {pages} {729} (\bibinfo {year} {2001})},\ \Eprint
  {https://arxiv.org/abs/math/0109120} {arXiv:math/0109120 [math.PR]}
  \BibitemShut {NoStop}%
\bibitem [{\citenamefont {Grimmett}\ and\ \citenamefont
  {Manolescu}(2013)}]{10.1214/11-AOP740}%
  \BibitemOpen
  \bibfield  {author} {\bibinfo {author} {\bibfnamefont {G.~R.}\ \bibnamefont
  {Grimmett}}\ and\ \bibinfo {author} {\bibfnamefont {I.}~\bibnamefont
  {Manolescu}},\ }\href {https://doi.org/10.1214/11-AOP740} {\bibfield
  {journal} {\bibinfo  {journal} {Ann. Probab.}\ }\textbf {\bibinfo {volume}
  {41}},\ \bibinfo {pages} {3261 } (\bibinfo {year} {2013})},\ \Eprint
  {https://arxiv.org/abs/1108.2784} {arXiv:1108.2784 [math.PR]} \BibitemShut
  {NoStop}%
\bibitem [{\citenamefont {Al'tshuler}\ and\ \citenamefont {Shklovski{\u
  i}}(1986)}]{altshuler1986repulsion}%
  \BibitemOpen
  \bibfield  {author} {\bibinfo {author} {\bibfnamefont {B.~L.}\ \bibnamefont
  {Al'tshuler}}\ and\ \bibinfo {author} {\bibfnamefont {B.~I.}\ \bibnamefont
  {Shklovski{\u i}}},\ }\href {http://jetp.ras.ru/cgi-bin/dn/e_064_01_0127.pdf}
  {\bibfield  {journal} {\bibinfo  {journal} {Sov. Phys. JETP}\ }\textbf
  {\bibinfo {volume} {64}},\ \bibinfo {pages} {127} (\bibinfo {year}
  {1986})}\BibitemShut {NoStop}%
\bibitem [{\citenamefont {Mehta}(2004)}]{mehta2004random}%
  \BibitemOpen
  \bibfield  {author} {\bibinfo {author} {\bibfnamefont {M.~L.}\ \bibnamefont
  {Mehta}},\ }\href@noop {} {\emph {\bibinfo {title} {{Random matrices}}}},\
  \bibinfo {edition} {3rd}\ ed.,\ \bibinfo {series} {Pure and Applied
  Mathematics}, Vol.\ \bibinfo {volume} {142}\ (\bibinfo  {publisher} {Academic
  Press},\ \bibinfo {address} {New York},\ \bibinfo {year} {2004})\BibitemShut
  {NoStop}%
\bibitem [{\citenamefont {Verbaarschot}\ and\ \citenamefont
  {Wettig}(2000)}]{Verbaarschot:2000dy}%
  \BibitemOpen
  \bibfield  {author} {\bibinfo {author} {\bibfnamefont {J.~J.~M.}\
  \bibnamefont {Verbaarschot}}\ and\ \bibinfo {author} {\bibfnamefont
  {T.}~\bibnamefont {Wettig}},\ }\href
  {https://doi.org/10.1146/annurev.nucl.50.1.343} {\bibfield  {journal}
  {\bibinfo  {journal} {Annu. Rev. Nucl. Part. Sci.}\ }\textbf {\bibinfo
  {volume} {50}},\ \bibinfo {pages} {343} (\bibinfo {year} {2000})},\ \Eprint
  {https://arxiv.org/abs/hep-ph/0003017} {arXiv:hep-ph/0003017 [hep-ph]}
  \BibitemShut {NoStop}%
%%CITATION = HEP-PH/0003017;%%
\bibitem [{\citenamefont {Shklovskii}\ \emph {et~al.}(1993)\citenamefont
  {Shklovskii}, \citenamefont {Shapiro}, \citenamefont {Sears}, \citenamefont
  {Lambrianides},\ and\ \citenamefont {Shore}}]{Shklovskii:1993zz}%
  \BibitemOpen
  \bibfield  {author} {\bibinfo {author} {\bibfnamefont {B.~I.}\ \bibnamefont
  {Shklovskii}}, \bibinfo {author} {\bibfnamefont {B.}~\bibnamefont {Shapiro}},
  \bibinfo {author} {\bibfnamefont {B.~R.}\ \bibnamefont {Sears}}, \bibinfo
  {author} {\bibfnamefont {P.}~\bibnamefont {Lambrianides}},\ and\ \bibinfo
  {author} {\bibfnamefont {H.~B.}\ \bibnamefont {Shore}},\ }\href
  {https://doi.org/10.1103/PhysRevB.47.11487} {\bibfield  {journal} {\bibinfo
  {journal} {Phys. Rev. B}\ }\textbf {\bibinfo {volume} {47}},\ \bibinfo
  {pages} {11487} (\bibinfo {year} {1993})}\BibitemShut {NoStop}%
\bibitem [{\citenamefont {Brody}(1973)}]{brody1973statistical}%
  \BibitemOpen
  \bibfield  {author} {\bibinfo {author} {\bibfnamefont {T.~A.}\ \bibnamefont
  {Brody}},\ }\href {https://doi.org/10.1007/BF02727859} {\bibfield  {journal}
  {\bibinfo  {journal} {Lett. Nuovo Cimento (1971-1985)}\ }\textbf {\bibinfo
  {volume} {7}},\ \bibinfo {pages} {482} (\bibinfo {year} {1973})}\BibitemShut
  {NoStop}%
\bibitem [{\citenamefont {Scaramazza}\ \emph {et~al.}(2016)\citenamefont
  {Scaramazza}, \citenamefont {Shastry},\ and\ \citenamefont
  {Yuzbashyan}}]{scaramazza2016integrable}%
  \BibitemOpen
  \bibfield  {author} {\bibinfo {author} {\bibfnamefont {J.~A.}\ \bibnamefont
  {Scaramazza}}, \bibinfo {author} {\bibfnamefont {B.~S.}\ \bibnamefont
  {Shastry}},\ and\ \bibinfo {author} {\bibfnamefont {E.~A.}\ \bibnamefont
  {Yuzbashyan}},\ }\href {https://doi.org/10.1103/PhysRevE.94.032106}
  {\bibfield  {journal} {\bibinfo  {journal} {Phys. Rev. E}\ }\textbf {\bibinfo
  {volume} {94}},\ \bibinfo {pages} {032106} (\bibinfo {year} {2016})},\
  \Eprint {https://arxiv.org/abs/1604.01691} {arXiv:1604.01691
  [cond-mat.mes-hall]} \BibitemShut {NoStop}%
\bibitem [{\citenamefont {Kalmeyer}\ \emph {et~al.}(1993)\citenamefont
  {Kalmeyer}, \citenamefont {Wei}, \citenamefont {Arovas},\ and\ \citenamefont
  {Zhang}}]{PhysRevB.48.11095}%
  \BibitemOpen
  \bibfield  {author} {\bibinfo {author} {\bibfnamefont {V.}~\bibnamefont
  {Kalmeyer}}, \bibinfo {author} {\bibfnamefont {D.}~\bibnamefont {Wei}},
  \bibinfo {author} {\bibfnamefont {D.~P.}\ \bibnamefont {Arovas}},\ and\
  \bibinfo {author} {\bibfnamefont {S.}~\bibnamefont {Zhang}},\ }\href
  {https://doi.org/10.1103/PhysRevB.48.11095} {\bibfield  {journal} {\bibinfo
  {journal} {Phys. Rev. B}\ }\textbf {\bibinfo {volume} {48}},\ \bibinfo
  {pages} {11095} (\bibinfo {year} {1993})}\BibitemShut {NoStop}%
\bibitem [{\citenamefont {Zhang}\ and\ \citenamefont
  {Arovas}(1994)}]{PhysRevLett.72.1886}%
  \BibitemOpen
  \bibfield  {author} {\bibinfo {author} {\bibfnamefont {S.-C.}\ \bibnamefont
  {Zhang}}\ and\ \bibinfo {author} {\bibfnamefont {D.~P.}\ \bibnamefont
  {Arovas}},\ }\href {https://doi.org/10.1103/PhysRevLett.72.1886} {\bibfield
  {journal} {\bibinfo  {journal} {Phys. Rev. Lett.}\ }\textbf {\bibinfo
  {volume} {72}},\ \bibinfo {pages} {1886} (\bibinfo {year} {1994})},\ \Eprint
  {https://arxiv.org/abs/cond-mat/9312010} {arXiv:cond-mat/9312010 [cond-mat]}
  \BibitemShut {NoStop}%
\bibitem [{\citenamefont {Liu}\ \emph {et~al.}(1999{\natexlab{a}})\citenamefont
  {Liu}, \citenamefont {Chen}, \citenamefont {Xiong},\ and\ \citenamefont
  {Xing}}]{PhysRevB.60.5295}%
  \BibitemOpen
  \bibfield  {author} {\bibinfo {author} {\bibfnamefont {W.-S.}\ \bibnamefont
  {Liu}}, \bibinfo {author} {\bibfnamefont {Y.}~\bibnamefont {Chen}}, \bibinfo
  {author} {\bibfnamefont {S.-J.}\ \bibnamefont {Xiong}},\ and\ \bibinfo
  {author} {\bibfnamefont {D.~Y.}\ \bibnamefont {Xing}},\ }\href
  {https://doi.org/10.1103/PhysRevB.60.5295} {\bibfield  {journal} {\bibinfo
  {journal} {Phys. Rev. B}\ }\textbf {\bibinfo {volume} {60}},\ \bibinfo
  {pages} {5295} (\bibinfo {year} {1999}{\natexlab{a}})}\BibitemShut {NoStop}%
\bibitem [{\citenamefont {Liu}\ \emph {et~al.}(1999{\natexlab{b}})\citenamefont
  {Liu}, \citenamefont {Chen},\ and\ \citenamefont
  {Xiong}}]{Wen-ShengLiu_1999}%
  \BibitemOpen
  \bibfield  {author} {\bibinfo {author} {\bibfnamefont {W.-S.}\ \bibnamefont
  {Liu}}, \bibinfo {author} {\bibfnamefont {T.}~\bibnamefont {Chen}},\ and\
  \bibinfo {author} {\bibfnamefont {S.-J.}\ \bibnamefont {Xiong}},\ }\href
  {https://doi.org/10.1088/0953-8984/11/36/306} {\bibfield  {journal} {\bibinfo
   {journal} {J. Phys. Condens. Matter}\ }\textbf {\bibinfo {volume} {11}},\
  \bibinfo {pages} {6883} (\bibinfo {year} {1999}{\natexlab{b}})}\BibitemShut
  {NoStop}%
\bibitem [{\citenamefont {Xie}\ \emph {et~al.}(1998)\citenamefont {Xie},
  \citenamefont {Wang},\ and\ \citenamefont {Liu}}]{xie1998kosterlitz}%
  \BibitemOpen
  \bibfield  {author} {\bibinfo {author} {\bibfnamefont {X.~C.}\ \bibnamefont
  {Xie}}, \bibinfo {author} {\bibfnamefont {X.~R.}\ \bibnamefont {Wang}},\ and\
  \bibinfo {author} {\bibfnamefont {D.~Z.}\ \bibnamefont {Liu}},\ }\href
  {https://doi.org/10.1103/PhysRevLett.80.3563} {\bibfield  {journal} {\bibinfo
   {journal} {Phys. Rev. Lett.}\ }\textbf {\bibinfo {volume} {80}},\ \bibinfo
  {pages} {3563} (\bibinfo {year} {1998})}\BibitemShut {NoStop}%
\bibitem [{\citenamefont {Zhang}\ \emph {et~al.}(2009)\citenamefont {Zhang},
  \citenamefont {Hu}, \citenamefont {Bernevig}, \citenamefont {Wang},
  \citenamefont {Xie},\ and\ \citenamefont {Liu}}]{zhang2009localization}%
  \BibitemOpen
  \bibfield  {author} {\bibinfo {author} {\bibfnamefont {Y.-Y.}\ \bibnamefont
  {Zhang}}, \bibinfo {author} {\bibfnamefont {J.}~\bibnamefont {Hu}}, \bibinfo
  {author} {\bibfnamefont {B.~A.}\ \bibnamefont {Bernevig}}, \bibinfo {author}
  {\bibfnamefont {X.~R.}\ \bibnamefont {Wang}}, \bibinfo {author}
  {\bibfnamefont {X.~C.}\ \bibnamefont {Xie}},\ and\ \bibinfo {author}
  {\bibfnamefont {W.~M.}\ \bibnamefont {Liu}},\ }\href
  {https://doi.org/10.1103/PhysRevLett.102.106401} {\bibfield  {journal}
  {\bibinfo  {journal} {Phys. Rev. Lett.}\ }\textbf {\bibinfo {volume} {102}},\
  \bibinfo {pages} {106401} (\bibinfo {year} {2009})},\ \Eprint
  {https://arxiv.org/abs/0810.1996} {0810.1996} \BibitemShut {NoStop}%
\bibitem [{\citenamefont {Lehoucq}\ \emph {et~al.}(1998)\citenamefont
  {Lehoucq}, \citenamefont {Sorensen},\ and\ \citenamefont
  {Yang}}]{lehoucq1998arpack}%
  \BibitemOpen
  \bibfield  {author} {\bibinfo {author} {\bibfnamefont {R.~B.}\ \bibnamefont
  {Lehoucq}}, \bibinfo {author} {\bibfnamefont {D.~C.}\ \bibnamefont
  {Sorensen}},\ and\ \bibinfo {author} {\bibfnamefont {C.}~\bibnamefont
  {Yang}},\ }\href@noop {} {\emph {\bibinfo {title} {{ARPACK users' guide:
  solution of large-scale eigenvalue problems with implicitly restarted Arnoldi
  methods}}}}\ (\bibinfo  {publisher} {SIAM},\ \bibinfo {address}
  {Philadelphia},\ \bibinfo {year} {1998})\BibitemShut {NoStop}%
\bibitem [{\citenamefont {Lepage}\ \emph {et~al.}(2002)\citenamefont {Lepage},
  \citenamefont {Clark}, \citenamefont {Davies}, \citenamefont {Hornbostel},
  \citenamefont {Mackenzie}, \citenamefont {Morningstar},\ and\ \citenamefont
  {Trottier}}]{Lepage:2001ym}%
  \BibitemOpen
  \bibfield  {author} {\bibinfo {author} {\bibfnamefont {G.~P.}\ \bibnamefont
  {Lepage}}, \bibinfo {author} {\bibfnamefont {B.}~\bibnamefont {Clark}},
  \bibinfo {author} {\bibfnamefont {C.~T.~H.}\ \bibnamefont {Davies}}, \bibinfo
  {author} {\bibfnamefont {K.}~\bibnamefont {Hornbostel}}, \bibinfo {author}
  {\bibfnamefont {P.~B.}\ \bibnamefont {Mackenzie}}, \bibinfo {author}
  {\bibfnamefont {C.}~\bibnamefont {Morningstar}},\ and\ \bibinfo {author}
  {\bibfnamefont {H.}~\bibnamefont {Trottier}},\ }\href
  {https://doi.org/10.1016/S0920-5632(01)01638-3} {\bibfield  {journal}
  {\bibinfo  {journal} {Nucl. Phys. B Proc. Suppl.}\ }\textbf {\bibinfo
  {volume} {106}},\ \bibinfo {pages} {12} (\bibinfo {year} {2002})},\ \Eprint
  {https://arxiv.org/abs/hep-lat/0110175} {arXiv:hep-lat/0110175} \BibitemShut
  {NoStop}%
\bibitem [{\citenamefont {James}\ and\ \citenamefont
  {Roos}(1975)}]{James:1975dr}%
  \BibitemOpen
  \bibfield  {author} {\bibinfo {author} {\bibfnamefont {F.}~\bibnamefont
  {James}}\ and\ \bibinfo {author} {\bibfnamefont {M.}~\bibnamefont {Roos}},\
  }\href {https://doi.org/10.1016/0010-4655(75)90039-9} {\bibfield  {journal}
  {\bibinfo  {journal} {Comput. Phys. Commun.}\ }\textbf {\bibinfo {volume}
  {10}},\ \bibinfo {pages} {343} (\bibinfo {year} {1975})}\BibitemShut
  {NoStop}%
\bibitem [{\citenamefont {Anderson}\ \emph {et~al.}(1999)\citenamefont
  {Anderson}, \citenamefont {Bai}, \citenamefont {Bischof}, \citenamefont
  {Blackford}, \citenamefont {Demmel}, \citenamefont {Dongarra}, \citenamefont
  {Du~Croz}, \citenamefont {Greenbaum}, \citenamefont {Hammarling},
  \citenamefont {McKenney},\ and\ \citenamefont
  {Sorensen}}]{anderson1999lapack}%
  \BibitemOpen
  \bibfield  {author} {\bibinfo {author} {\bibfnamefont {E.}~\bibnamefont
  {Anderson}}, \bibinfo {author} {\bibfnamefont {Z.}~\bibnamefont {Bai}},
  \bibinfo {author} {\bibfnamefont {C.}~\bibnamefont {Bischof}}, \bibinfo
  {author} {\bibfnamefont {S.}~\bibnamefont {Blackford}}, \bibinfo {author}
  {\bibfnamefont {J.}~\bibnamefont {Demmel}}, \bibinfo {author} {\bibfnamefont
  {J.}~\bibnamefont {Dongarra}}, \bibinfo {author} {\bibfnamefont
  {J.}~\bibnamefont {Du~Croz}}, \bibinfo {author} {\bibfnamefont
  {A.}~\bibnamefont {Greenbaum}}, \bibinfo {author} {\bibfnamefont
  {S.}~\bibnamefont {Hammarling}}, \bibinfo {author} {\bibfnamefont
  {A.}~\bibnamefont {McKenney}},\ and\ \bibinfo {author} {\bibfnamefont
  {D.}~\bibnamefont {Sorensen}},\ }\href@noop {} {\emph {\bibinfo {title}
  {LAPACK Users' Guide}}}\ (\bibinfo  {publisher} {SIAM},\ \bibinfo {address}
  {Philadelphia},\ \bibinfo {year} {1999})\BibitemShut {NoStop}%
\bibitem [{\citenamefont {Ashcroft}\ and\ \citenamefont
  {Mermin}(1976)}]{ashcroft_mermin1976}%
  \BibitemOpen
  \bibfield  {author} {\bibinfo {author} {\bibfnamefont {N.~W.}\ \bibnamefont
  {Ashcroft}}\ and\ \bibinfo {author} {\bibfnamefont {N.~D.}\ \bibnamefont
  {Mermin}},\ }\href@noop {} {\emph {\bibinfo {title} {Solid State Physics}}}\
  (\bibinfo  {publisher} {Saunders College Publishing},\ \bibinfo {address}
  {Philadelphia},\ \bibinfo {year} {1976})\BibitemShut {NoStop}%
\end{thebibliography}%

\end{document}